\pdfoutput=1
\documentclass[11pt,twoside,a4paper,cmspaper,final,collab]{cms-tdr}
\def\svnVersion{642c877-D}\def\svnDate{2021/07/30}\def\cmsCernNoTag{CERN-EP-2020-202}\def\cmsCernDate{\today}\def\cmsMessage{"Published in Physical Review D as \href{http://dx.doi.org/10.1103/PhysRevD.104.012015}{\doi{10.1103/PhysRevD.104.012015}.}"}
\begin{document}\cmsNoteHeader{EXO-19-021}

\newlength\cmsTabSkip\setlength{\cmsTabSkip}{1ex}
\ifthenelse{\boolean{cms@external}}{\newlength\cmsFigWidth\setlength{\cmsFigWidth}{0.49\textwidth}}{\newlength\cmsFigWidth\setlength{\cmsFigWidth}{0.75\textwidth}}
\ifthenelse{\boolean{cms@external}}{\providecommand{\cmsLeft}{Upper\xspace}}{\providecommand{\cmsLeft}{Left\xspace}}
\ifthenelse{\boolean{cms@external}}{\providecommand{\cmsRight}{Lower\xspace}}{\providecommand{\cmsRight}{Right\xspace}}
\newcommand{\ip}{\ensuremath{\mathrm{IP_{2D}}}\xspace}
\newcommand{\ipsig}{\ensuremath{\mathrm{Sig[IP_{2D}]}}\xspace}
\providecommand{\cmsTable}[1]{\resizebox{\textwidth}{!}{#1}}
\cmsNoteHeader{EXO-19-021} 
\title{Search for long-lived particles using displaced jets in proton-proton collisions at \texorpdfstring{$\sqrt{s} = 13\TeV$}{sqrt(s) = 13 TeV}}

\date{\today}

\abstract{
An inclusive search is presented for long-lived particles using displaced jets. The search uses a data sample collected with the CMS detector at the CERN LHC in 2017 and 2018, from 
proton-proton collisions at a center-of-mass energy of 13\TeV. The results of this search are combined with those of a previous search using a data sample collected with the CMS 
detector in 2016, yielding a total integrated luminosity of 132\fbinv. The analysis searches for the distinctive topology of 
displaced tracks and displaced vertices associated with a dijet system. For a simplified model, where pair-produced long-lived neutral particles decay into quark-antiquark pairs, 
pair production cross sections larger than 0.07\unit{fb} are excluded at 95\% confidence level (\CL) for long-lived particle masses larger than 500\GeV and mean proper decay 
lengths between 2 and 250\mm. For a model where the standard model-like Higgs boson decays to two long-lived scalar particles that each 
decays to a quark-antiquark pair, branching fractions larger than 1\% are excluded at 95\%~\CL for mean proper decay 
lengths between 1\mm and 340\mm. A group of supersymmetric models with pair-produced long-lived gluinos or top squarks decaying into various final-state topologies 
containing displaced jets is also tested. Gluino masses up to 2500\GeV and top squark masses up to 1600\GeV are excluded at 95\%~\CL for mean proper decay lengths 
between 3 and 300\mm. The highest lower bounds on mass reach 2600\GeV for long-lived gluinos and 1800\GeV for long-lived top squarks. These are the most stringent 
limits to date on these models. 

}

\hypersetup{
pdfauthor={CMS Collaboration},
pdftitle={Search for long-lived particles decaying into displaced jets},
pdfsubject={CMS},
pdfkeywords={CMS, BSM, long-lived particles, displaced jets}}

\maketitle 

\section{Introduction}

The existence of long-lived particles (LLPs) that have macroscopic decay lengths is a common feature in both the standard model (SM) and beyond-the-SM (BSM) 
scenarios. There are numerous alternative BSM physics cases for the production of LLPs at the CERN LHC. Examples include, but are not limited to: split 
supersymmetry (SUSY)~\cite{ArkaniHamed:2004fb,Giudice:2004tc,Hewett:2004nw,ArkaniHamed:2004yi,Gambino:2005eh,Arvanitaki:2012ps,ArkaniHamed:2012gw}, where the gluino decays are suppressed by heavy scalars; SUSY with weak $R$-parity violation (RPV)~\cite{Fayet:1974pd,Farrar:1978xj,Weinberg:1981wj,Hall:1983id,Barbier:2004ez}, where the decays of the lightest supersymmetric particle are 
suppressed by small RPV couplings; SUSY with gauge-mediated SUSY breaking (GMSB)~\cite{GIUDICE1999419,Meade:2008wd,Buican:2008ws}, where the decays of the next-to-lightest supersymmetric particle are suppressed by a large SUSY breaking 
scale; ``stealth SUSY"~\cite{Fan:2011yu,Fan:2012jf}; ``hidden valley" models~\cite{Strassler:2006im,Strassler:2006ri,Han:2007ae}; dark matter models~\cite{Kaplan:2009ag,Hall:2009bx,Kim:2013ivd,Cui:2014twa,Co:2015pka,Calibbi:2018fqf,Cui:2011ab,Cui:2012jh}; models with heavy neutral leptons that have small mixing parameters~\cite{Atre:2009rg,Drewes:2013gca,Deppisch:2015qwa,Cai:2017mow}; and 
models incorporating neutral naturalness~\cite{Chacko:2005pe,Cai:2008au,Craig:2014aea,Craig:2015pha,Curtin:2015fna,Alipour-fard:2018mre}. In the examples listed above, it is very common 
for the LLPs to further decay into final states containing jets, giving rise to displaced-jets signatures.

Given the large variety of the BSM scenarios that lead to displaced-jets signatures, it is important to make the displaced-jets 
search as model independent as possible. In this paper, we present an inclusive search for LLPs decaying into jets, 
with at least one LLP having a decay vertex within the tracker acceptance, but which is displaced from the production vertex 
by up to 550\mm in the plane transverse to the beam direction. 
The search looks for a pair of jets known as dijets, where the jets are clustered from energy deposits in the calorimeters. For jets arising from the decay of an LLP, the associated tracks are usually 
displaced from the primary vertices (PVs), and the decay vertex can be reconstructed from the displaced tracks. The properties of the tracks and the decay 
vertex can provide discrimination power to distinguish long-lived signals from SM backgrounds. As mentioned above, a large number of models predict LLPs decaying 
into displaced jets. Our tests for some of these will be discussed in detail in Section~\ref{sec: trigger}. 

Events used in this analysis were collected with the CMS detector~\cite{Chatrchyan:2008aa} at the
LHC from proton-proton ($\Pp\Pp$) collisions at a center-of-mass energy of 13\TeV in 2017 and 2018, corresponding to an
integrated luminosity of 95.9\fbinv. The results
are combined with those of a previous displaced-jets search using the events collected in 2016~\cite{Sirunyan:2018vlw}, yielding a total integrated luminosity
of 132\fbinv. For the models that were not studied in the 2016 displaced-jets search, additional simulated signal samples have been produced following the 2016 run condition 
of the CMS detector.  These additional samples are then processed with the reconstruction and selection procedures described in Ref.~\cite{Sirunyan:2018vlw} to compute the additional signal yields 
and systematic uncertainties for the 2016 data that are used in the combination.   

Compared to the 2016 displaced-jets search, a set of new 
techniques that significantly improves the sensitivity to long-lived signatures is implemented in this analysis. The new techniques include one additional dedicated trigger aimed at selecting events containing displaced jets to 
recover efficiencies for high-mass LLPs, an auxiliary nuclear interactions (NIs) veto map to improve background rejection, a dedicated variable based on 
the sum of signed impact parameters of the tracks assigned to the displaced vertex, and the use of machine learning techniques to improve signal-to-background discrimination. With these new techniques, 
compared to the 2016 search, we have reduced the background rate by approximately a factor of three, while significantly increasing the signal efficiencies for almost all signal points 
in different LLP models. Results of searches for 
similar LLP signatures with hadronic decays at $\sqrt{s}=13\TeV$ have also been reported by ATLAS~\cite{Aaboud:2017iio,Aaboud:2018aqj,Aaboud:2019opc,Aad:2020srt,Aad:2019xav} and CMS~\cite{Sirunyan:2017jdo,Sirunyan:2018pwn,Sirunyan:2019gut}. 

The paper is organized as follows. A brief description of the CMS detector is introduced in Section~\ref{sec: det}. The data and the simulated samples are described in Section~\ref{sec: trigger}. 
Section~\ref{sec: presel} details the event reconstruction and the preselection criteria. Section~\ref{sec: back} describes the event selections and the background estimation methods. 
The systematic uncertainties are summarized in Section~\ref{sec: sys}. The observation and the interpretation of the results are described in Section~\ref{sec: results}. The paper is 
summarized in Section~\ref{sec: summary}. 

\section{The CMS detector}\label{sec: det}

The central feature of the CMS apparatus is a superconducting solenoid of 6\unit{m} internal diameter, providing a magnetic field of 3.8\unit{T}. Within the
solenoid volume are a silicon pixel and strip tracker, a lead tungstate crystal electromagnetic calorimeter (ECAL), and a brass and scintillator
hadron calorimeter (HCAL), each composed of a barrel and two endcap detectors. Muons are detected in gas-ionization chambers embedded in the
steel flux-return yoke outside the solenoid.

The silicon tracker measures charged particles within the pseudorapidity range $\abs{\eta} < 2.5$. During the LHC run in 2017 and 2018, the silicon tracker consisted of 1856 silicon pixel and 15\,148 silicon strip detector modules, and it occupies a cylindrical volume around the interaction point with a length of 5.8\unit{m} and a diameter of 2.6\unit{m}. For nonisolated particles with $1<\pt<10\GeV$ and $\abs{\eta}<1.4$, the track resolutions are typically 1.5\% in \pt, and
25--75\mum in the transverse impact parameter~\cite{Chatrchyan:2014fea}.

In the region $\abs{\eta}<1.74$, the HCAL cells have widths of $\Delta\eta=0.087$ in pseudorapidity and $\Delta\phi=0.087$ in azimuth. In the $\eta$-$\phi$ plane,
and for $\abs{\eta}<1.48$, the HCAL cells map on to $5{\times}5$ arrays of ECAL crystals to form calorimeter towers projecting radially outward from the nominal interaction point. For $\abs{\eta}>1.74$, the coverage of the towers increases progressively to a maximum of 0.174 in
$\Delta\eta$ and $\Delta\phi$. Within each tower, the energy deposits in ECAL and HCAL cells are summed to define the calorimeter tower
energies, and are subsequently used to provide the energies and directions of hadronic jets.

Events of interest are selected using a two-tiered trigger system \cite{Khachatryan:2016bia}. The first level, composed of custom hardware processors, uses
information from the calorimeters and muon detectors to select events at a rate of around 100\unit{kHz} within a time interval of less than 4\mus.
The second level, known as the high-level trigger (HLT), consists of a farm of processors running a version of the full event reconstruction
software optimized for fast processing, and reduces the event rate to around 1\unit{kHz} before data storage.

A more detailed description of the CMS detector, together with a definition of the coordinate system used and the relevant kinematic variables,
can be found in Ref. \cite{Chatrchyan:2008aa}.

\section{Datasets and simulated samples}\label{sec: trigger}
Data were collected with two dedicated triggers aimed at selecting events containing displaced jets. At the HLT, jets are reconstructed from the energy deposits in the calorimeter towers, 
clustered using the anti-\kt algorithm~\cite{Cacciari:2008gp,Cacciari:2011ma} with a distance parameter of 0.4. In this process, the contribution from each calorimeter tower is assigned a momentum, the magnitude and the direction
of which are given by the energy measured in the tower and the coordinates of the tower. The raw jet energy is obtained from the sum of the
tower energies, and the raw jet momentum from the vector sum of the tower momenta, which results in a nonzero jet mass. The raw jet energies are
then corrected~\cite{Khachatryan:2016kdb} to establish a uniform relative response of the calorimeter in $\eta$ and a calibrated absolute response in transverse momentum \pt.

Identification of the PV is a prerequisite for the selection of displaced jets at the HLT. Events may contain multiple PVs, corresponding to multiple $\Pp\Pp$ collisions occurring in the same bunch crossing.
The candidate vertex with the largest value of summed physics-object $\pt^2$ is taken to be the primary $\Pp\Pp$ interaction vertex. The physics objects are the jets, clustered using the jet finding algorithm~\cite{Cacciari:2008gp,Cacciari:2011ma} with the tracks assigned to candidate vertices as inputs, and the associated missing transverse momentum, taken as the negative vector sum of the \pt of those jets. More details are given in Section~9.4.1 of Ref.~\cite{CMS-TDR-15-02}.

The first trigger, referred to as the ``displaced" trigger, requires $\HT>430\GeV$, where \HT is the scalar sum of the jet \pt for all jets that have $\pt>40\GeV$ and $\abs{\eta}<2.5$ in the event. The trigger also
requires the presence of at least two jets, with the following requirements 
satisfied for each jet:
\begin{itemize}
\item $\pt>40\GeV$ and $\abs{\eta}<2.0$;
\item at most two associated prompt tracks with $\pt>1\GeV$, where 
prompt tracks are those having a transverse impact parameter (\ip) with respect to the leading PV smaller than 1.0\mm; and
\item at least one associated displaced track with $\pt>1\GeV$, where a displaced track is a track having an \ip larger than 0.5\mm and an
impact parameter significance (\ipsig) larger than 5.0. The significance is defined as the ratio between the impact parameter and its uncertainty.
\end{itemize}
The second trigger, referred to as the ``inclusive" trigger, requires $\HT>650\GeV$ and the presence of at least two jets, each of them satisfying:
\begin{itemize}
\item $\pt>60\GeV$ and $\abs{\eta}<2.0$; and 
\item at most two associated prompt tracks with $\pt>1\GeV$.
\end{itemize}

{\tolerance=800 The displaced trigger is more efficient for low-mass LLPs, while the inclusive trigger is 
designed to recover the trigger efficiency for high-mass LLPs with small (${\lesssim}3\mm$) or large 
(${\gtrsim}300\mm$) mean proper decay lengths ($c\tau_{0}$). \par}

The background sources in this search include NIs between outgoing particles and detector material, long-lived SM hadrons, and misreconstructed displaced vertices 
formed by accidentally crossing tracks. The background events mainly arise from SM events containing jets produced through the 
strong interaction, referred to as quantum chromodynamics (QCD) multijet events. The QCD multijet Monte Carlo (MC) sample is simulated at leading order 
with \MGvATNLO~2.4.2~\cite{Alwall:2014hca}. Parton showering and hadronization  
are simulated with \PYTHIA~8.226~\cite{Sjostrand:2014zea}. The matching 
of jets from the matrix element calculations and parton shower jets is achieved 
using the MLM algorithm~\cite{Alwall:2007fs}. The \PYTHIA parameters for the 
underlying event modeling are set to be the CP5 tune~\cite{Sirunyan:2019dfx}. The set of parton
distribution functions (PDFs) used for the production is the NNPDF3.1~NNLO PDF set~\cite{Ball:2017nwa}. The QCD multijet MC sample is mainly used to inspire the analysis strategy and 
to estimate systematic uncertainties, while the background estimation for this search is purely determined from data.

\begin{figure*}[hbtp]
\centering
        \includegraphics[width=0.44\textwidth]{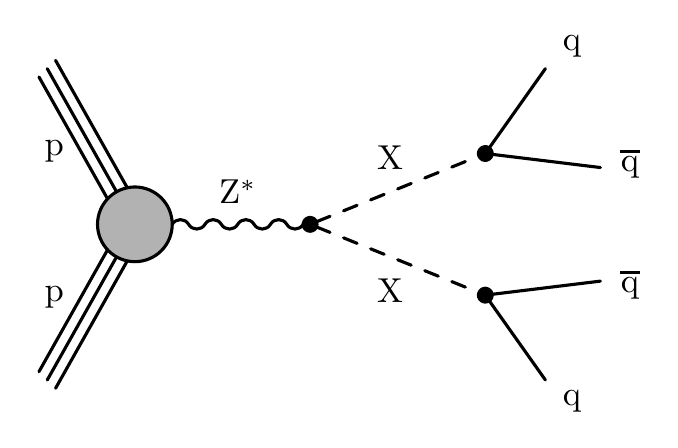}
        \includegraphics[width=0.44\textwidth]{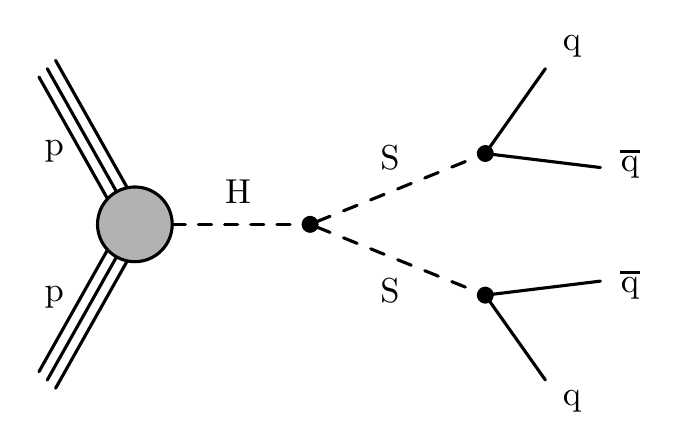}\\
        \includegraphics[width=0.44\textwidth]{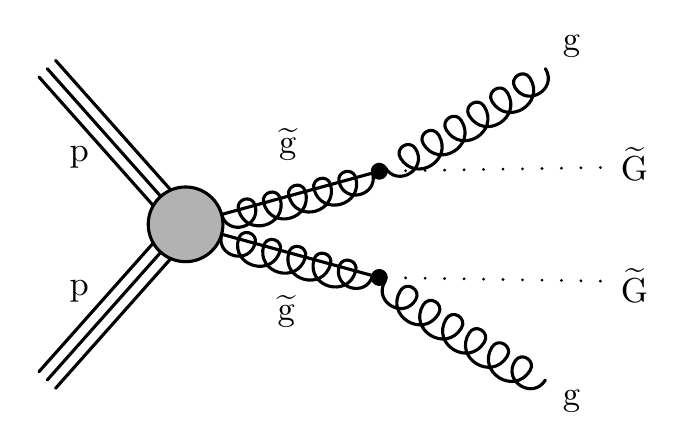}
        \includegraphics[width=0.44\textwidth]{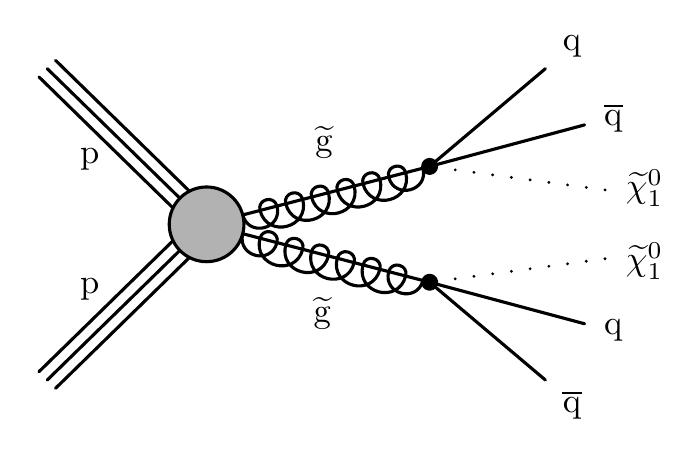}\\
        \includegraphics[width=0.44\textwidth]{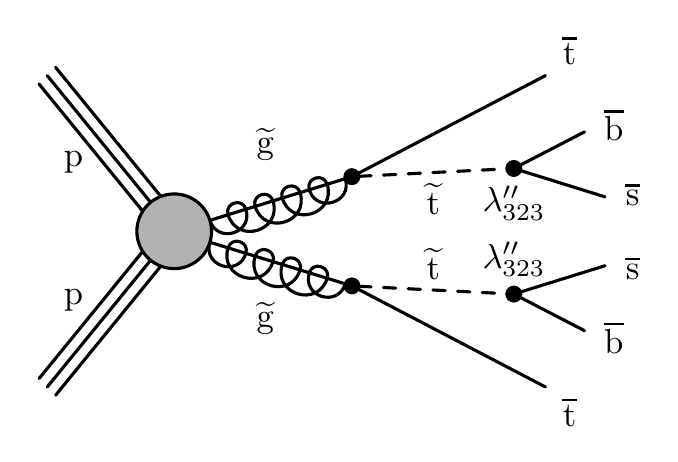}
        \includegraphics[width=0.44\textwidth]{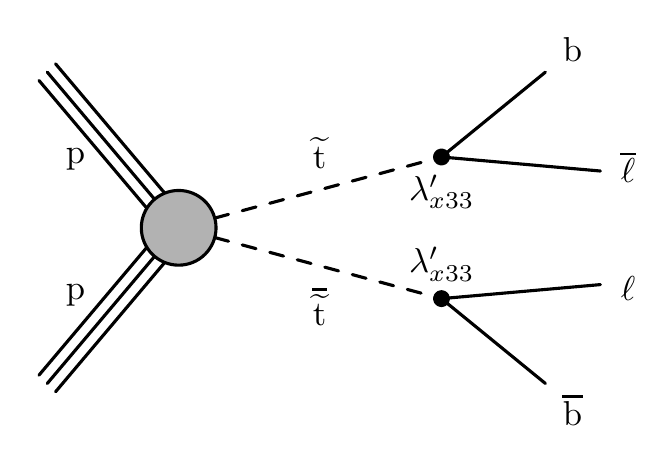}\\
        \includegraphics[width=0.44\textwidth]{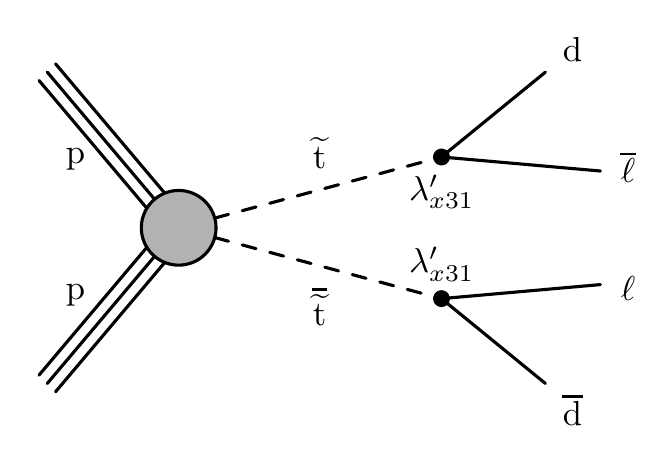}
        \includegraphics[width=0.44\textwidth]{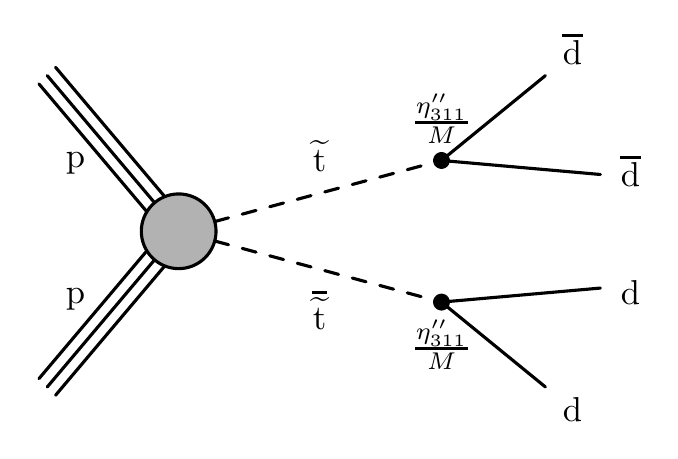}
        \caption{The Feynman diagrams for the different long-lived models
        considered, including the jet-jet model (upper left), models with an exotic decay of the SM-like Higgs boson (upper right), general gauge mediation models with $\sGlu\to\Glu\sGra$ decay (second row, left), mini-split SUSY with $\sGlu\to \cPq\cPaq\widetilde{\chi}^{0}_{1}$
decay (second row, right), RPV SUSY with $\sGlu\to\PQt\PQb\cPqs$ decay (third row, left), RPV SUSY with $\sTop\to\PQb\ell$ decay (third row, right), RPV SUSY with $\sTop\to\PQd\ell$ decay (lower left), and dRPV SUSY with $\sTop\to\cPaqd\cPaqd$ decay (lower right).}
\label{fig: feynman}
\end{figure*}

Feynman diagrams for the benchmark models studied in this paper are summarized in Fig.~\ref{fig: feynman}. One of the benchmark signal models is a simplified model, where long-lived scalar neutral particles $\mathrm{X}$ are pair produced through a 
scattering process mediated by an off-shell \PZ boson. In this model, each $\mathrm{X}$ particle decays to a quark-antiquark pair, assuming equal branching fractions to \cPqu, \PQd, \cPqs, \cPqc, and \PQb quark pairs. The decays to top quark pairs are excluded to provide a simple 
final-state topology for this model, but it was important that the analysis strategy would still be sensitive to a variety of other models. We checked the impact on 
the signal efficiencies of excluding 
decays to top quark pairs, and found it to be small, where the relative changes of the signal efficiencies are generally at the order of a few percents. This model is referred to as the jet-jet model.  
The samples are produced with different resonance masses ranging from 50 to 1500\GeV, and with different proper decay
lengths ranging from 1 to $10^{4}\mm$.

Another signature we consider is the case where LLPs arise from exotic decays of an SM-like Higgs boson, 
which can happen in many BSM scenarios (a review can be found in the 
Section IV.6.6 of Ref.~\cite{deFlorian:2016spz}), including ``hidden Valley" models~\cite{Strassler:2006im,Strassler:2006ri}, twin Higgs models~\cite{Craig:2015pha}, and the folded SUSY model~\cite{Burdman:2006tz}. For the simulation, we use $\mathrm{POWHEG}$~2.0~\cite{Nason:2004rx,Frixione:2007vw,Alioli:2010xd,Bagnaschi:2011tu} to generate events containing a 125\GeV
Higgs boson produced through gluon-gluon fusion. The 125\GeV Higgs boson then decays to two long-lived scalar particles $\mathrm{S}$, and 
each scalar particle 
then decays to a quark-antiquark pair. Two scenarios are considered; in the first scenario the scalar particle has a branching fraction of
100\% to decay to a down quark-antiquark pair, while in the second one the scalar particle has a branching fraction of 100\% to decay to
a bottom quark-antiquark pair. The samples are produced with the scalar particle mass $m_{\mathrm{S}}$ set to be 15, 40, or 55\GeV, while
the $c\tau_{0}$ of $\mathrm{S}$ varies from 1 to 3000\mm.

We also consider a group of SUSY models with different final-state topologies. The first one is a GMSB SUSY model~\cite{Liu:2015bma} in the general gauge mediation scenario~\cite{Meade:2008wd,Buican:2008ws}, where gluinos are pair produced and the gravitino is the lightest SUSY particle, while the gluino is the next-to-lightest supersymmetric particle. After the gluino is produced, 
it decays to a gluon and a gravitino, producing a single displaced jet and missing transverse momentum. This decay is suppressed by the SUSY-breaking scale, and therefore the gluino is
long lived. The model in which this process occurs is referred to as the $\sGlu\to\Glu\sGra$ model.
The
samples are produced with gluino masses from 800 to 2500\GeV, and with the $c\tau_{0}$ of the gluino varying from 1 to $10^{4}\mm$.

The second SUSY model we consider is a mini-split SUSY model~\cite{ArkaniHamed:2012gw,Arvanitaki:2012ps}, referred to as the $\sGlu\to \cPq\cPaq\widetilde{\chi}^{0}_{1}$ model. In this model the gluino decays to
a quark-antiquark pair and the lightest neutralino ($\widetilde{\chi}^{0}_{1}$), with equal branching fractions to \cPqu, \PQd, \cPqs, and \cPqc quark pairs. This decay is mediated by a squark, which is much heavier than the gluino. 
The squark's large mass suppresses the gluino decay, making it long lived. The mass of
the neutralino is assumed to be 100\GeV, the samples are produced with gluino masses from 1400 to 3000\GeV, and the $c\tau_{0}$ of
the gluino varies from 1 to $10^{4}\mm$.
 
The third SUSY model is an RPV SUSY model~\cite{Csaki:2011ge} with minimal flavor violation, where gluinos are pair produced and long lived. Each long-lived 
gluino decays to top, bottom, and strange antiquarks through the RPV coupling $\lambda^{\prime\prime}_{323}$ and the mediation of a virtual top squark~\cite{Barbier:2004ez}, leading to a multijet final-state topology. This model is referred to as the $\sGlu\to\PQt\PQb\cPqs$ model. The samples
are produced with gluino masses from 1200 to 3000\GeV, and a $c\tau_{0}$ varying from 1 to $10^{4}\mm$.

We also consider two other RPV SUSY models~\cite{Graham2012} with semileptonic decays, in which long-lived top squarks are pair produced, and each top squark decays to a bottom quark (down quark) and a charged lepton
via RPV couplings $\Lamp_{133}$, $\lambda^{\prime}_{233}$, and $\lambda^{\prime}_{333}$ ($\Lamp_{131}$, $\Lamp_{231}$, and $\Lamp_{331}$)~\cite{Barbier:2004ez}. The decay rate to each of the three lepton flavors is assumed to be equal. The two models are referred to as the $\sTop\to\PQb\ell$ ($\sTop\to\PQd\ell$) models. The samples are produced with different top squark masses from 600 to 2000\GeV, and a $c\tau_{0}$ varying from 1 to $10^{4}\mm$.

Finally, we consider another SUSY model, referred to as the $\sTop\to\cPaqd\cPaqd$ model, motivated by dynamical RPV (dRPV)~\cite{Csaki:2013jza, Csaki:2015fea}, where long-lived top
squarks are pair produced, and each top squark decays to two down antiquarks via a nonholomorphic RPV coupling $\eta^{\prime\prime}_{311}$~\cite{Csaki:2015uza}. The nonholomorphic RPV 
coupling is suppressed by some large scale $M$, thus giving rise to long lifetimes. The samples
are produced with different top squark masses from 800 to 1800\GeV, and a $c\tau_{0}$ 
varying from 1 to $10^{4}\mm$.

\PYTHIA~8.226 is used for the production of the signal samples, and the PDF set used for the production is NNPDF3.1~LO. For 
SUSY-particle production, the \PYTHIA~8.226 simulation is cross checked with \MGvATNLO~2.4.2 for representative signal points, where 
the \MGvATNLO simulation is performed at LO with up to two additional outgoing partons. The 
resulting signal efficiencies are found to be consistent within the statistical uncertainties. The \PYTHIA parameters for the 
underlying event modeling are set to be the CP2 tune~\cite{Sirunyan:2019dfx}. In the
SUSY models, a long-lived gluino or top squark can form a hadronic
state through strong interactions, an $R$-hadron~\cite{Farrar:1978xj,Farrar:1978rk,Fairbairn:2006gg},
which is simulated with \PYTHIA. The interactions of the $R$-hadron with matter were studied following the
simulation described in Ref.~\cite{Mackeprang:2006gx,Mackeprang:2009ad}, and were found to have negligible impact on this
analysis, since they have very little influence on the vertex reconstruction. 

The simulated background and signal events are processed with a \GEANTfour-based~\cite{Agostinelli:2002hh} simulation for the detailed CMS detector
response. To take into account the effects of additional $\Pp\Pp$ interactions within the same or nearby bunch crossings (``pileup"), additional minimum-bias
events are overlaid on the simulated events to match the pileup distribution observed in the data.

\section{Event reconstruction and preselection}\label{sec: presel}
This search examines dijet candidates in a given event. The algorithms for the offline jet reconstruction and PV selection are the same as those applied at the HLT (as described in Section~\ref{sec: trigger}), except that the full offline information is used. To make sure that the online \HT and jet \pt requirements in 
the displaced-jet triggers reach full efficiency, we apply selections on the offline \HT of the event as well as on the \pt and $\eta$ of each jet. After the trigger 
selection, if an event passes the displaced trigger, we require the event to have offline $\HT>500\GeV$, and dijet candidates 
are formed from all possible pairs of jets in the event, with the jets satisfying $\pt>50\GeV$ and $\abs{\eta}<2.0$. On the other hand, if an event only passes the inclusive trigger, it is required to have offline $\HT>700\GeV$, 
and the dijet candidates are formed from all possible pairs of jets in the event, with the jets satisfying $\pt>80\GeV$ and 
$\abs{\eta}<2.0$. 

In this search, the track candidates are required to have $\pt>1\GeV$ and to be high-purity tracks. The 
high-purity selection is based on track information (such as the normalized $\chi^{2}$ of the track fit, the impact 
parameters, and the number of hits in different tracker layers) to reduce the fraction of misreconstructed tracks, and the selection is optimized separately for each 
iteration of the tracking~\cite{Chatrchyan:2014fea}, so that it is efficient for selecting tracks with different 
displacements. More details of the high-purity selection can be found in Ref.~\cite{Chatrchyan:2014fea}. 
The $\eta$ and $\phi$ of a given track are determined by the direction of its momentum vector at the closest approach point 
to the leading PV. For a given dijet candidate, we associate track candidates with each jet by requiring that 
$\Delta R<0.5$, where $\Delta R = \sqrt{\smash[b]{(\Delta\eta)^2+(\Delta\phi)^2}}$ is the angular distance between the 
jet axis and the track direction. When a track satisfies $\Delta R<0.5$ for both jets, it is associated with the jet 
giving the smaller $\Delta R$.

After associating track candidates with each jet, the next step is to reconstruct a secondary vertex (SV) for each dijet candidate. From all the tracks associated with a dijet candidate, we select 
displaced tracks that satisfy $\ip>0.5\mm$ and $\ipsig>5.0$. We then attempt to reconstruct 
an SV from these displaced tracks using an adaptive vertex fitter algorithm~\cite{Fruhwirth:2007hz}. 
The reconstructed SV is not required to have associated tracks from both jets, so that the search can 
be sensitive to the models where the LLP decays to a single displaced jet. To improve the 
signal-to-background discrimination, we implement a set of preselection criteria on the dijet/SV candidates, which are described in 
the rest of this section. 

To ensure the quality of the vertex reconstruction, the SV is selected only if it is reconstructed with a $\chi^{2}$ per 
degree of freedom ($\chi^{2}/\mathrm{n_{dof}}$) of less than 5.0. In order to suppress long-lived SM mesons and baryons, the invariant mass of the 
vertex is required to be larger than 4\GeV, and the transverse momentum of the vertex is required to be larger than 8\GeV, 
where the four-momentum of the vertex is calculated assuming the charged pion mass for all assigned tracks.

We only consider dijet candidates that have a reconstructed SV satisfying the above requirements.  Furthermore, SVs in background events tend to have only one track with a high value for 
\ip, corresponding to the tail of the impact parameter distribution. We 
therefore consider the track with the second-highest \ipsig among the tracks 
that are assigned to the SV, since this provides a more sensitive discriminant for identifying 
displaced jets. We require the second-highest \ipsig to be larger than 15. 

We also compute another quantity $\epsilon$, 
which is the ratio between the sum of energy for all the tracks
assigned to the SV and the sum of the energy for all the tracks associated with the two jets:
\begin{linenomath}
\begin{equation}
\epsilon=\frac{\sum_{\mathrm{track\in SV}} E_{\mathrm{track}}}{\sum_{\mathrm{track\in dijet}} E_{\mathrm{track}}}.
\end{equation}
\end{linenomath}
Since $\epsilon$ is expected to be large for displaced-jet signatures, dijet candidates with $\epsilon$ smaller than 0.15 are rejected.

An additional variable, $\zeta$, is defined to characterize the contribution of prompt activity to the jets. For each track associated with a jet, we identify the PV (including the leading PV and the pileup vertices) with the minimum three-dimensional (3D) impact parameter significance to the track. If this minimum value is smaller than 5, we assign the track to this PV. Then for each jet, we compute the track energy contribution from each PV, and the PV with the largest track energy contribution to the jet is chosen. Finally, we define $\zeta$ as
\begin{linenomath}
\begin{equation}
\zeta=\frac{\sum_{\mathrm{track\in PV_{1}}}E_{\mathrm{track}}^{\mathrm{Jet_{1}}}+\sum_{\mathrm{track\in PV_{2}}}E_{\mathrm{track}}^{\mathrm{Jet_{2}}}}{E_{\mathrm{Jet_{1}}}+E_{\mathrm{Jet_{2}}}},
\end{equation}
\end{linenomath}
where $\sum_{\mathrm{track}\in \mathrm{PV}_{i}}E_{\mathrm{track}}^{\mathrm{Jet}_{i}}$ is the sum of the track energy coming from the most compatible PV for a given jet, while $E_{\mathrm{Jet}_{i}}$ 
is the energy of a given jet, thus 
$\zeta$ is the charged energy fraction of the dijet associated with the most compatible PVs. For displaced-jet signatures, $\zeta$ tends to be small since the jets are not compatible with PVs. Dijet candidates with $\zeta$ larger than 0.2 are rejected. 

To suppress the background events arising from NIs in the tracker material, we compare the positions of the SVs with a map of the distribution of material in the inner 
tracker. The map was obtained from the distribution of NI candidate vertices, which are 
reconstructed using the adaptive vertex fitter on a sample of events collected with isolated 
single-muon triggers.  
The NI candidates are required 
to satisfy the following criteria:

\begin{itemize}
\item The tracks are required to be associated with dijet candidates, which are formed from the jets having $\pt>10\GeV$ and $\abs{\eta}<2.0$;
\item The associated tracks must have $\pt>0.2\GeV$, high purity, $\ip>0.5\mm$, and $\ipsig>5.0$;
\item vertex track multiplicity is larger than 3; 
\item the ratio $\epsilon$ of the energy sum for SV tracks to that for all tracks is less than 0.15 for the dijet candidate;
\item vertex $L_{xy}$ significance is larger than 200, where the transverse decay length $L_{xy}$ is the distance between the SV and the leading PV; 
and 
\item vertex $\chi^{2}/\mathrm{n_{dof}}$ is smaller than 3.0.
\end{itemize}

\begin{figure}[hbtp]
\centering
\includegraphics[width=0.49\textwidth]{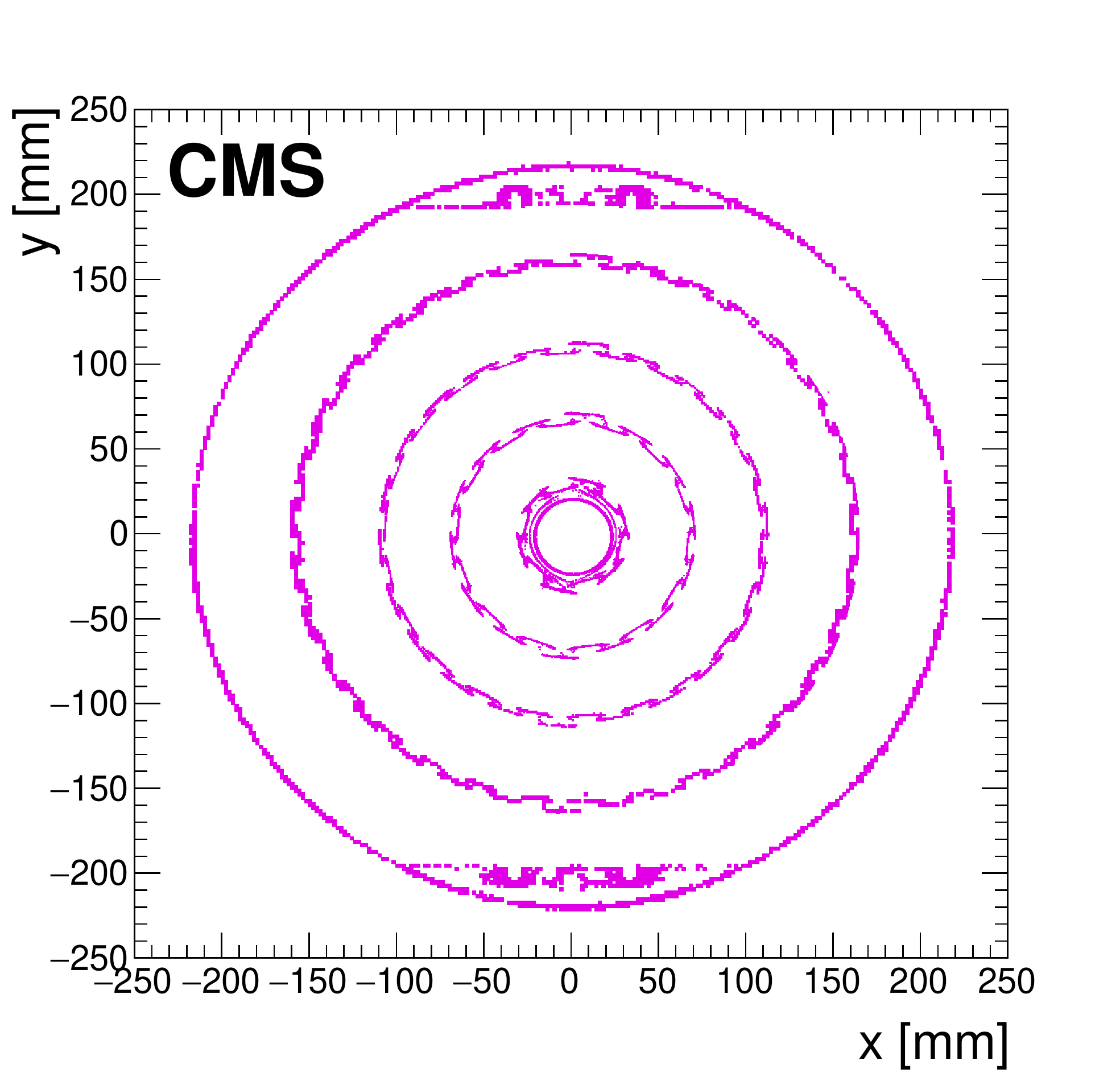}
\includegraphics[width=0.49\textwidth]{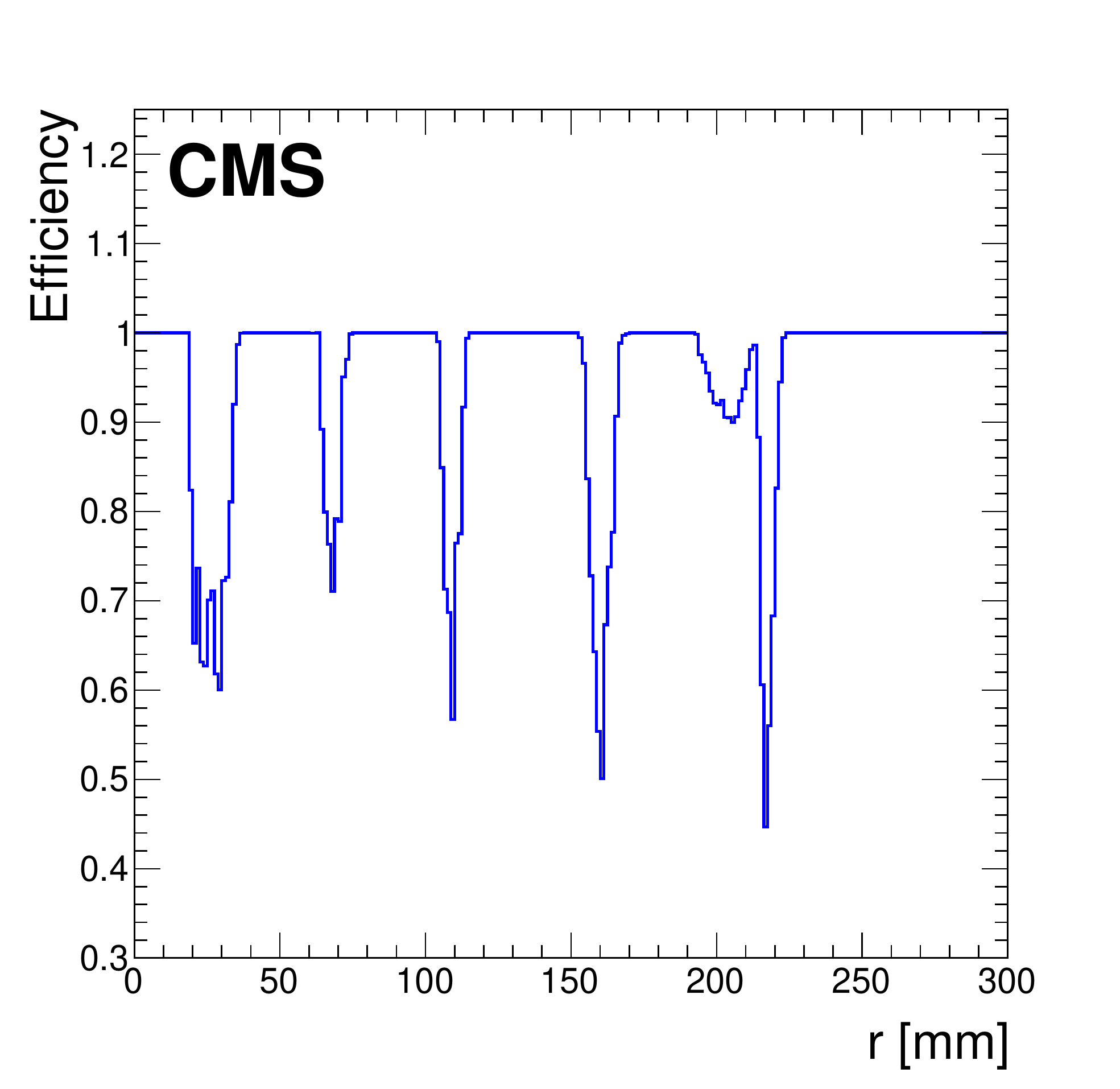}
\caption{\cmsLeft: the NI-veto map based on the NI vertex reconstruction in the 2017 and 2018 data collected by the CMS detector, the map corresponds to the geometry of the CMS 
pixel detector used in 2017--2018 data taking~\cite{CMS:2012sda}. The structures of the different pixel layers can be clearly seen. \cmsRight: the efficiency for 
a given vertex candidate to pass the NI-veto as a function of radius $r$. }
\label{fig: tracker_map}
\end{figure}

After these selections, the distribution of the NI vertex candidates is transferred to an NI-veto map in the transverse plane, with 
$\abs{x}$ and $\abs{y}<25\cm$, as shown in Fig.~\ref{fig: tracker_map}. To suppress misreconstructed vertices and the displaced vertices produced by decaying 
long-lived SM mesons and baryons, we only select 
the region where the NI vertex density is above a threshold that varies for different layers of the pixel 
detector. In the NI-veto map, we can clearly see the structures of the beam pipe (at $r=\sqrt{\smash[b]{x^{2}+y^{2}}}\approx23\mm$), 
the four pixel layers (at $r\approx29$, $\approx68$, $\approx109$, and $\approx160\mm$), and the support rails (at $r\approx200\mm$). In our search, any SV 
candidate that overlaps with the NI-veto map is rejected. The loss of the fiducial volume within $r<300\mm$ due to the veto is around 4\%, and the 
efficiencies for signal events to pass this selection are generally well above 90\%. In the veto no requirement is placed on the $z$ coordinates 
of the SVs, but the impact of restricting the veto to the barrel region of the pixel detector ($\abs{z}<27\cm$) is negligible on the signal efficiencies. A similar study on the structure of the CMS inner tracking system using a more sophisticated 
NI reconstruction technique with 2016 data has been reported in Ref.~\cite{Sirunyan:2018icq}.

The preselection criteria for this search, summarized in Table~\ref{tab: presel}, are 
efficient for a wide range of long-lived models with different final-state topologies. 

\begin{table*}[htb]
\centering
\topcaption{Summary of the preselection criteria.}
\begin{scotch}{lc}
SV/dijet variable & Requirement \\\hline
 Vertex $\chi^{2}/\mathrm{n_{dof}}$ & $<$5.0 \\
 Vertex invariant mass & $>$$4\GeV$ \\
 Vertex transverse momentum & $>$$8\GeV$ \\
Second largest \ip significance & $>$15 \\
$\epsilon$ (SV track energy fraction in the dijet) & $>$0.15 \\
$\zeta$ (energy fraction from compatible PVs)   & $<$0.20 \\
Vertex position in the $x$-$y$ plane  & no overlap with the NI-veto map \\
\end{scotch}
\label{tab: presel}
\end{table*}

\section{Event selection and background prediction}\label{sec: back}
After reconstructing the SV using the adaptive vertex fitter, we employ an auxiliary algorithm to check the consistency between the SV system and the dijet system. For each displaced
track (having $\ip>0.5\mm$, $\ipsig>5.0$) associated with the dijet, we determine an expected decay point consistent with the displaced dijet hypothesis by finding the crossing point
of the track helix and the dijet direction in the transverse ($x$-$y$) plane. The dijet direction is the space direction of the four-momentum-sum of the two jets, for 
which the production vertex of the two jets is taken to be the SV. For each crossing point, an expected transverse decay length ($L_{xy}^{\mathrm{exp}}$) is computed with respect to the leading PV. 
The $L_{xy}^{\mathrm{exp}}$ is positive if the crossing point is at the same side of the dijet direction, otherwise it is negative. The associated displaced tracks are then clustered based on their $L_{xy}^{\mathrm{exp}}$, using a hierarchical clustering algorithm~\cite{Johnson1967}. During the clustering, two clusters are merged when the smallest $L_{xy}^{\mathrm{exp}}$ difference between the two clusters is smaller than $15\%$ of the $L_{xy}$ of the SV.
After the clustering procedure is finished, if more than one cluster is formed, the one closest to the SV is selected.
The cluster root-mean-square (RMS), taken to be the relative RMS of individual tracks $L_{xy}^{\mathrm{exp}}$ with respect to the SV $L_{xy}$, is computed to
provide signal-to-background discrimination:
\begin{linenomath}
\begin{equation}
\mathrm{RMS}_{\mathrm{cluster}}=\sqrt{\frac{1}{N_{\mathrm{tracks}}}\sum_{i=1}^{N_{\mathrm{tracks}}}\frac{(L_{xy}^{\mathrm{exp}}(i)-L_{xy})^{2}}{L_{xy}^{2}}},
\end{equation}
\end{linenomath}
where $N_{\mathrm{tracks}}$ is the number of tracks in the selected cluster.
 
For each track assigned to the SV, a sign is given to the \ip and \ipsig based on the angle between the dijet direction and the impact parameter vector 
that points from the leading PV to the closest approach point (with respect to the leading PV) of the track in the transverse plane. The sign is positive if this angle is 
smaller than $\pi/2$; otherwise the sign is negative. A new variable, $\kappa$, is then introduced as the signed \ipsig sum of the six leading tracks from the 
SV (where the tracks are ordered by the absolute values of their \ipsig): 
\begin{linenomath}
\begin{equation}
\kappa=\sum_{i=1}^{6}\mathrm{Sig[IP_{2D}(track_{\emph{i}})]}.
\end{equation}
\end{linenomath}
If the track multiplicity is smaller than six, the sum is taken over all the tracks from the SV. For background processes, since the tracks assigned to the SV are uncorrelated with the dijet direction, the signed 
\ipsig of different tracks tend to cancel each other, therefore $\kappa$ peaks sharply around zero. On 
the other hand, for displaced jets that originate from the SV, the directions of the tracks will be highly 
correlated with the dijet direction, therefore $\kappa$ is significantly different from zero and $\abs{\kappa}$ tends to be large. 
 
To improve signal-to-background discrimination and to define a region with signal events enriched, we proceed to construct a multivariate discriminant based on the following variables for the vertex/dijet candidates:

\begin{itemize}
\item Vertex track multiplicity;
\item Vertex $L_{xy}$ significance;
\item Cluster RMS;
\item The magnitude of the signed \ipsig sum of the six leading tracks $\abs{\kappa}$.
\end{itemize}

\begin{figure*}[hbtp]
\centering
\includegraphics[width=0.49\textwidth]{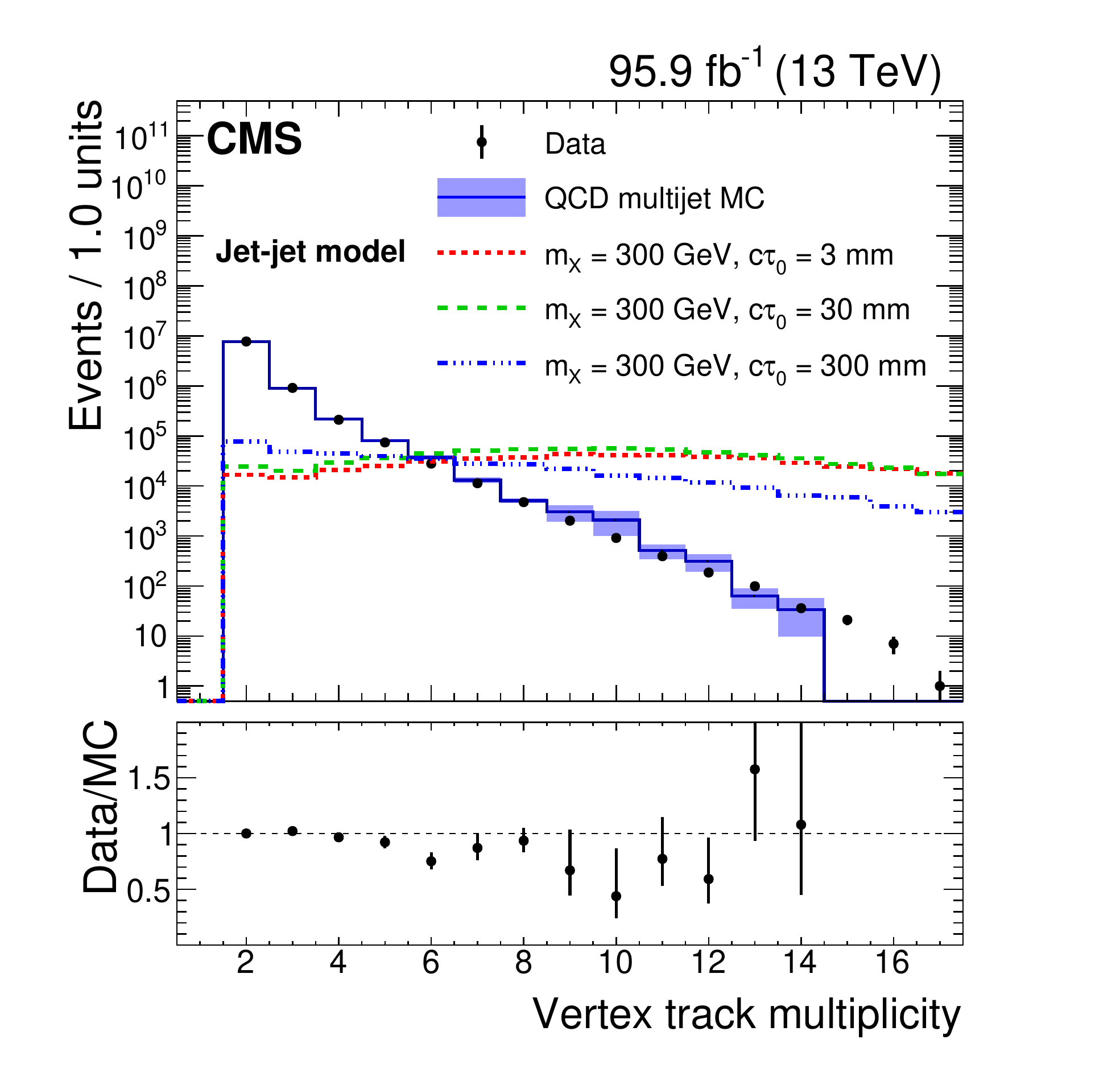}
\includegraphics[width=0.49\textwidth]{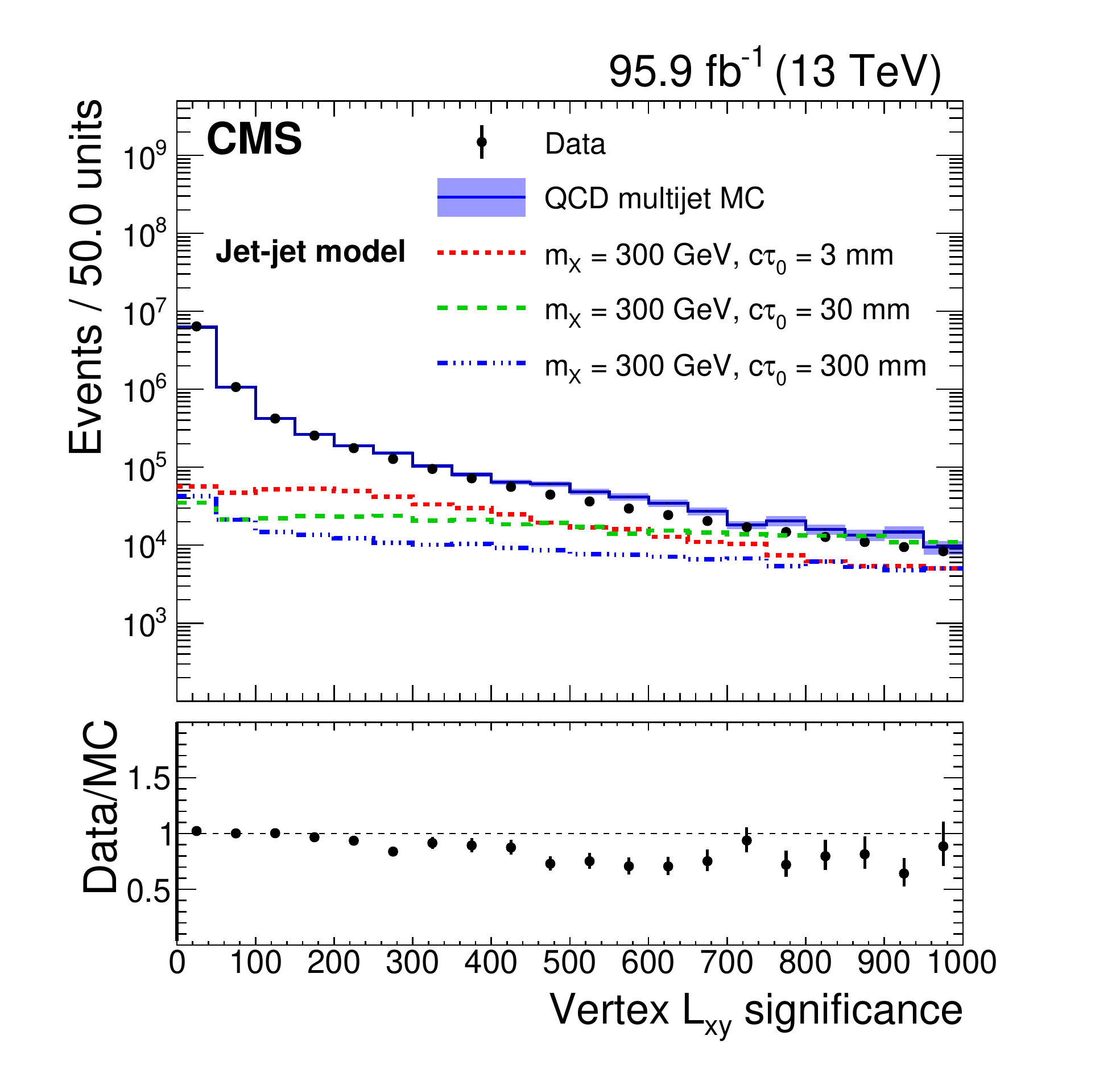}\\
\includegraphics[width=0.49\textwidth]{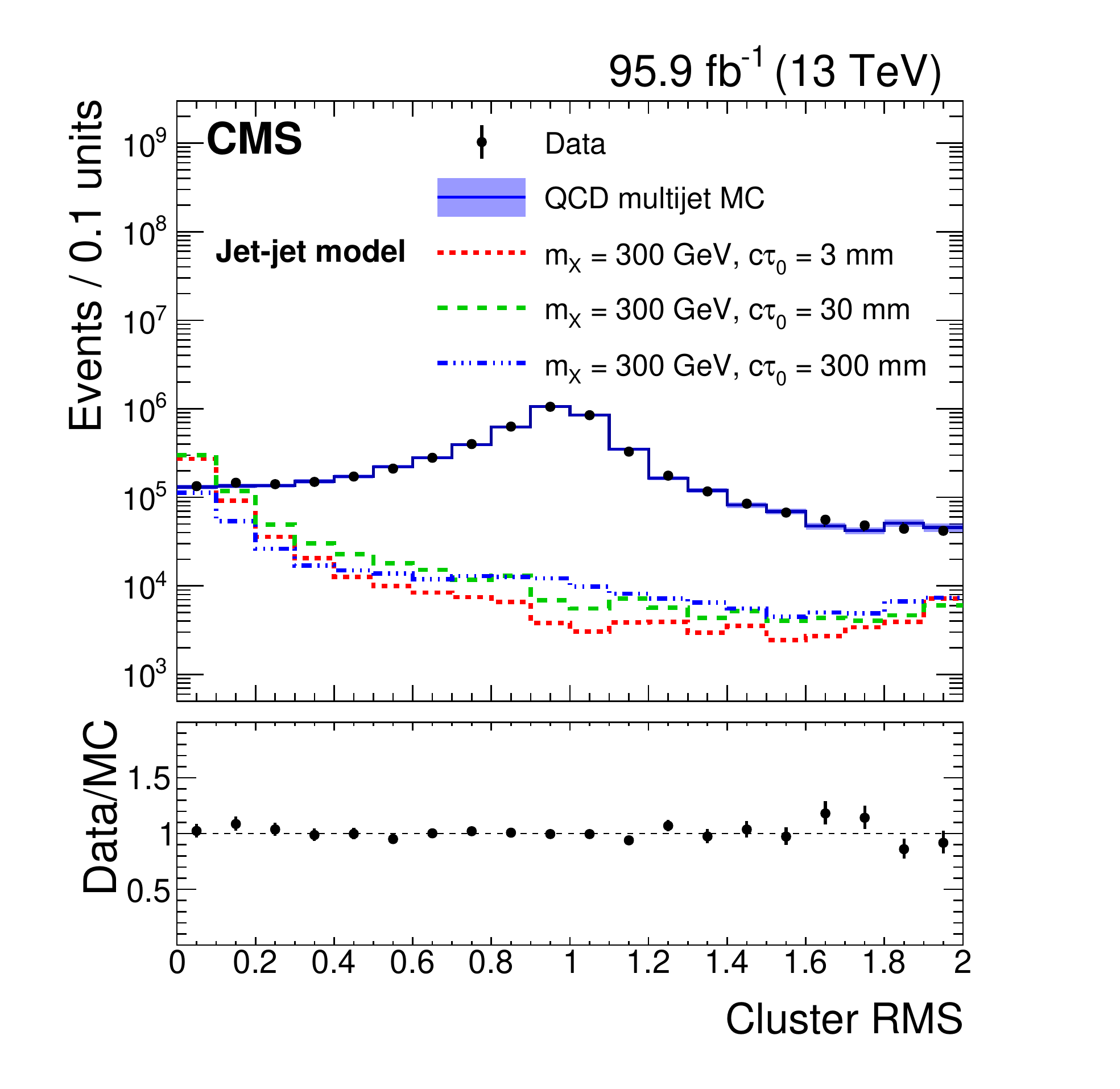}
\includegraphics[width=0.49\textwidth]{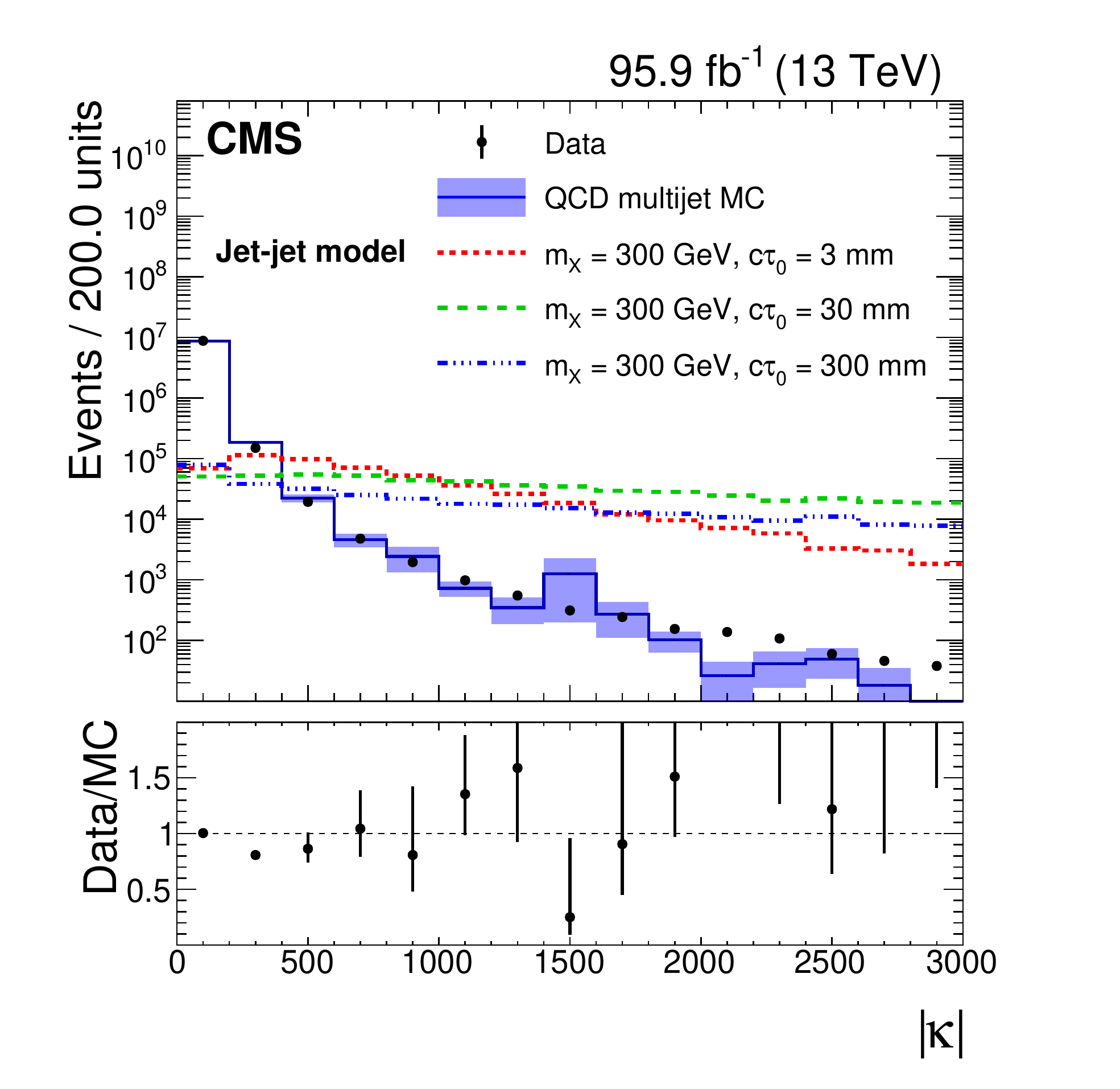}
\caption{The distributions of the vertex track multiplicity (upper left), vertex $L_{xy}$ significance (upper right), 
cluster RMS (lower left), and the magnitude of the signed \ipsig sum $\abs{\kappa}$ (lower right), for data, simulated QCD multijet events, and simulated signal events. 
Data and simulated events are selected with the displaced-jet triggers and with the offline \HT, jets \pt, and $\eta$ selections 
applied. For a given event, if there is more than one SV candidate being reconstructed, the one with the largest vertex track 
multiplicity is chosen. If the track multiplicities are the same, the one with the smallest $\chi^{2}/\mathrm{n_{dof}}$ is chosen. The lower panels show the ratios between the data and the simulated QCD multijet events. The blue shaded error bands and vertical bars represent the statistical uncertainties. Three benchmark signal distributions are 
shown (dashed lines) for the jet-jet model with $m_{\mathrm{X}}=300\GeV$ and varying $c\tau_{0}$. For visualization purposes, each signal process is given a cross section that yields 
$10^{6}$ events produced in the analyzed data sample.}
\label{fig: vtx_vars}
\end{figure*}

The distributions of the four variables are shown in Fig.~\ref{fig: vtx_vars}, with displaced-jet 
triggers, offline \HT, and offline jet kinematic variables selections (described in Section~\ref{sec: presel}) applied. For the multivariate discriminant we utilize the 
gradient boosted decision tree (GBDT) algorithm~\cite{Freund:1996:ENB:3091696.3091715,friedman2000,friedman2001}, with cross entropy 
as the loss function. The GBDT algorithm is implemented using the \textsc{TMVA} (toolkit for multivariate data analysis) package~\cite{Hocker:2007ht} interfaced with
\textsc{Scikit-learn}~\cite{Pedregosa:2011:SML:1953048.2078195}. Given the large cross section of the QCD multijet process and the relatively low \HT threshold of our 
displaced-jet triggers, the event count of the simulated QCD multijet sample 
(after preselections) is 
insufficient for the GBDT training, since it is much smaller than the number of expected QCD multijet events in the analyzed data sample. 
Therefore, for the background sample in the GBDT training, we use the data in the 
following region:
\begin{itemize}
\item events are selected by the displaced-jet triggers, and pass the offline \HT and jet kinematic variables selections;
\item $\epsilon<0.12$ for the dijet candidate, making this region orthogonal to the signal region;
\item the veto using the NI-veto map is not applied;
\item all the other preselection criteria are satisfied. 
\end{itemize} 
For the signal sample in the GBDT training, simulated jet-jet model events that pass the preselection criteria are used, with $m_{\mathrm{X}}=100$, 300, and 1000\GeV, 
and with $c\tau_{0}=1$, 10, 100, 1000\mm. If there is more than one dijet/SV candidate passing the selection criteria in a given event, the one with the largest track 
multiplicity is chosen for the training. If the track multiplicities are the same, the one with the smallest SV $\chi^{2}/\mathrm{n_{dof}}$ is chosen. An event weight is assigned separately to each signal point with a given $m_{\mathrm{X}}$ and $c\tau_{0}$ such that the sum of 
weights is identical for each point, thus each signal point has the same priority in the training. Twenty percent of the events in the signal and background samples 
are randomly selected to validate the performance of the GBDT and to make sure it is not overtrained. 

\begin{figure}
\centering
\includegraphics[width=\cmsFigWidth]{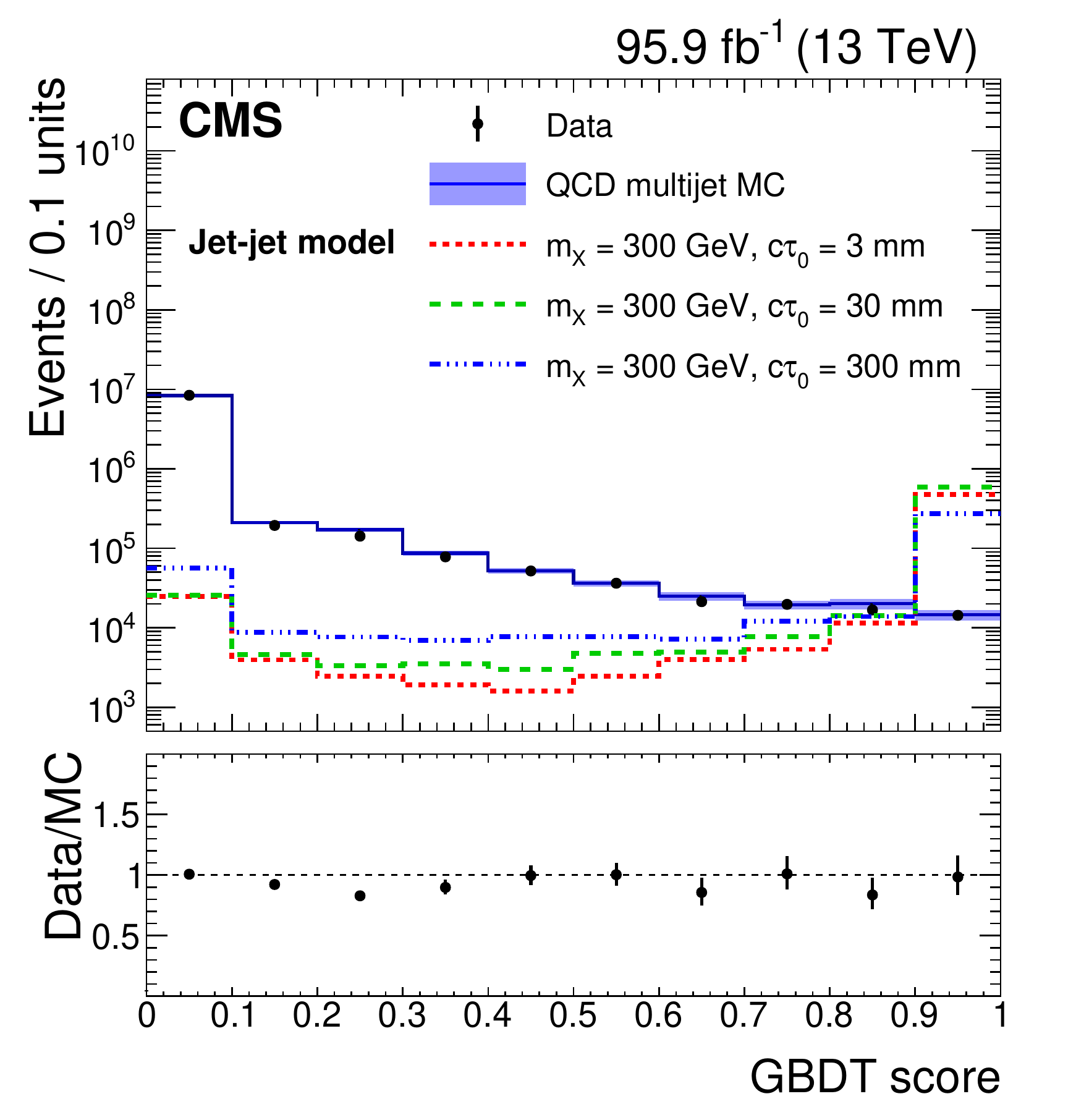}
\caption{The distributions of the GBDT output score for data, simulated QCD multijet events, and simulated signal events. 
Data and simulated events are selected with the displaced-jet triggers and with the offline \HT, jets \pt, and $\eta$ selections 
applied. For a given event, if there is more than one SV candidate being reconstructed, the one with the largest vertex track
multiplicity is chosen. If the track multiplicities are the same, the one with the smallest $\chi^{2}/\mathrm{n_{dof}}$ is chosen. 
The lower panel shows the ratio between the data and the simulated QCD multijet events. The blue shaded error bands and vertical bars represent the statistical uncertainties. Three benchmark signal distributions are 
shown (dashed lines) for the jet-jet model with $m_{\mathrm{X}}=300\GeV$ and varying $c\tau_{0}$. For visualization purposes, each signal process is given a cross section corresponding to 
$10^{6}$ events produced in the analyzed data sample. The signal events shown in this plot are not used in the GBDT training.}
\label{fig: GBDT}
\end{figure}

The GBDT output values or scores for data, simulated QCD multijet events, and 
simulated signal events are shown in Fig.~\ref{fig: GBDT}. The signal efficiencies for this search are measured 
with simulated signal events produced separately. The background prediction is purely based on some other control samples in 
data, which are different from the one used for the GBDT training. Although only the jet-jet model is used as the signal sample in the GBDT training, the GBDT is highly 
efficient in selecting signatures of other LLP models with different final-state topologies, since we have explicitly chosen the input variables to make the GBDT as model-independent as possible.  

In addition to the GBDT score $g$, we use another variable $N_{\mathrm{tracks}}^{\mathrm{3D}}$ in the final event selection, which is the number of 
3D prompt tracks in a single jet, where the 3D prompt tracks are the tracks that have 
3D impact parameters with respect to the leading PV smaller than 0.3\mm. 

If more than one dijet candidate passes the preselection criteria described in Section~\ref{sec: presel}, the one with the largest 
GBDT score is selected. If the GBDT scores are the same, the one with the smallest SV $\chi^{2}/\mathrm{n_{dof}}$ is selected. 
In the final signal region, the candidate is further required to pass three final selection criteria, which are:
\begin{itemize}
\item Selection 1: for the leading jet in \pt, $N_{\mathrm{tracks}}^{\mathrm{3D}}$ is smaller than 3;
\item Selection 2: for the subleading jet, $N_{\mathrm{tracks}}^{\mathrm{3D}}$ is smaller than 3; and 
\item Selection 3: the GBDT score $g$ is larger than 0.988. 
\end{itemize}
The numerical values for the selection criteria are chosen by 
optimizing the discovery potential of 5 standard deviations based on the Punzi significance~\cite{Punzi:2003bu} for the jet-jet model and the $\sGlu\to\Glu\sGra$ model across different LLP masses and lifetimes. The chosen models encompass the displaced-dijet and displaced-single-jet signatures, with $m_{\mathrm{X}}=100$, 300, and 
1000\GeV for the jet-jet model, and with $m_{\sGlu}=600$, 1000, and 1600\GeV for the $\sGlu\to\Glu\sGra$ model, while $c\tau_{0}$ is taken to be 1, 10, 100, and 1000\mm. Thus there are 24 signal 
points considered in total for the selection optimization. 
 
Based on the three selection criteria, we can define eight nonoverlapping regions A--H, which include the 
final signal region. The event counts in different 
regions are $N_{fff}$, $N_{pff}$, $\cdots$, and $N_{ppp}$, for regions A, B, $\cdots$, and H, respectively, as shown in Table~\ref{tab: abcdefgh}. The region H is the region where the events pass all the three selection criteria, and thus is the final signal 
region. Events in the remaining regions (A--G) fail one or more of the three selection criteria. 
Since the three selection criteria have little correlation between them for the background events, a property that 
has been verified with simulated QCD multijet events, the background yield 
in the signal region H can be estimated by different ratios of event counts in regions A--G,  where the ratio $b_{\mathrm{nominal}}=N_{ppf}(N_{ffp}+N_{fpp}+N_{pfp})/(N_{fff}+N_{pff}+N_{fpf})$ uses the fraction of events passing to those failing the GBDT selection
(Selection 3) and is taken as the central value of the predicted background yields. Three additional ratios are computed 
using the events failing either one of the first two selections:

\begin{itemize}
\item Cross-check 1: $N_{ppf}(N_{ffp}+N_{fpp})/(N_{fff}+N_{fpf})$;
\item Cross-check 2: $N_{ppf}(N_{ffp}+N_{pfp})/(N_{fff}+N_{pff})$; 
\item Cross-check 3: $N_{ppf}(N_{fpp}+N_{pfp})/(N_{fpf}+N_{pff})$.
\end{itemize}  

\begin{table*}[htb]
\centering
\topcaption{The definitions of the different regions used in the background estimation.}
\label{tab: abcdefgh}
\begin{scotch}{ccccc}
Region &Selection 1 &Selection 2 &Selection 3 & Event count\\ \hline
A      &Fail        &Fail        &Fail        & $N_{fff}$\\
B      &Pass        &Fail        &Fail        & $N_{pff}$\\
C      &Fail        &Pass        &Fail        & $N_{fpf}$ \\
D      &Fail        &Fail        &Pass        & $N_{ffp}$ \\
E      &Fail        &Pass        &Pass        & $N_{fpp}$\\
F      &Pass        &Fail        &Pass        & $N_{pfp}$\\
G      &Pass        &Pass        &Fail        & $N_{ppf}$\\
H      &Pass        &Pass        &Pass        & $N_{ppp}$\\
\end{scotch}
\end{table*} 

These cross-checks provide an important test of the robustness of the background prediction and the assumption that the three selection criteria
are minimally correlated. Differences between the predictions obtained with the nominal ratio $b_{\mathrm{nominal}}$ and the cross-checks are used to estimate the systematic uncertainties in the
background prediction.

\begin{figure}[hbtp]
\centering
\includegraphics[width=\cmsFigWidth]{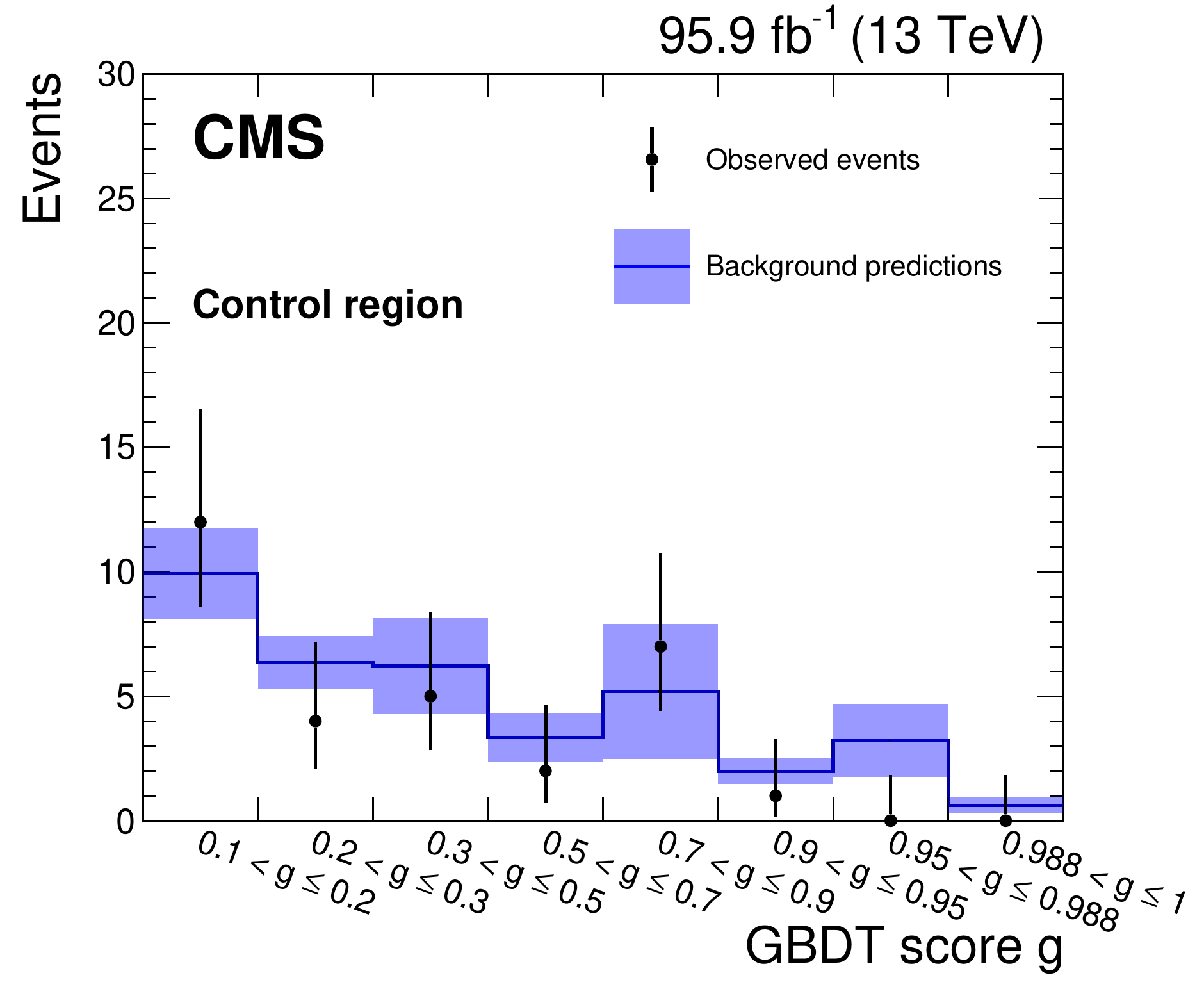}
\caption{The predicted background yields and the numbers of observed events in the control region, for different bins of the GBDT scores. The 
background predictions in different bins are correlated, since the events that are used for background predictions in lower bins are also used in 
the background predictions in higher bins. The error bands for the predictions 
represent statistical uncertainties and systematic uncertainties added in quadrature. The error bars for the observed events represent statistical uncertainties, assuming 
Poisson statistics.}
\label{fig: back_control}
\end{figure}

The method described above can be generalized to predict the background yield in arbitrary intervals of the GBDT score $g$, 
where the first two selections are also satisfied. In this way we can verify our background prediction method in the different bins of GBDT score $g$. 
The background prediction method is first tested with simulated QCD multijet events and simulated signal events, and is found to be robust both with and 
without signal contaminations. We then test the background prediction in data, for which we define a control region by inverting 
the selection on $\epsilon$, requiring $\epsilon$ to be smaller than 0.15. In order 
to improve the statistical precision, we remove the selection requirement that uses the NI-veto map, although the background 
prediction method is still robust with the presence of the NI-veto map. The predicted background 
yields and observed events in the control region are shown in Fig.~\ref{fig: back_control}, which shows a good agreement between the predicted 
and observed yields. The predicted background yields in the different bins of GBDT score are correlated, since the events that are used for background predictions in lower bins are also used in the background predictions in higher bins.

\section{Systematic uncertainties}~\label{sec: sys}
The systematic uncertainty in the background prediction is taken to be the largest deviation of the three cross checks from the nominal prediction $b_{\mathrm{nominal}}$, and is 
found to be 52\% in the final signal region where the GBDT score $g$ is larger than 0.988.

The systematic uncertainties in the integrated luminosity for 13\TeV $\Pp\Pp$ collision data are 2.3\% and 2.5\% for the 2017 and 2018~\cite{CMS:2018elu,CMS:2019jhq} data taking periods, respectively, and are modeled as uncorrelated nuisance parameters 
between the years.

The systematic uncertainty in the signal yields arising from the pileup modeling is estimated by varying the inelastic $\Pp\Pp$ cross section~\cite{Sirunyan:2018nqx} used in the pileup distribution determination by
4.6\%. The resulting variations of the signal yields are generally smaller than 1\% and are well within the statistical fluctuations, therefore the impact of the pileup modeling is negligible and no 
corresponding systematic uncertainty is assigned. 

The systematic uncertainty in the signal yields due to the online \HT requirements of the displaced-jet triggers is estimated by comparing the 
efficiency of the online \HT requirements measured in data with the one measured in the QCD multijet MC sample. The efficiencies are measured using the events 
collected with an isolated single-muon trigger. The discrepancies between the measurements in data and MC simulation 
are parameterized as functions of offline \HT, and corresponding corrections are applied to the simulated signal events. The signal yields are
then recalculated for different masses and mean proper decay lengths. The largest correction in the signal yield for a given 
model is taken as the corresponding systematic uncertainty, and is found to be smaller than 2\%. 

The systematic uncertainty in the signal yields due to the online jet \pt requirements of the displaced-jet triggers is estimated by measuring 
the per-jet efficiencies of the online jet \pt requirements in data and in the QCD multijet MC sample, using events collected with a 
prescaled \HT trigger that requires $\HT>425\GeV$. Corrections for the discrepancies between the measurements in data and MC simulation 
are applied to the simulated signal events, and the signal yields are recalculated. The correction is more significant for 
low-mass LLPs (with $m\sim50\GeV$), and is smaller than 8\%, which is taken as the corresponding systematic 
uncertainty.

To estimate the systematic uncertainty due to the online tracking requirements of the displaced-jet triggers, we measure the 
per-jet efficiencies of the online tracking requirements as functions of the number of prompt tracks and displaced tracks, 
using events collected with the prescaled \HT trigger. The efficiencies obtained in data are found to be consistent with the efficiencies obtained in MC simulations. Therefore no corresponding systematic uncertainty is assigned.

To estimate the impact of the possible mismodeling of the GBDT score in MC simulation on the signal yields, we compare the distribution of the 
GBDT score in simulated QCD multijet events with the distribution measured in data, using events collected with the prescaled \HT 
trigger. The discrepancies between data and MC simulation are taken into account by varying the GBDT scores in the simulated 
samples by the same magnitude. The largest variation in the signal efficiency for a given model is taken as the corresponding systematic 
uncertainty in signal yields, and is found to be in the range of 4\%--15\%. 

Similarly, the uncertainty in the signal yields due to impact parameter modeling is estimated by comparing the distribution of the impact parameters 
in simulated QCD multijet events with the distribution measured in data, also using events collected with the prescaled \HT 
trigger. The impact parameters in simulated QCD multijet events are varied until the discrepancies between data and MC simulation are covered by the variation. The impact parameters in the simulated signal samples are then varied by the same magnitude. The largest variation in the signal efficiency for a given model is taken as the corresponding systematic      
uncertainty, and is found to be in the range of 8\%--18\%.

The impact of the jet energy scale uncertainty on the signal yields is estimated by varying the jet energy and \pt by one standard deviation of 
the jet energy scale uncertainty~\cite{Khachatryan:2016kdb}. The variations of the signal efficiencies are smaller than 3\%, which is taken as the 
corresponding systematic uncertainty.

The impact of the PDF uncertainty on the signal acceptance is estimated by reweighting the simulated signal events with NNPDF, 
CT14~\cite{Dulat:2015mca}
and MMHT14~\cite{Harland-Lang:2014zoa} PDF sets, and their associated uncertainty sets~\cite{Butterworth:2015oua,Buckley:2014ana}, 
following the PDF4LHC recommendation~\cite{Butterworth:2015oua}. The uncertainty in signal efficiency for a given signal model is quantified by
comparing the efficiencies calculated with alternative PDF sets and the ones with the nominal NNPDF set, and is found to
be in the range of 4--6\%.

The uncertainty in the signal yields due to the selection of the PV is estimated by replacing the leading PV with
the subleading PV when calculating impact parameters and vertex displacement. 
The largest variation of the signal efficiency for a given signal model is taken as the corresponding 
systematic uncertainty, and is found to be in the range of 8--15\%.

The various systematic uncertainties in the signal yields are summarized in Table~\ref{tab: sys_unc}.  

\begin{table*}[htb]
\centering
\topcaption{Summary of the systematic uncertainties in the signal yields.}
\label{tab: sys_unc}
\begin{scotch}{lc}
Source & Uncertainties (\%) \\\hline 
Integrated luminosity           & 2.3--2.5 \\          
Online \HT requirement        & 0--2 \\ 
Online jet \pt requirement     & 0--8 \\ 
Offline vertexing               & 4--15\\
Track impact parameter modeling & 8--18 \\
Jet energy scale                & 0--3 \\
PDF                             & 4--6 \\
Primary vertex selection        & 8--15\\[\cmsTabSkip]
Total                           & 17--25\\
\end{scotch}
\end{table*}

\section{Results}\label{sec: results}

\subsection{Data in the signal region}

\begin{figure}[hbtp]
\centering
\includegraphics[width=\cmsFigWidth]{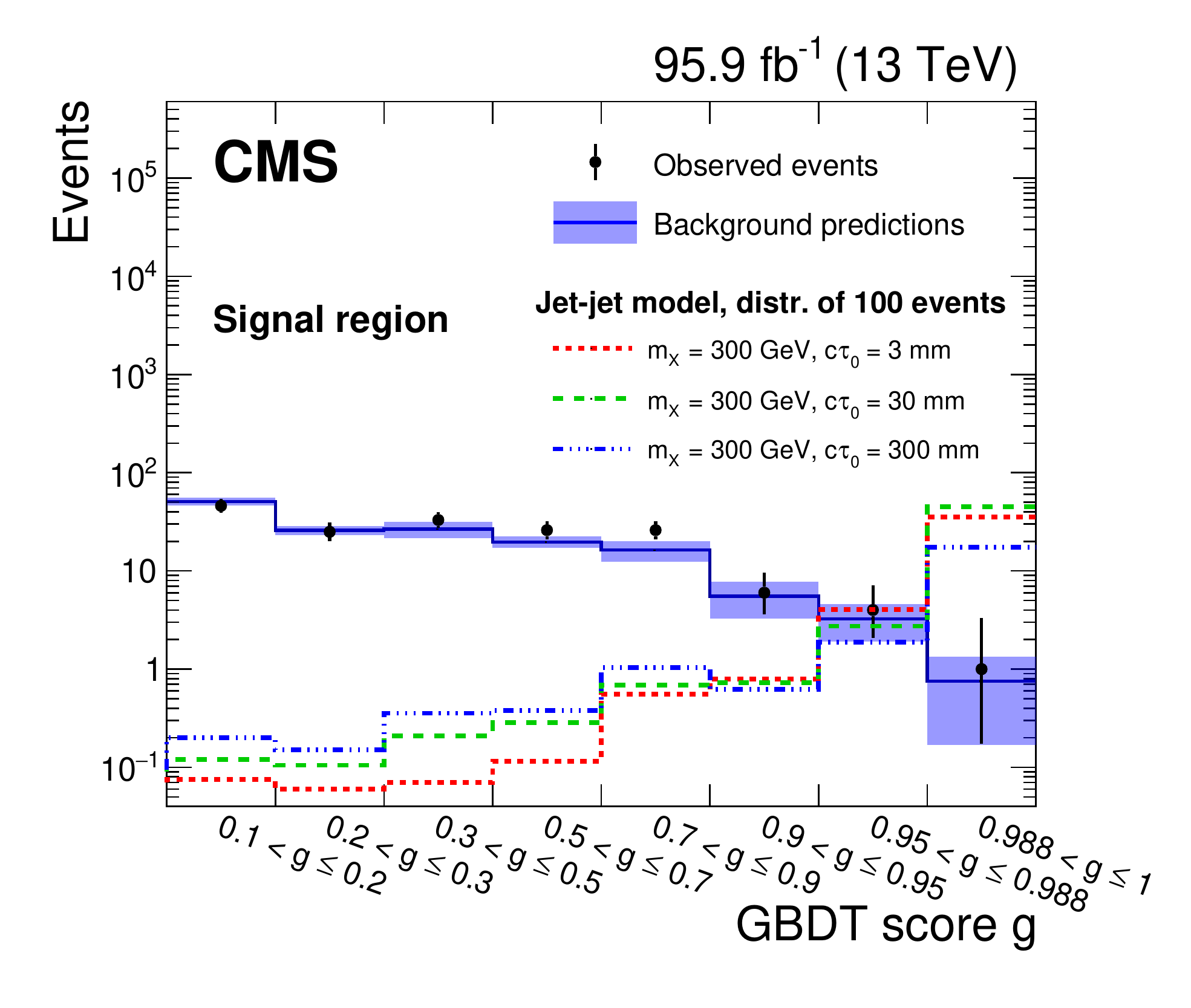}
\caption{The predicted background yields and the number of observed events for the data in the signal region, with  $N_{\mathrm{tracks}}^{\mathrm{3D}}$ smaller 
than 3 for both jets, shown for different bins of the GBDT scores. The background
predictions in different bins are correlated, since the events that are used for background predictions in lower bins
are also used in the background predictions in higher bins. For comparison,
three benchmark signal points are also shown (dashed lines) for the jet-jet model with $m_{\mathrm{X}}=300\GeV$ and different lifetimes.
For visualization purposes, each signal process is given a cross section that yields 100 events produced in the analyzed data sample.}
\label{fig: back_sig}
\end{figure}

The predicted background yields and the numbers of observed events in different GBDT ranges, after all the preselection criteria are applied, are shown in Fig.~\ref{fig: back_sig}, with $N_{\mathrm{tracks}}^{\mathrm{3D}}$ smaller than 3 for both jets. 
The final signal region is defined by a GBDT score larger than 0.988, and the predicted background yield is $0.75\pm0.44\stat\pm0.39\syst$ events in the data 
samples collected in 2017 and 2018. After all the preselection criteria are applied, the efficiencies for signal events to have at least one SV/dijet candidate satisfying $\mathrm{GBDT}>0.988$ are generally above 80--90\% for the different LLP models considered with $m_{\mathrm{LLP}}\gtrsim300\GeV$ and $1\lesssim c\tau_{0}\lesssim1000\mm$, while the 
background rate has been reduced by a factor of around $3\times10^{3}$ by the GBDT selection. 

Event yields in data after different selection requirements have been applied are shown in Table~\ref{tab: cut_flow}. We observe one event in the final signal region, which is consistent with the predicted background yield. The observed event 
has an offline \HT of 570\GeV, and an SV candidate with an $L_{xy}$ of ${\approx}26\cm$ and 
8 associated tracks. The position of the SV is close to one of the silicon strip layers, and was likely 
produced by NIs with the silicon strip detector.  

\begin{table*}[hbt]
\topcaption{Event yields after different selection requirements have been applied for data collected in 2017 and 2018. Signal
efficiencies for the jet-jet model with $m_{\mathrm{X}}=1000\GeV$ and different $c\tau_{0}$ are also shown for comparison. 
Selection requirements are cumulative from the first row to the last.}
\label{tab: cut_flow}
\centering
\cmsTable{
\begin{scotch}{lrccc}
\multirow{4}*{Selections} & \multirow{4}*{Observed events} & \multicolumn{3}{c}{Signal efficiency ($\%$)} \\
                          &                                & \multicolumn{3}{c}{$m_{\mathrm{X}}=1000\GeV$}\\
                          &                                       &\multicolumn{3}{c}{$c\tau_{0}$}\\

                          &                                       & $3\mm$ & $30\mm$ & $300\mm$ \\\hline
Displaced-jet triggers, & \multirow{2}*{17758640} & \multirow{2}*{$69.4$} & \multirow{2}*{$91.2$} & \multirow{2}*{$80.5$} \\
offline $\HT$ selections &  &  &\\[\cmsTabSkip]
Offline jet $\pt$ and $\eta$ selections, &\multirow{2}*{8387775} &\multirow{2}*{$68.9$} & \multirow{2}*{$90.7$} & \multirow{2}*{$73.5$} \\
vertex $\chi^{2}/\mathrm{n_{dof}}<5.0$  & & & & \\[\cmsTabSkip]
Vertex $\pt>8\GeV$ & 3794960& $68.2$ & $90.3$ & $69.4$ \\
Vertex invariant mass $>4\GeV$ & 1129531 & $66.5$ & $89.3$ & $61.6$ \\
Second largest $\ipsig>15$ & 422449 & $66.0$ & $89.0$ & $60.9$ \\
Charged energy fraction from the SV $\epsilon>0.15$ & 93873 & $64.3$ & $87.6$ & $58.4$ \\
Energy fraction from the PVs $\zeta<0.20$ & 15891 & $63.6$ & $86.9$ & $57.9$ \\
Veto using the NI-veto map & 13721 & $63.6$ & $84.9$ & $55.4$ \\[\cmsTabSkip]
$N_{\mathrm{tracks}}^{\mathrm{3D}}<3$ for each jet & 2753 & $54.6$ & $76.4$ & $48.4$ \\
GBDT $>0.988$ & 1 & $51.5$ & $73.5$ & $42.5$ \\
\end{scotch}
}
\end{table*}

\subsection{Interpretation of the results}

The signal yields in the final signal region are used to set limits on a variety of models. The signal efficiencies for representative signal points in different models can be found in Tables~\ref{tab: eff_JetJet}--\ref{tab: eff_StopToDD} and Fig.~\ref{fig: eff_maps} of 
Appendix~\ref{sec: appen}. The results obtained with the 2017--2018 analysis are further combined in this paper 
with the results from the displaced-jets search using the $\Pp\Pp$ collision data collected with 
the CMS experiment in 2016~\cite{Sirunyan:2018vlw}, for which the systematic uncertainties arising from the same source are taken 
to be fully correlated, while the other systematic uncertainties and the statistical uncertainties in signal yields or expected background yields are taken to be uncorrelated. The total integrated luminosity is 132\fbinv. For the models that were not studied in the 2016 displaced-jets search, we have produced additional signal MC samples following the 2016 run condition of the CMS detector. We then 
process these samples with the reconstruction and selection procedures implemented in the 2016 search to compute the additional signal yields and systematic uncertainties for the 2016 data that are used in the combination. 

Upper limits on the cross section for a given model are presented with different masses and lifetimes by computing the 95\% confidence level 
(\CL) associated with each signal point using the \CLs prescription~\cite{Junk:1999kv,Read:2002hq,Cowan:2010js,CMS-NOTE-2011-005}, for which an LHC-style profile likelihood ratio~\cite{Cowan:2010js,CMS-NOTE-2011-005} is taken as the
test statistic. Systematic uncertainties are incorporated through the use of nuisance parameters treated 
according to the frequentist paradigm. The asymptotic approximation~\cite{Cowan:2010js} is used for the calculation of the \CLs values, which have been verified with
full-frequentist results for representative signal points. Since the background yields of this search are small, the impact of the 
associated statistical or systematic 
uncertainties on the upper limits are also small.  

\begin{figure}[hbtp]
\centering
\includegraphics[width=0.47\textwidth]{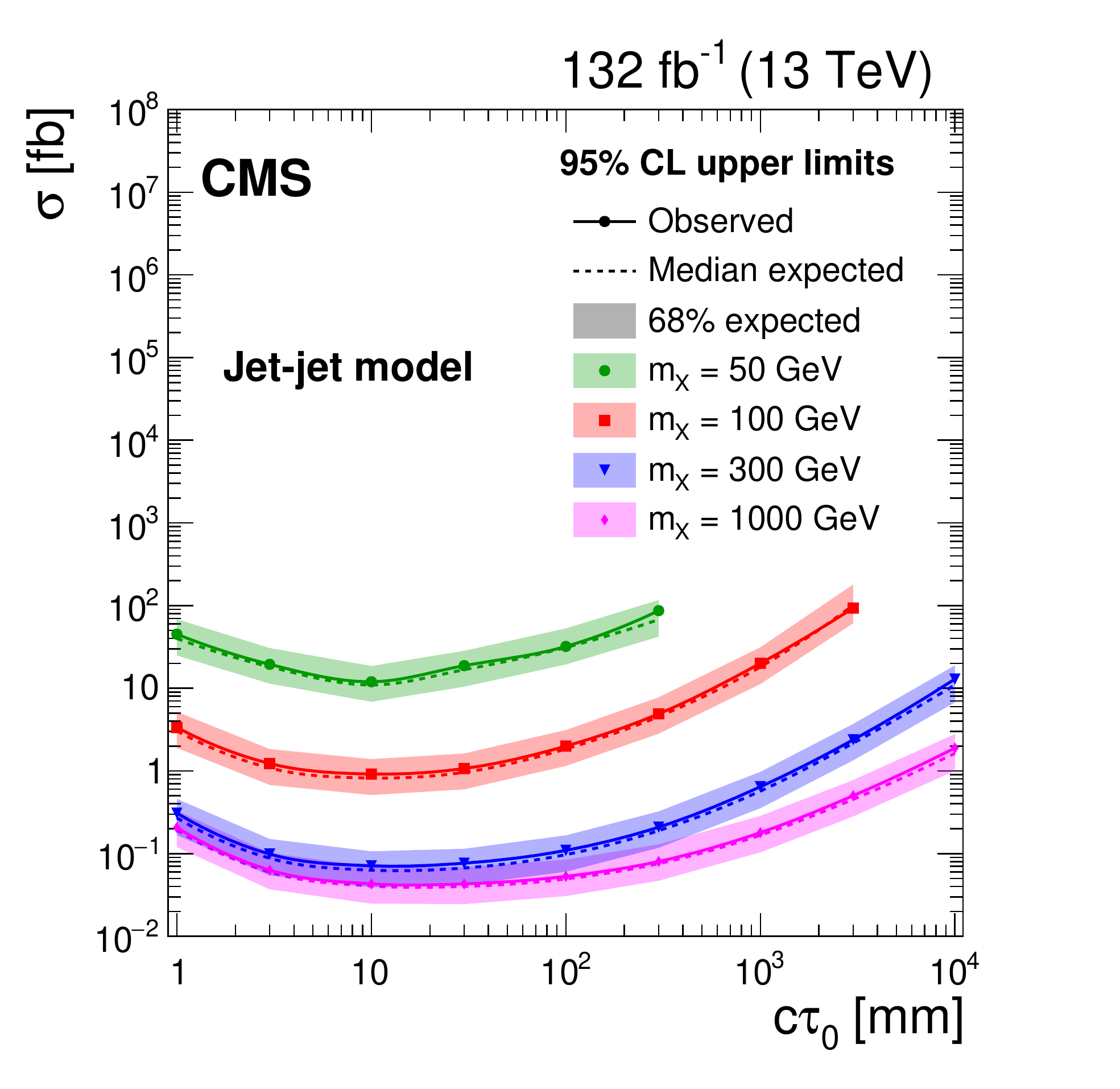}
\includegraphics[width=0.47\textwidth]{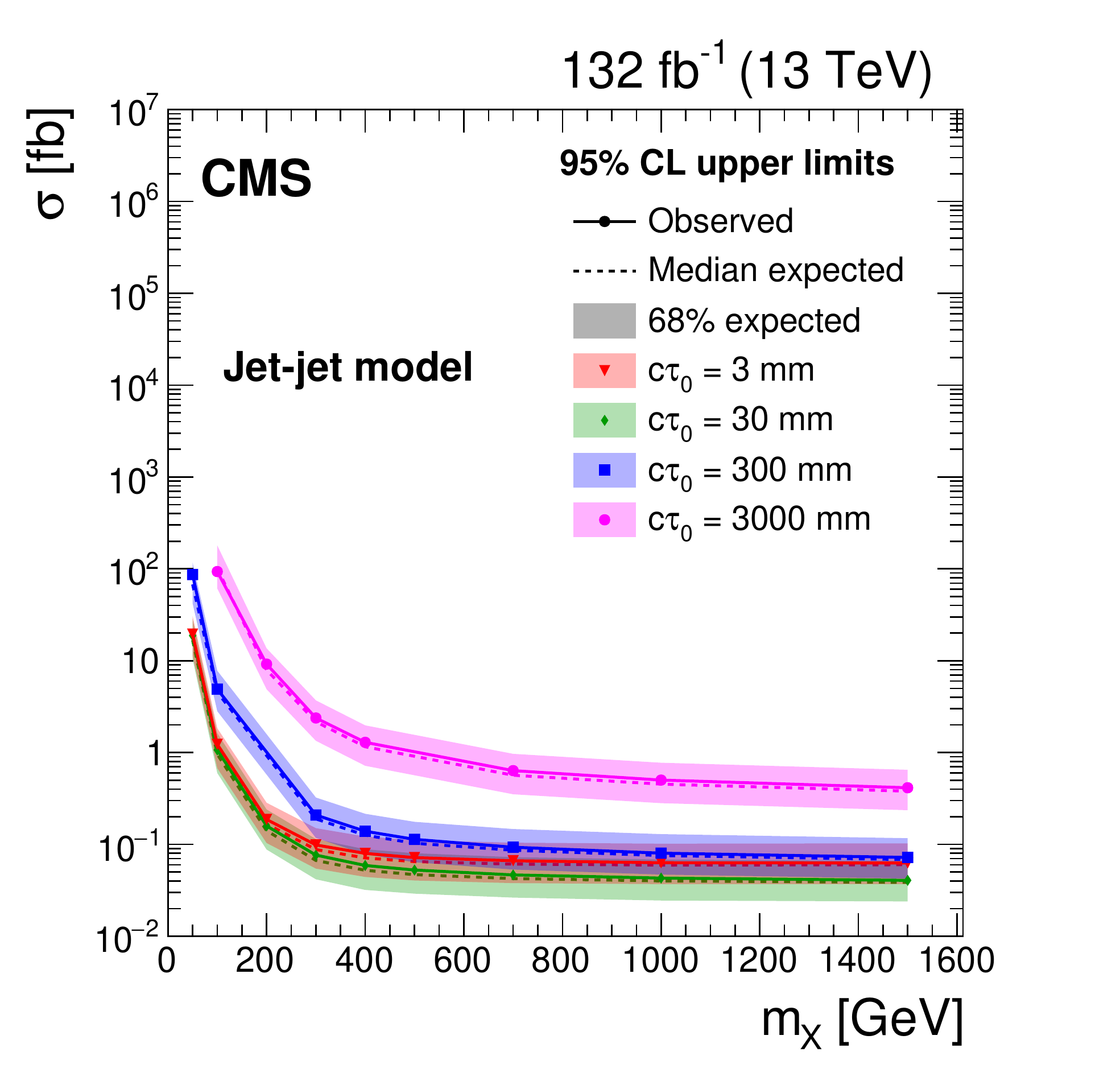}
\caption{The 95\% \CL upper limits on the pair production cross section of the LLP $\mathrm{X}$, where 
a 100\% branching fraction for $\mathrm{X}$ to decay to a quark-antiquark pair is assumed. \cmsLeft: the upper limits as functions of $c\tau_{0}$ for different masses. \cmsRight: the upper limits as functions of the particle mass for different $c\tau_{0}$. The solid (dashed) curves show the observed (median expected) limits. The shaded bands indicate the regions
containing 68\% of the distributions of the limits expected under the background-only hypothesis.}
\label{fig: limits_XXTo4J}
\end{figure}

The expected and observed upper limits on the pair production cross section for the jet-jet model
at different values of $c\tau_{0}$ and LLP mass $m_{\mathrm{X}}$ are shown in Fig.~\ref{fig: limits_XXTo4J}, where a branching fraction of $100\%$ for $\mathrm{X}$ to decay into a quark-antiquark pair is assumed. For a fixed LLP mass $m_{\mathrm{X}}$, the limits are most restrictive for $c\tau_{0}$ between 3 and 300\mm.
For smaller $c\tau_{0}$, the limits become less stringent, because we only select displaced tracks to reconstruct SVs and we veto dijet candidates with a large 
number of prompt tracks. The limits also become less restrictive for larger $c\tau_{0}$ ($c\tau_{0}>300\mm$), because the tracking efficiency 
becomes worse with large displacement of the SV. For high-mass LLPs ($m_{\mathrm{X}}>500\GeV$), 
pair production cross sections larger than 0.07\unit{fb} are excluded for $c\tau_{0}$ between 2 and 250\mm.
The lowest pair production cross section excluded is 0.04\unit{fb}, at $c\tau_{0}=30\mm$ and $m_{\rm{X}}>1000\GeV$.

\begin{figure}[hbtp]
\centering
\includegraphics[width=0.47\textwidth]{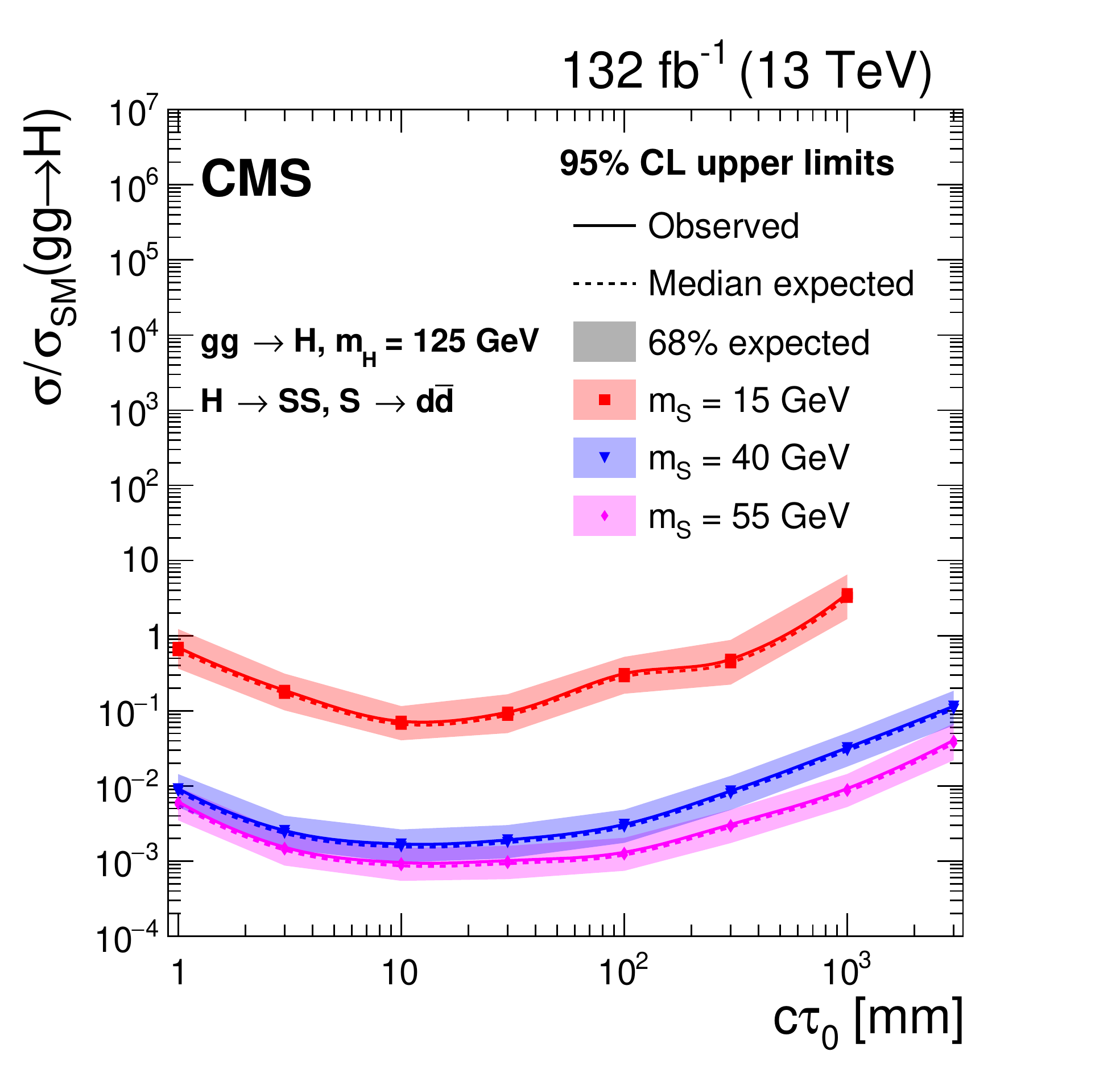}
\includegraphics[width=0.47\textwidth]{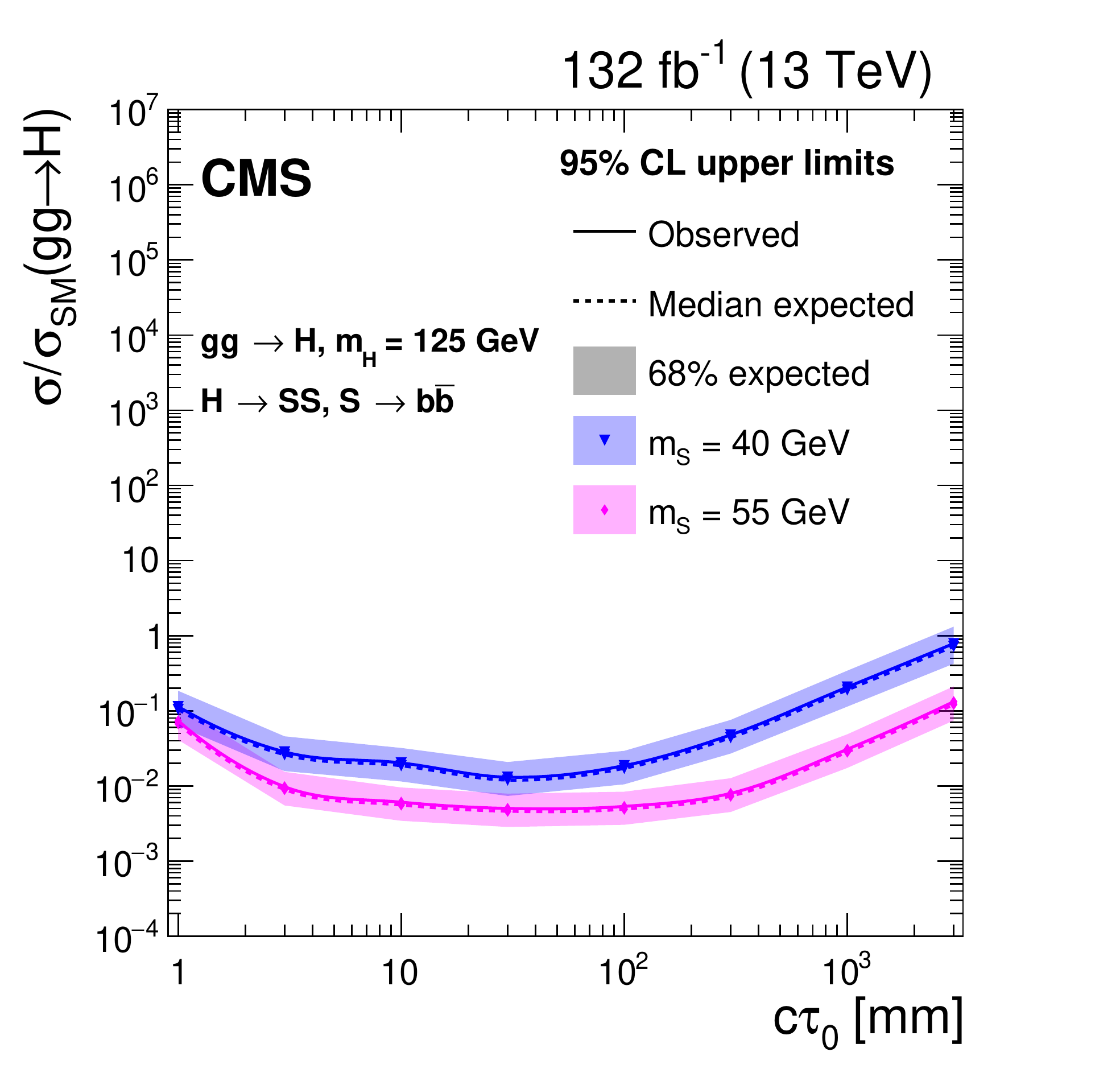}
\caption{The expected and observed 95$\%$ $\CL$ upper limits on the branching fraction of the SM-like Higgs boson to decay to 
two long-lived scalar particles, assuming
the gluon-gluon fusion Higgs boson production cross section of 49\unit{pb} at 13\TeV with $m_{\PH}=125\GeV$, shown at different masses and
$c\tau_{0}$ for the scalar particle $\mathrm{S}$. \cmsLeft: the upper limits when each scalar particle decays to a down quark-antiquark pair. \cmsRight: the upper limits when each
scalar particle decays to a bottom quark-antiquark pair. The solid (dashed) curves represent the observed (median expected) limits. The shaded bands represent the regions
containing 68\% of the distributions of the expected limits under the background-only hypothesis.}
\label{fig: limits_Higgs}
\end{figure}

Figure~\ref{fig: limits_Higgs} presents the expected and observed upper limits on the branching fraction 
of the SM-like Higgs boson to decay to two long-lived scalar particles $\mathrm{S}$, each of which decays 100\% of times to a quark-antiquark pair of 
a specific flavor. The upper limits on the branching fraction are calculated assuming the 
gluon-gluon fusion production cross section of a 125\GeV Higgs boson at 13\TeV~\cite{deFlorian:2016spz}. When the long-lived scalar particle decays to a light-flavor 
quark-antiquark pair, branching fractions larger than 1\% are excluded for $c\tau_{0}$ between 1 and 340\mm with 
$m_{\mathrm{S}}\geq40\GeV$. When the long-lived scalar particle decays to a bottom quark-antiquark pair, branching fractions larger 
than $10\%$ are excluded for $c\tau_{0}$ between 1 and 530\mm with $m_{\mathrm{S}}\geq40\GeV$. These are the most stringent limits to date 
on this model for $c\tau_{0}$ between 1 and 1000\mm. For $m_{\mathrm{S}}=15\GeV$, 
where the track multiplicity of the SV is small, and the tracks are collimated due to 
the boost of $\mathrm{S}$, 
the limits become worse. The limits are also worse for 
the case where the scalar particle decays to a bottom quark-antiquark pair, because the decays of \PQb hadrons can produce 
tertiary vertices, which can be missed by the SV reconstruction we deploy in this search. 

\begin{figure}[hbtp]
\centering
\includegraphics[width=0.45\textwidth]{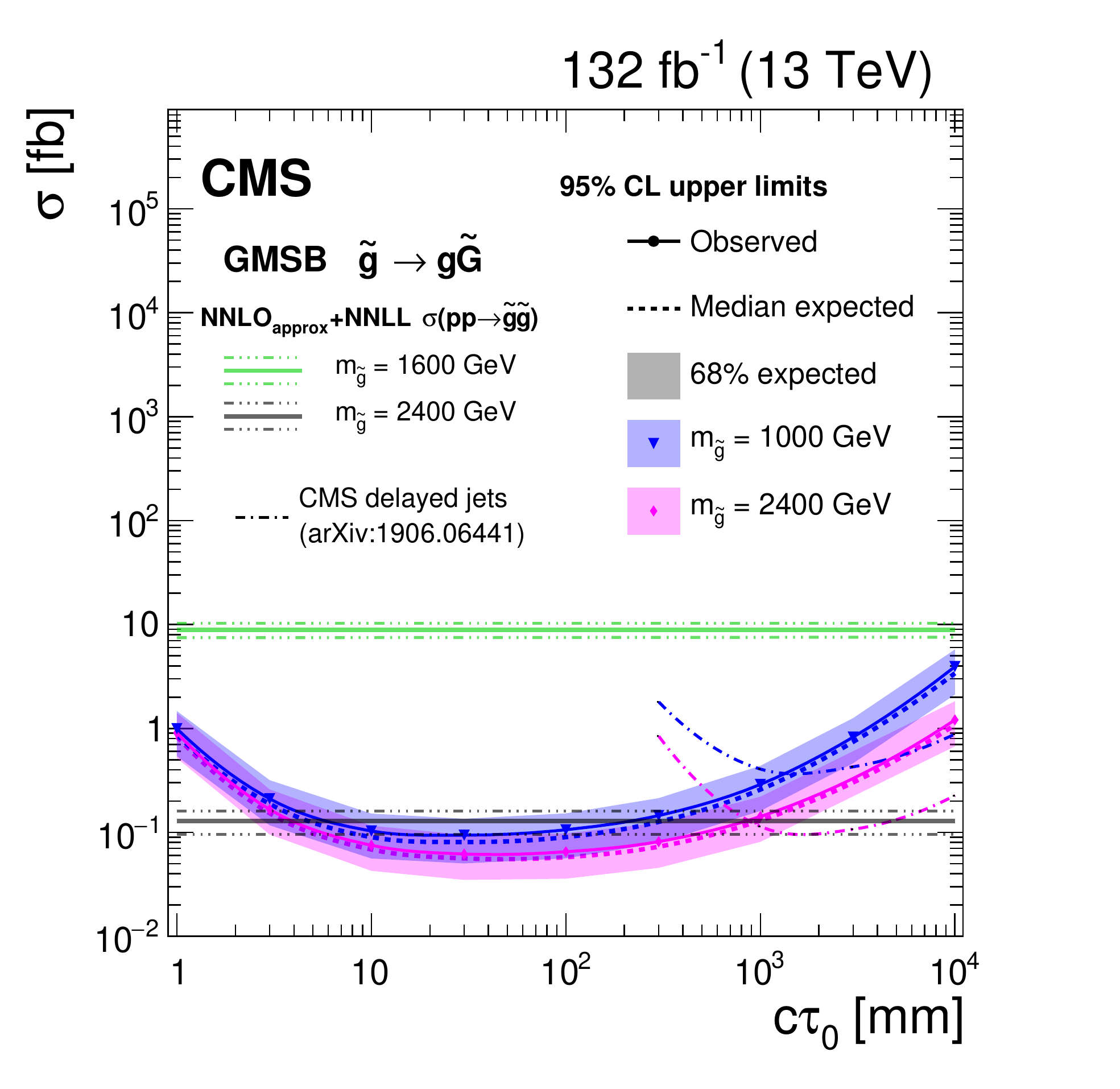}
\includegraphics[width=0.45\textwidth]{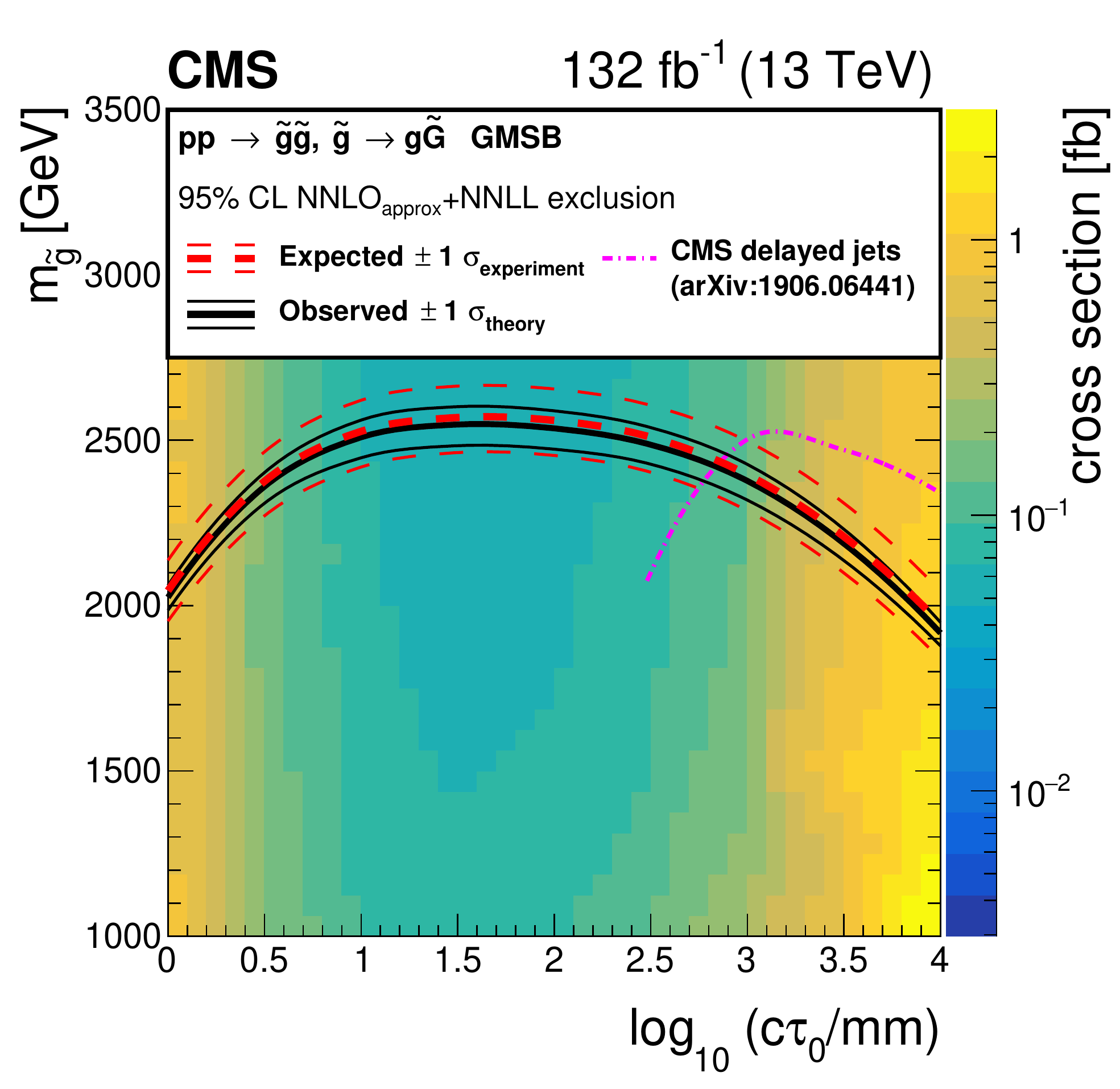}
\caption{\cmsLeft: the 95\% \CL upper limits on the pair production cross section for the long-lived gluinos with $m_{\sGlu}=2400$ and 1600\GeV, where a
100\% branching fraction for $\sGlu\to\Glu\sGra$ decays is assumed. The $\mathrm{NNLO_{approx}}$+NNLL gluino pair production cross sections for $m_{\sGlu}=2400$ and 1600\GeV, as well as their variations due to the theoretical uncertainties, are shown as horizontal lines. The solid (dashed) curves show the observed (median expected) limits, and the shaded bands indicate the regions containing 68\% of
the distributions of the limits expected under the background-only hypothesis. The observed limits from the CMS search utilizing the 
timing capabilities of the ECAL system~\cite{Sirunyan:2019gut} are also shown for comparison. 
\cmsRight: the 95\% \CL upper limits on the pair production cross section for the $\sGlu\to\Glu\sGra$ model as a function of the mean proper decay length $c\tau_{0}$ and the gluino mass $m_{\sGlu}$. 
The thick solid black
(dashed red) curve shows the observed (median expected) 95\% \CL limit on the gluino mass as a function of $c\tau_{0}$, assuming the $\mathrm{NNLO_{approx}}$+NNLL cross sections. The thin dashed red curves indicate the region containing 68\% of
the distribution of the limits expected under the background-only hypothesis. The thin solid black curves represent the change in 
the observed limit when the signal cross sections are varied according to their theoretical uncertainties.}
\label{fig: limits_gluinoGMSB}
\end{figure}

The expected and observed upper limits on the pair production cross section of long-lived gluinos in
the GMSB $\sGlu\to\Glu\sGra$ model are shown in Fig.~\ref{fig: limits_gluinoGMSB}, where a branching fraction of $100\%$ for the gluino to decay into a gluon and a gravitino is assumed.
Since we do not require the reconstructed SV to have associated tracks from both jets, the two separate displaced single jets produced by 
the decays of the two long-lived gluinos in the  $\sGlu\to\Glu\sGra$ model can be paired
together and pass the selections, therefore the search is sensitive to the models with similar signatures. When the gluino mass is
2400\GeV, signal efficiencies are around 21, 53, and 41\% in the 2017 and 2018 analysis for $c\tau_{0}=3$, 30, and 300\mm, respectively.
With the data samples collected in 2016--2018, gluino pair production cross sections larger than 0.1\unit{fb} are excluded for $c\tau_{0}$ between 7 and 600\mm at $m_{\sGlu}=2400\GeV$. 
We then compute the upper limits on the gluino mass for different $c\tau_{0}$ according to the upper limits on the pair production cross section, and a calculation at next-to-next-to-leading logarithmic precision matched to the approximated next-to-next-to-leading order
predictions ($\mathrm{NNLO_{approx}}$+NNLL) of the gluino pair production cross section at $\sqrt{s}=13\TeV$~\cite{Beenakker:1996ch,Kulesza:2008jb,Kulesza:2009kq,Beenakker:2011fu,Borschensky:2014cia,Beenakker:2016lwe}. Gluino masses up to 2450\GeV
are excluded for $c\tau_{0}$ between 6 and 550\mm. The largest gluino mass excluded is 2560\GeV with a $c\tau_{0}$ of 30\mm. These limits are the most restrictive to date on this model for $c\tau_{0}$ between 1 and 1000\mm. 

\begin{figure}[hbtp]
\centering
\includegraphics[width=0.45\textwidth]{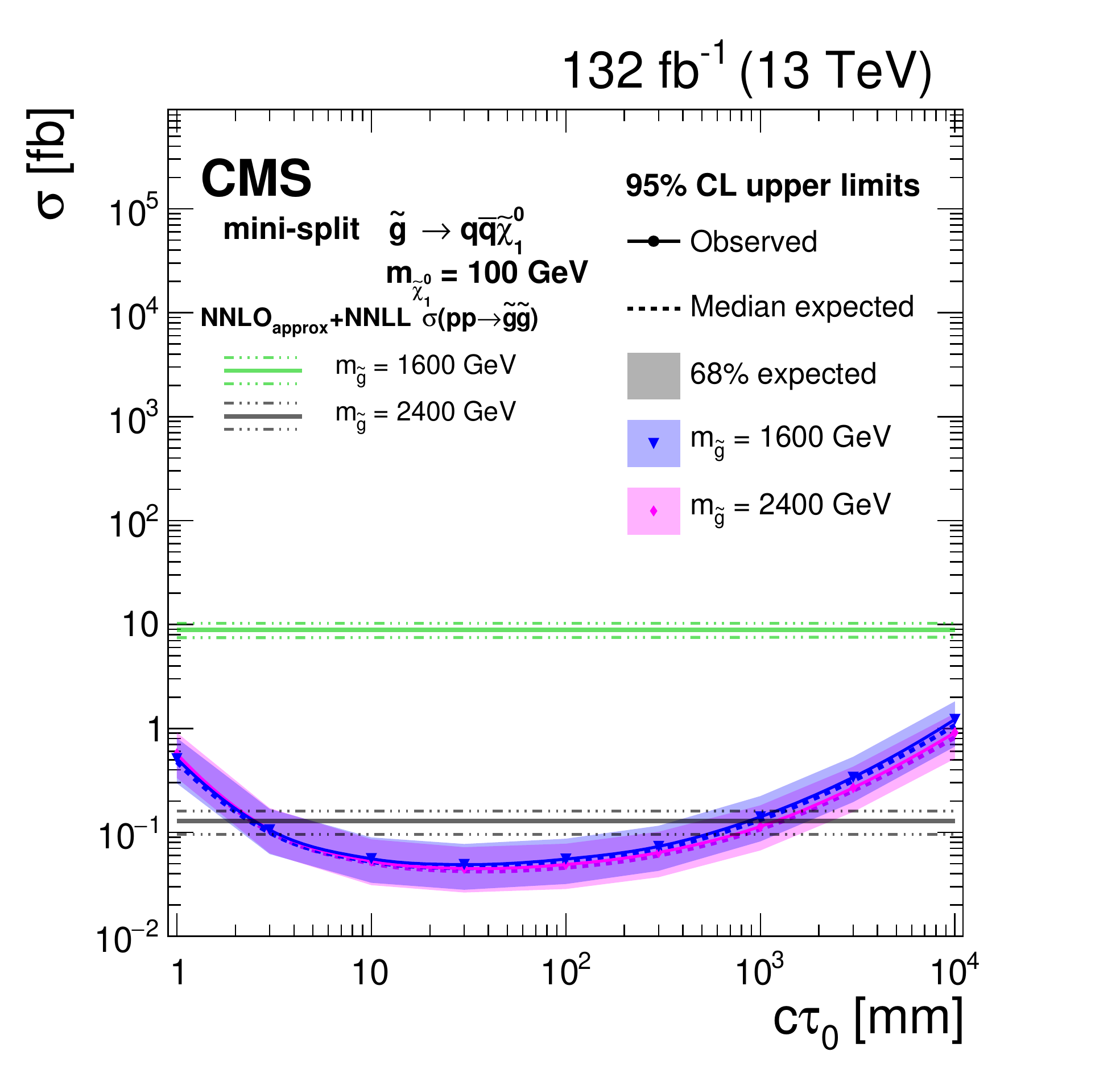}
\includegraphics[width=0.45\textwidth]{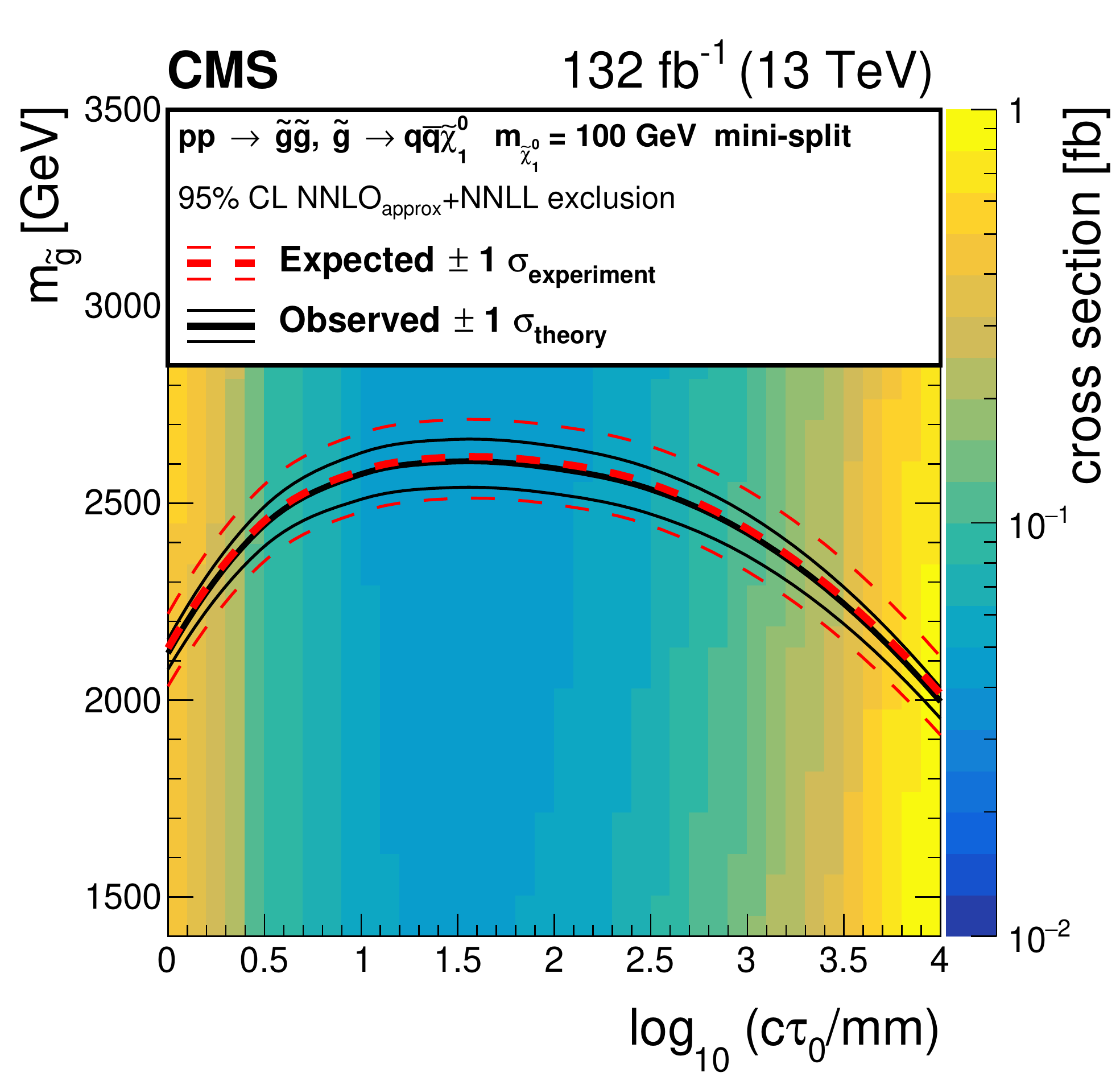}
\caption{\cmsLeft: the 95\% \CL upper limits on the pair production cross section for the long-lived gluinos with $m_{\sGlu}=2400\GeV$ and 1600\GeV, where a
100\% branching fraction for $\sGlu\to\cPq\cPaq\widetilde{\chi}^{0}_{1}$ decays is assumed. The $\mathrm{NNLO_{approx}}$+NNLL gluino pair production cross sections for $m_{\sGlu}=2400$ and 1600\GeV, as well as their variations due to the theoretical uncertainties, are shown as horizontal lines. The solid (dashed) curves show the observed (median expected) limits, and the shaded bands indicate the regions containing 68\% of
the distributions of the limits expected under the background-only hypothesis.
\cmsRight: the 95\% \CL limits on the pair production cross section for the $\sGlu\to\cPq\cPaq\widetilde{\chi}^{0}_{1}$ model as a function of the mean proper decay length $c\tau_{0}$ and the 
gluino mass $m_{\sGlu}$. The thick solid black
(dashed red) curve shows the observed (median expected) 95\% \CL limits on the gluino mass as a function of $c\tau_{0}$, assuming the $\mathrm{NNLO_{approx}}$+NNLL cross sections. The thin 
dashed red curves indicate the region containing 68\% of
the distribution of the limits expected under the background-only hypothesis. The thin solid black curves represent the change in 
the observed limit when the signal cross sections are varied according to their theoretical uncertainties.} 
\label{fig: limits_SplitSUSY}
\end{figure}

Figure~\ref{fig: limits_SplitSUSY} shows the expected and observed upper limits on the pair production cross section 
of the long-lived gluinos in the minisplit $\sGlu\to\cPq\cPaq\widetilde{\chi}^{0}_{1}$ model, assuming a branching fraction of 100\% for 
the gluino to decay into a quark-antiquark pair and the lightest neutralino. The neutralino mass is assumed to be 100\GeV. When 
the gluino mass is 2400\GeV, signal efficiencies are around 31, 69, and 51\% in the 2017 and 2018 analysis for $c\tau_{0}=3$, 30, and 300\mm, 
respectively. With the data samples collected in 2016--2018, gluino pair production cross sections larger than 0.1\unit{fb} are excluded for proper decay 
lengths between 3 and 900\mm. The upper limits on the pair production cross sections are then translated into upper 
limits on the gluino mass for different $c\tau_{0}$, based on the $\mathrm{NNLO_{approx}}$+NNLL gluino pair 
production cross sections. Gluino masses up to 2500\GeV are excluded for $c\tau_{0}$ between 7 and 360\mm. 
The largest gluino mass excluded is 2610\GeV with a $c\tau_{0}$ of 30\mm. These bounds are the most stringent 
to date on this model for $c\tau_{0}$ between 10 and 1000\mm. 

\begin{figure}[hbtp]
\centering
\includegraphics[width=0.45\textwidth]{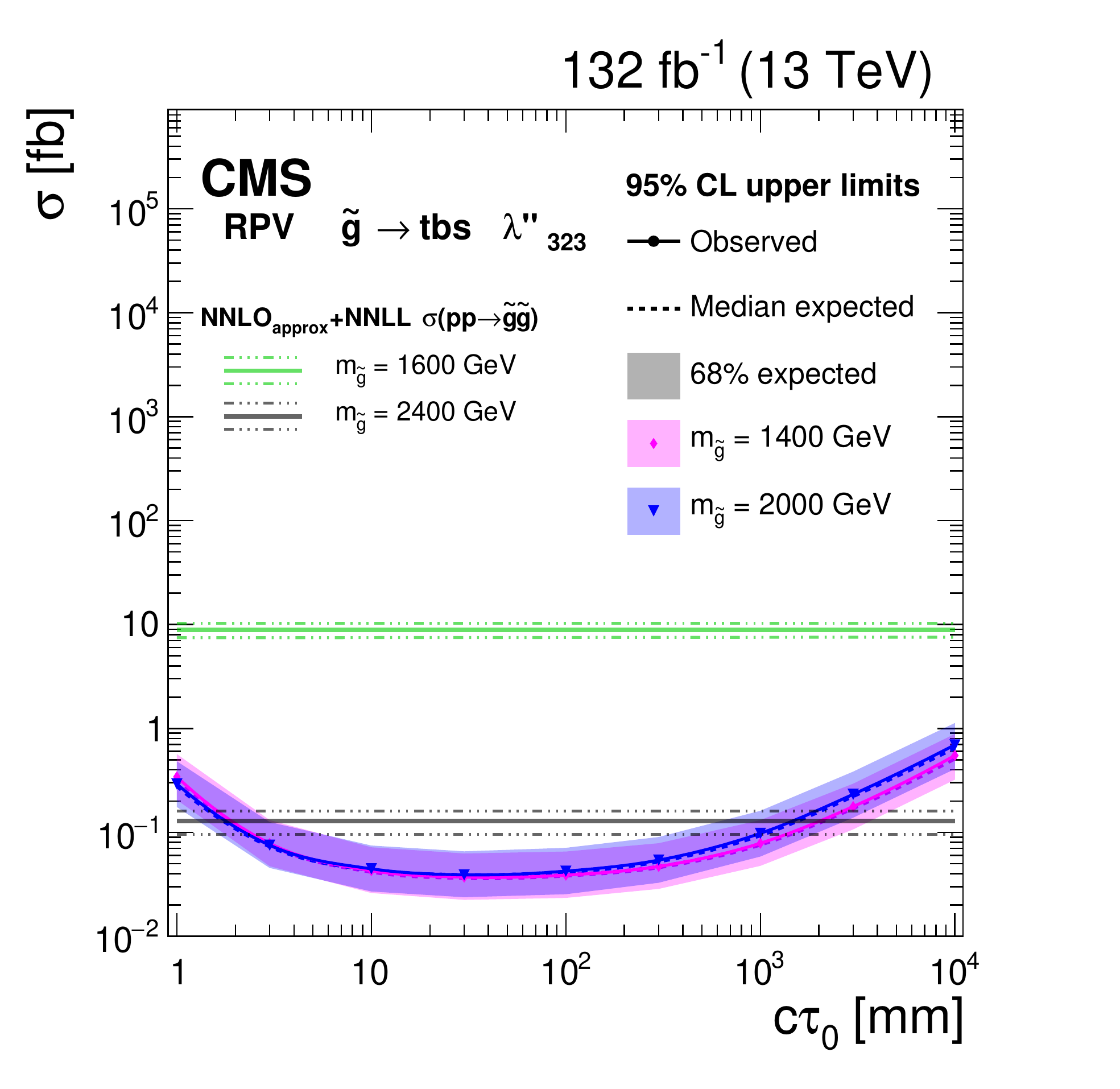}
\includegraphics[width=0.45\textwidth]{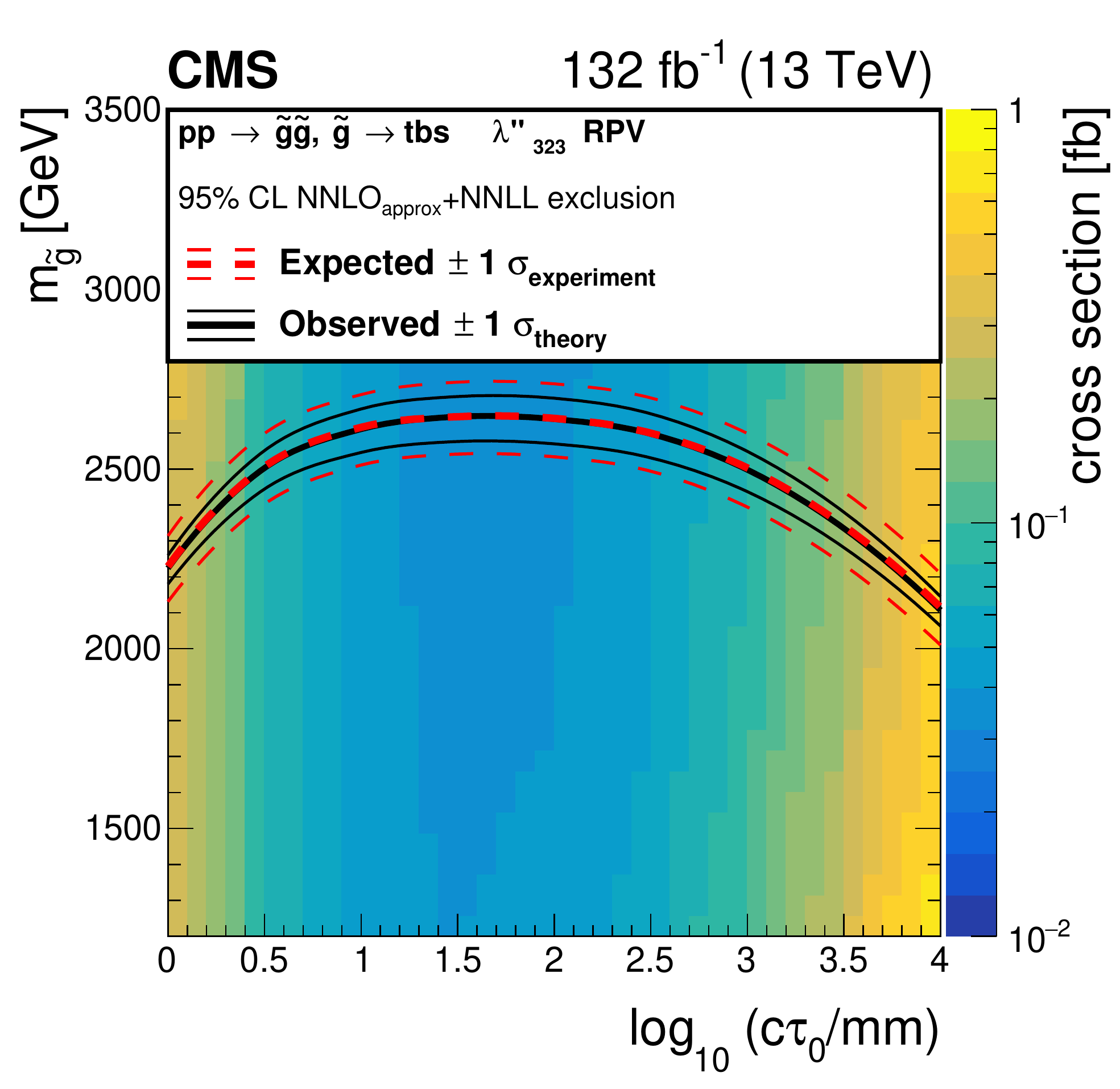}
\caption{\cmsLeft: the 95\% \CL upper limits on the pair production cross section for the long-lived gluinos with $m_{\sGlu}=2000\GeV$ and 1400\GeV, where a
$100\%$ branching fraction for $\sGlu\to\PQt\PQb\cPqs$ decays is assumed. The $\mathrm{NNLO_{approx}}$+NNLL gluino pair production cross sections for $m_{\sGlu}=2400$ and 1600\GeV, as well as their variations due to the theoretical uncertainties, are shown as horizontal lines. The solid (dashed) curves show the observed (median expected) limits, and the shaded bands indicate the regions containing 68\% of
the distributions of the limits expected under the background-only hypothesis.
\cmsRight: the 95\% \CL limits on the pair production cross section for the $\sGlu\to\PQt\PQb\cPqs$ model as a function of the mean proper decay length $c\tau_{0}$ and the 
gluino mass $m_{\sGlu}$. The thick solid black
(dashed red) curve shows the observed (median expected) 95\% \CL limits on the gluino mass as a function of $c\tau_{0}$, assuming the $\mathrm{NNLO_{approx}}$+NNLL cross sections. The thin 
dashed red curves indicate the region containing $68\%$ of
the distribution of the limits expected under the background-only hypothesis. The thin solid black curves represent the change in 
the observed limit when the signal cross sections are varied according to their theoretical uncertainties.} 
\label{fig: limits_GGToNN}
\end{figure}

The expected and observed upper limits on the pair production cross section of the long-lived gluinos in the $\sGlu\to\PQt\PQb\cPqs$ 
model are shown in Fig.~\ref{fig: limits_GGToNN} , where a branching fraction of 100\% for the gluino to decay into top, bottom, and strange quarks is assumed. When the gluino mass is 2400\GeV, signal efficiencies are around 41, 81, and 66\% in 
the 2017 and 2018 analysis for $c\tau_{0}=3$, 30, and 300\mm, respectively. With the data samples collected in 2016--2018, 
gluino pair production cross 
sections larger than 0.1\unit{fb} are excluded for $c\tau_{0}$ between 3 and 1490\mm at $m_{\sGlu}=2400\GeV$. We then compute the upper limits on the gluino mass for different $c\tau_{0}$ according to the upper limits on the pair production cross section and the calculation of the $\mathrm{NNLO_{approx}}$+NNLL gluino pair production cross sections. Gluino masses 
up to 2500\GeV are excluded for $c\tau_{0}$ between 3 and 1000\mm. The largest gluino mass excluded is 2640\GeV with a $c\tau_{0}$ of 30\mm. These 
limits are the most stringent to date on this model for $c\tau_{0}$ between 30 and $10^{4}\mm$.

\begin{figure}[hbtp]
\centering
\includegraphics[width=0.45\textwidth]{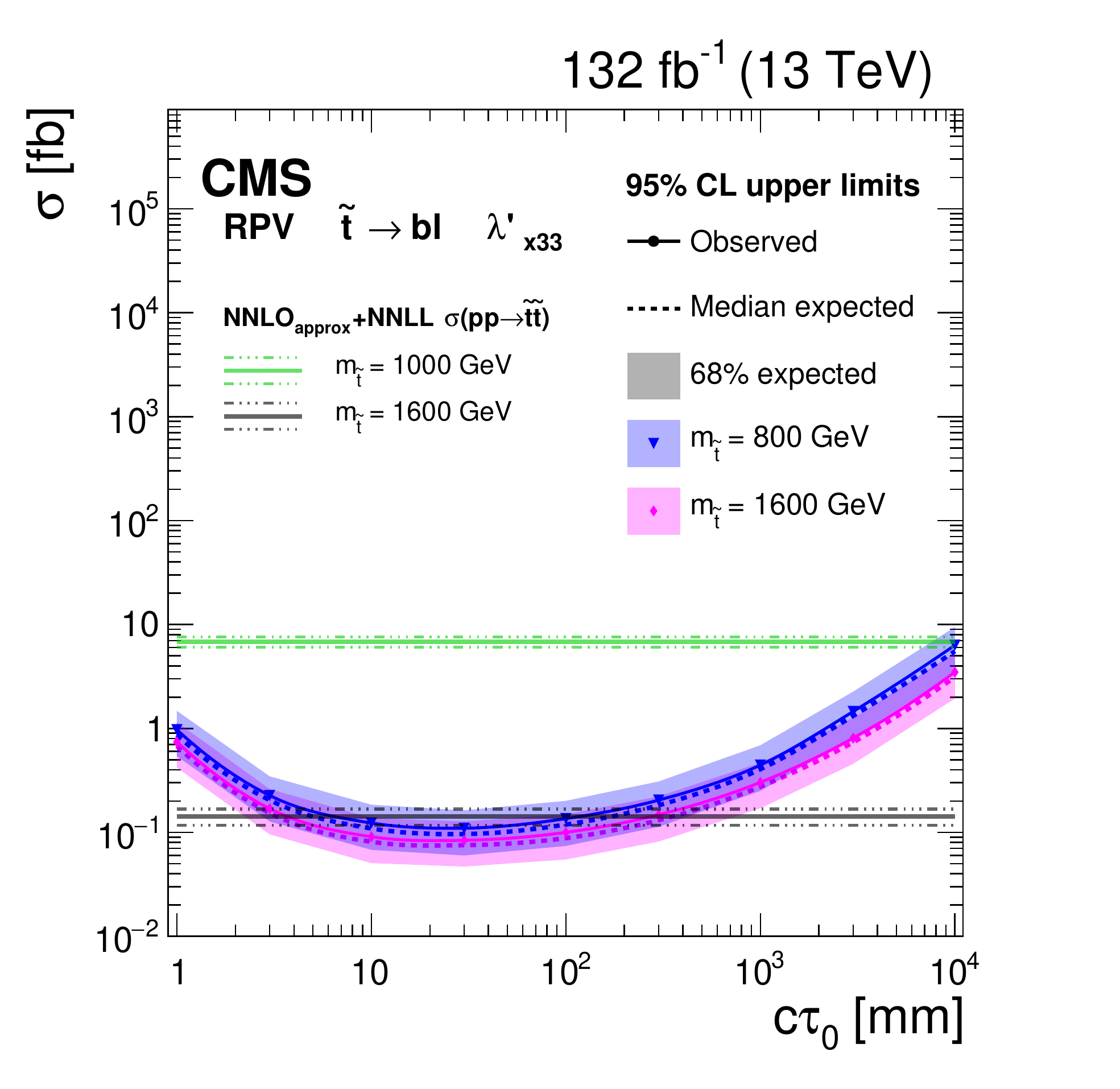}
\includegraphics[width=0.45\textwidth]{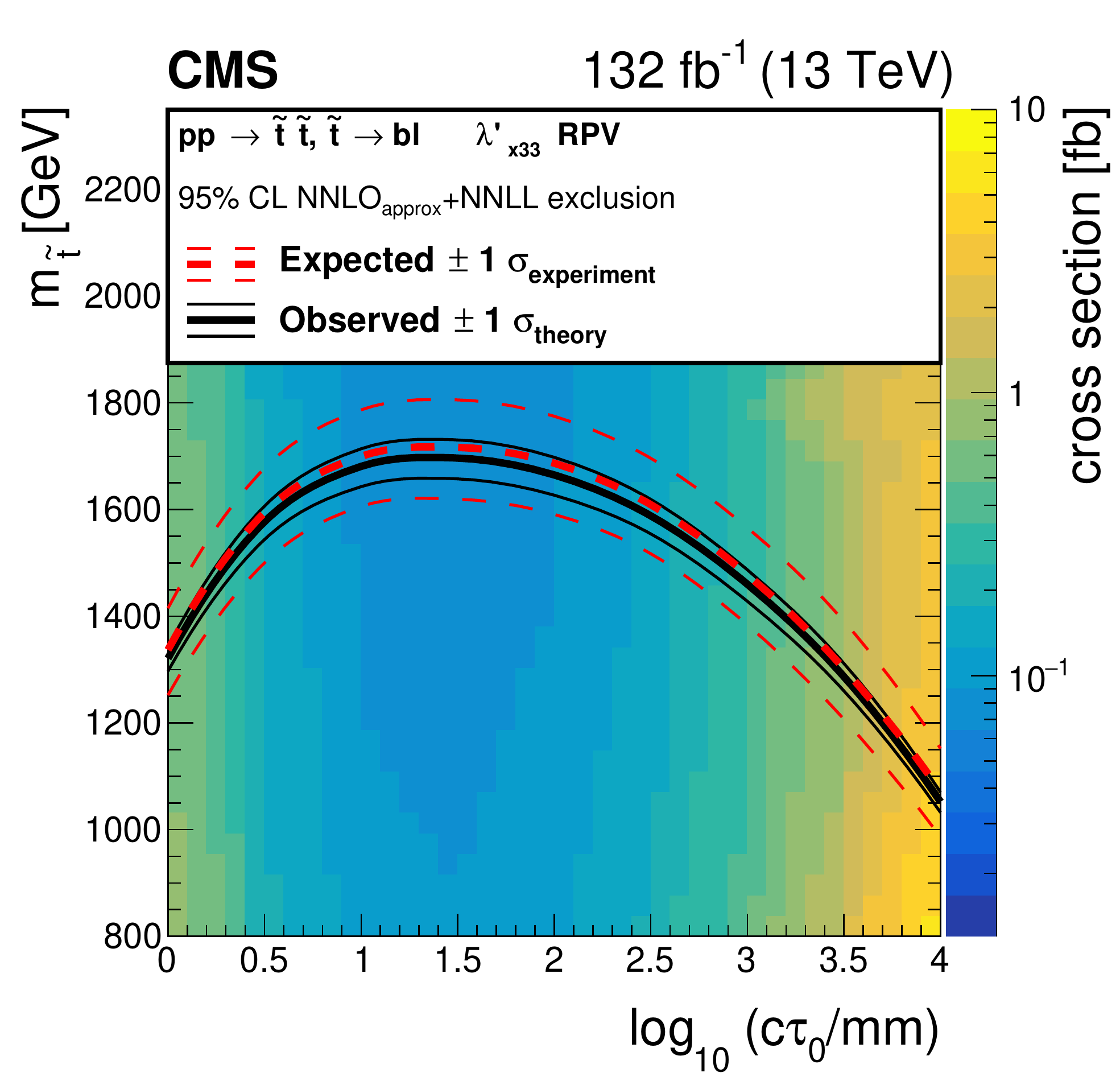}
\caption{\cmsLeft: the 95\% \CL upper limits on the pair production cross section for the long-lived top squarks with $m_{\sTop}=1600\GeV$ and 800\GeV, where a
100\% branching fraction for $\sTop\to\PQb\ell$ decays is assumed, with equal branching fractions for \Pe, $\mu$, and $\tau$. The $\mathrm{NNLO_{approx}}$+NNLL top squark pair production cross sections for $m_{\sTop}=1600$ and 1000\GeV, as well as their variations due to the theoretical uncertainties, are shown as horizontal lines. The solid (dashed) curves show the observed (median expected) limits, and the shaded bands indicate the regions containing 68\% of
the distributions of the limits expected under the background-only hypothesis.
\cmsRight: the 95\% \CL limits on the pair production cross section for the $\sTop\to\PQb\ell$ model as a function of the mean proper decay length $c\tau_{0}$ and the
top squark mass $m_{\sTop}$. The thick solid black
(dashed red) curve shows the observed (median expected) 95\% \CL limits on the top squark mass as a function of $c\tau_{0}$, assuming the $\mathrm{NNLO_{approx}}$+NNLL cross sections. The thin
dashed red curves indicate the region containing 68\% of
the distribution of the limits expected under the background-only hypothesis. The thin solid black curves represent the change in
the observed limit when the signal cross sections are varied according to their theoretical uncertainties.}
\label{fig: limits_stopToLB}
\end{figure}

The expected and observed upper limits on the pair production cross section of the
long-lived top squarks in the RPV $\sTop\to\PQb\ell$ model are shown in Fig.~\ref{fig: limits_stopToLB}, where a branching fraction of $100\%$ for the
top squark to decay into a bottom quark and a charged lepton is assumed, with equal branching fractions for \Pe, $\mu$, and $\tau$. When the top squark mass is 1600\GeV, signal efficiencies 
are around 22, 43, and 26\% in the 2017 and 2018 analysis for $c\tau_{0}=3$, 30, and 300\mm, respectively. With the data samples collected in 2016--2018, top squark pair production cross
sections larger than 0.1\unit{fb} are excluded for $c\tau_{0}$ between 8 and 160\mm at $m_{\sTop}=1600\GeV$. We then compute the upper limits on the top squark mass for different $c\tau_{0}$ according 
to the upper limits on the pair production cross section and the calculation of the $\mathrm{NNLO_{approx}}$+NNLL top squark pair production cross sections. Top squark masses
up to 1600\GeV are excluded for $c\tau_{0}$ between 5 and 240\mm. The largest top squark mass excluded is
1720\GeV with a $c\tau_{0}$ of 30\mm. These limits are the most stringent to date on this model in the tested
$c\tau_{0}$ range. 

\begin{figure}[hbtp]
\centering
\includegraphics[width=0.45\textwidth]{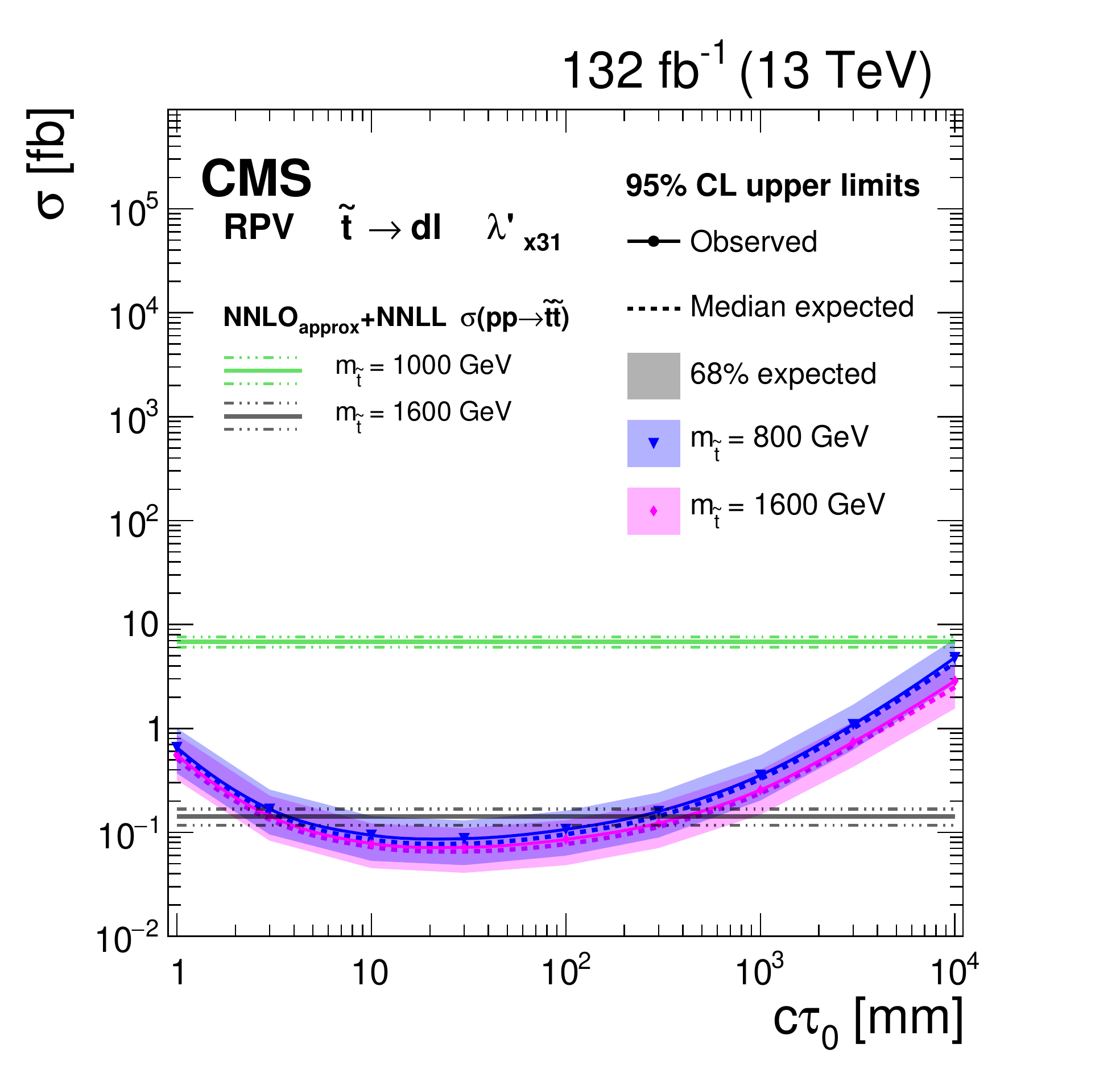}
\includegraphics[width=0.45\textwidth]{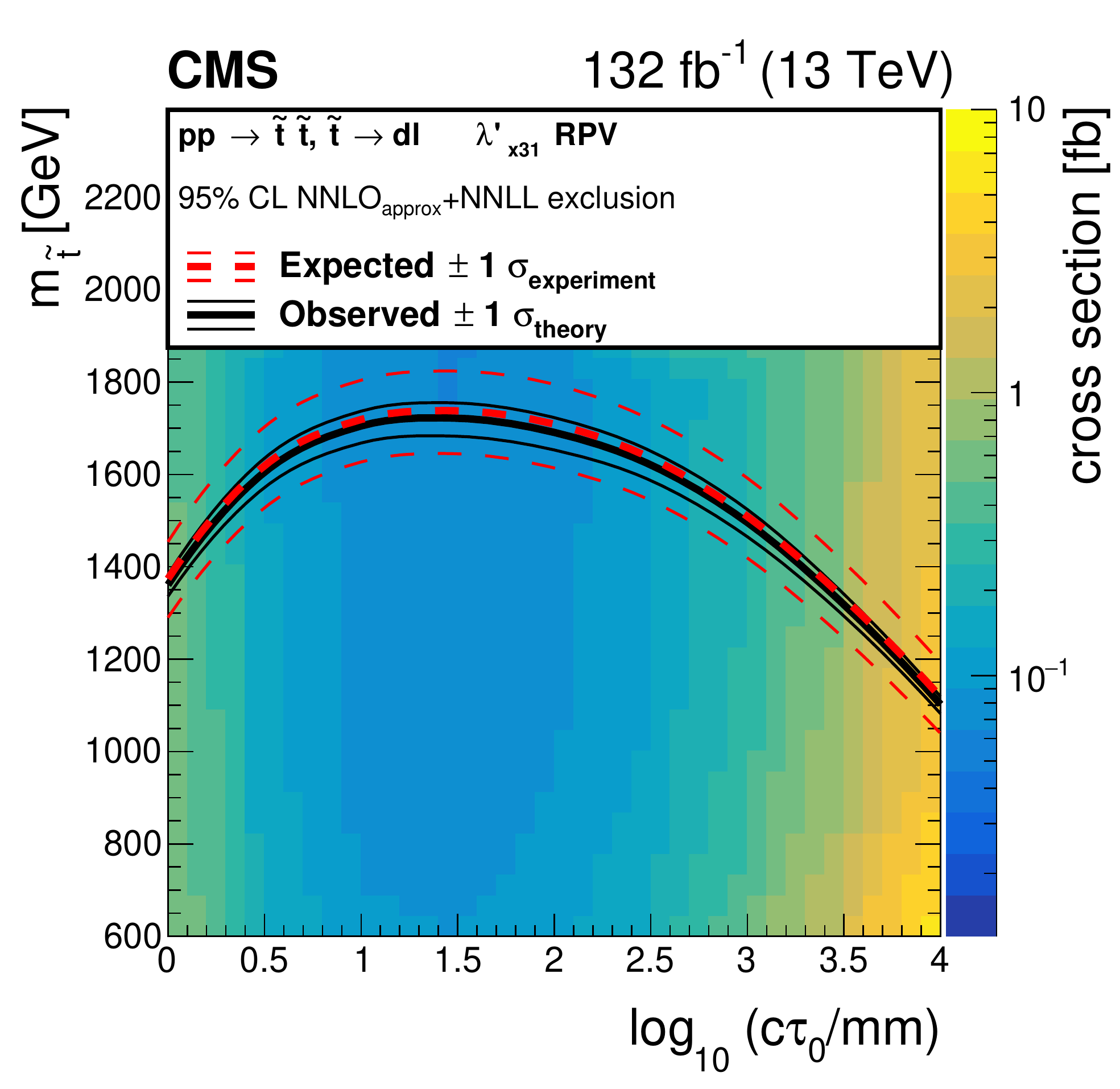}
\caption{\cmsLeft: the 95\% \CL upper limits on the pair production cross section for the long-lived top squarks with $m_{\sTop}=1600\GeV$ and 800\GeV, where a
100\% branching fraction for $\sTop\to\PQd\ell$ decays is assumed, with equal branching fractions for \Pe, $\mu$, and $\tau$. The $\mathrm{NNLO_{approx}}$+NNLL top squark pair production cross sections for $m_{\sTop}=1600$ and 1000\GeV, as well as their variations due to the theoretical uncertainties, are shown as horizontal lines. The solid (dashed) curves show the observed (median expected) limits, and the shaded bands indicate the regions containing 68\% of
the distributions of the limits expected under the background-only hypothesis.
\cmsRight: the 95\% \CL limits on the pair production cross section for the $\sTop\to\PQd\ell$ model as a function of the mean proper decay length $c\tau_{0}$ and the
top squark mass $m_{\sTop}$. The thick solid black
(dashed red) curve shows the observed (median expected) 95\% \CL limits on the top squark mass as a function of $c\tau_{0}$, assuming the $\mathrm{NNLO_{approx}}$+NNLL cross sections. The thin
dashed red curves indicate the region containing 68\% of
the distribution of the limits expected under the background-only hypothesis. The thin solid black curves represent the change in
the observed limit when the signal cross sections are varied according to their theoretical uncertainties.}
\label{fig: limits_stopToLD}
\end{figure}

The expected and observed upper limits on the pair production cross section of the
long-lived top squarks in the RPV $\sTop\to\PQd\ell$ model are shown in Fig.~\ref{fig: limits_stopToLD}, where a branching fraction of 100\% for the
top squark to decay into a down quark and a charged lepton, with equal branching fractions for \Pe, $\mu$, and $\tau$. When the top squark mass is 1600\GeV, signal efficiencies 
are around 25, 48, and 29\% in the 2017 and 2018 analysis for $c\tau_{0}=3$, 30, and 300\mm, respectively. With the data samples collected in 2016--2018, and top squark pair production cross
sections larger than 0.1\unit{fb} are excluded for $c\tau_{0}$ between 7 and 220\mm. We then compute the upper limits on the top squark mass for different $c\tau_{0}$ according to the upper limits on the pair production cross section and the calculation of the $\mathrm{NNLO_{approx}}$+NNLL top squark pair production cross sections. Top squark masses
up to 1600\GeV are excluded for $c\tau_{0}$ between 3 and 360\mm. The largest top squark mass excluded is
1740\GeV with a $c\tau_{0}$ of 30\mm. These limits are the most restrictive to date on this model in the tested
$c\tau_{0}$ range. 

\begin{figure}[hbtp]
\centering
\includegraphics[width=0.45\textwidth]{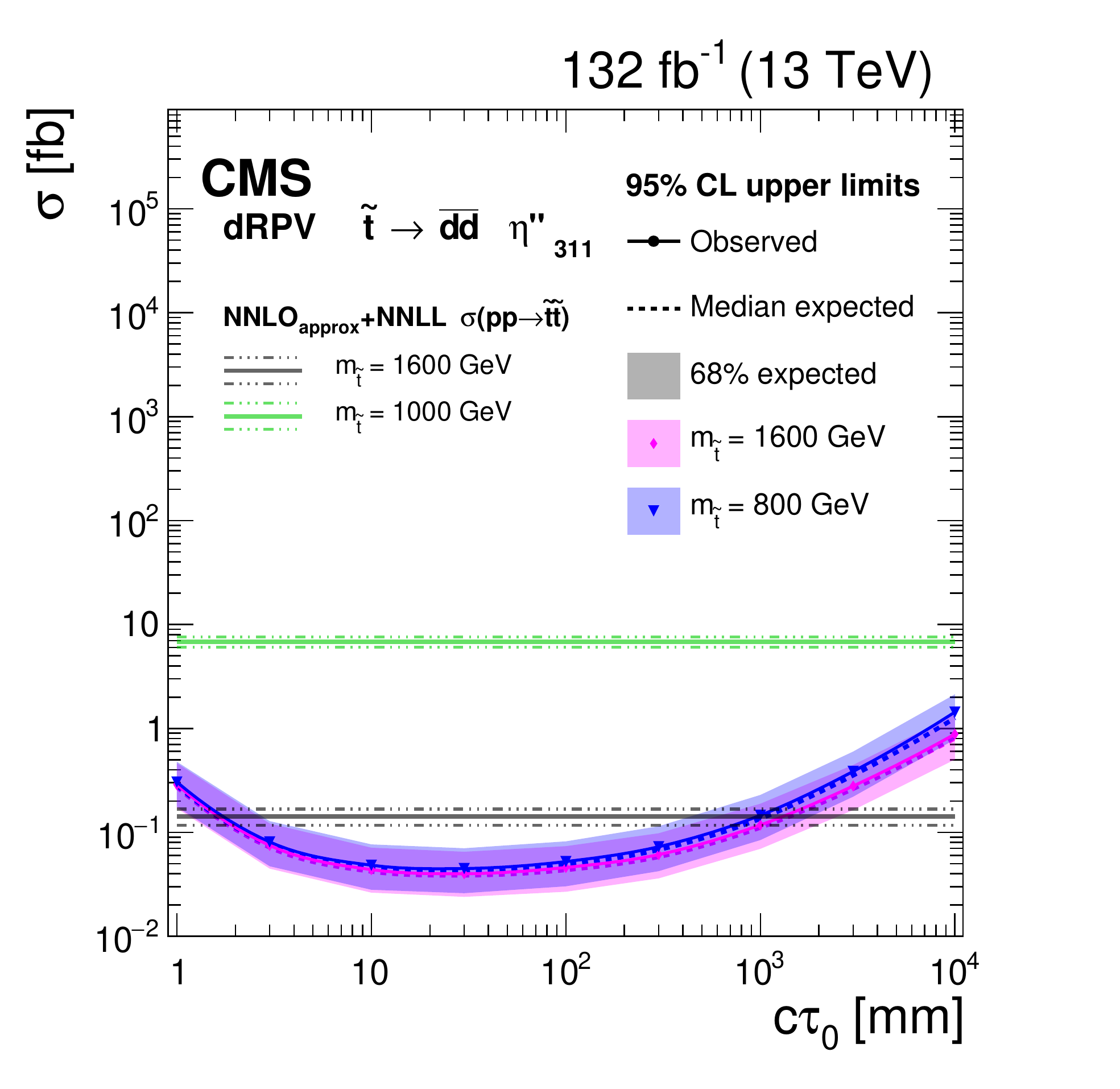}
\includegraphics[width=0.45\textwidth]{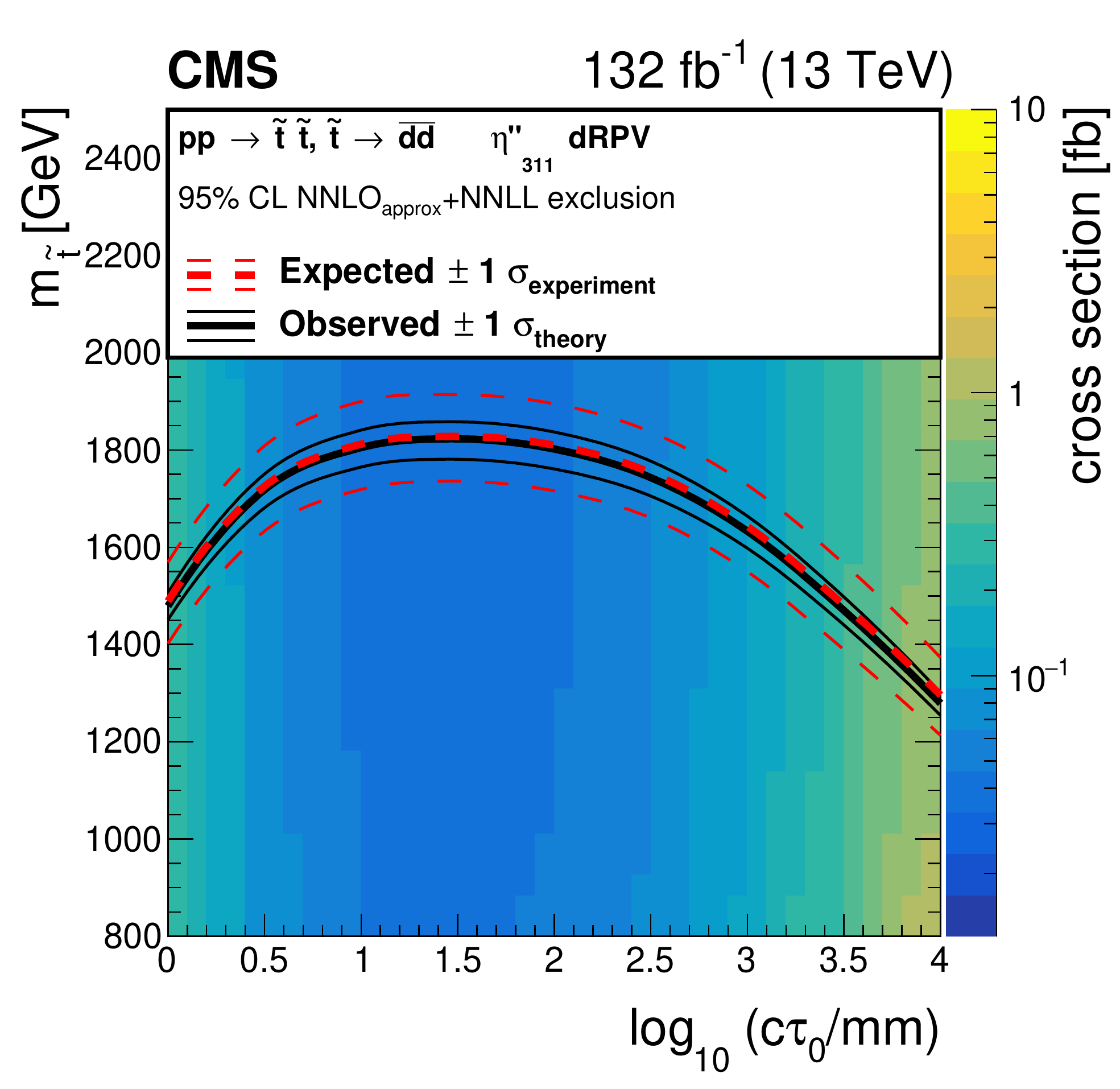}
\caption{\cmsLeft: the 95\% \CL upper limits on the pair production cross section for the long-lived top squarks with $m_{\sTop}=1600\GeV$ and 800\GeV, where a
100\% branching fraction for $\sTop\to\cPaqd\cPaqd$ decays is assumed. The $\mathrm{NNLO_{approx}}$+NNLL top squark pair production cross sections for $m_{\sTop}=1600$ and 1000\GeV, as well as their variations due to the theoretical uncertainties, are shown as horizontal lines. The solid (dashed) curves show the observed (median expected) limits, and the shaded bands indicate the regions containing 68\% of
the distributions of the limits expected under the background-only hypothesis.
\cmsRight: the 95\% \CL limits on the pair production cross section for the $\sTop\to\cPaqd\cPaqd$ model as a function of the mean proper decay length $c\tau_{0}$ and the
top squark mass $m_{\sTop}$. The thick solid black
(dashed red) curve shows the observed (median expected) 95\% \CL limits on the top squark mass as a function of $c\tau_{0}$, assuming the $\mathrm{NNLO_{approx}}$+NNLL cross sections. The thin
dashed red curves indicate the region containing 68\% of
the distribution of the limits expected under the background-only hypothesis. The thin solid black curves represent the change in
the observed limit when the signal cross sections are varied according to their theoretical uncertainties.}
\label{fig: limits_stopToDD}
\end{figure}

The expected and observed upper limits on the pair production cross section of the 
long-lived top squarks in the dRPV $\sTop\to\cPaqd\cPaqd$ model are shown in Fig.~\ref{fig: limits_stopToDD}, where a branching fraction of 100\% for the 
top squark to decay into two down antiquarks is assumed. When the top squark mass is 1600\GeV, signal efficiencies are 
around 43, 76, and 53\% in the 2017 and 2018 analysis for $c\tau_{0}=3$, 30, and 300\mm, respectively. With the data samples collected in 2016--2018, top squark pair production cross 
sections larger than 0.1\unit{fb} are excluded for $c\tau_{0}$ between 3 and 820\mm at $m_{\sTop}=1600\GeV$. We then compute the upper limits on the top squark mass for different $c\tau_{0}$ according to the upper limits on the pair production cross section and the calculation of the $\mathrm{NNLO_{approx}}$+NNLL top squark pair production cross sections. Top squark masses 
up to 1600\GeV are excluded for $c\tau_{0}$ between 2 and 1320\mm. The largest top squark mass excluded is 
1820\GeV with a $c\tau_{0}$ of 30\mm. These bounds are the most stringent to date on this model for $c\tau_{0}$ between 10 and $10^{4}\mm$.
 
\section{Summary}\label{sec: summary}
A search has been presented for long-lived particles decaying to jets, using proton-proton collision data collected with the CMS 
experiment at a center-of-mass energy of 13\TeV in 2017 and 2018. The results are combined with those of a previous CMS search for 
displaced jets using proton-proton collision data from 2016, accumulating to a total integrated luminosity of 132\fbinv. 
After all selections, one event is observed in the data collected in 2017 and 2018, which is consistent with the predicted 
background yield. The search is designed to be model independent, and is sensitive to a large number of models predicting 
displaced-jets signatures with different final-state topologies.  

The best current limits are set on a variety of models that have long-lived particles 
with mean proper decay lengths between 1\mm and 10\unit{m}. All limits are 
computed at the 95\% confidence level. 
For a simplified model where pair-produced 
long-lived neutral particles decay to quark-antiquark pairs, pair production cross sections larger than 0.07\unit{fb} are 
excluded for mean proper decay lengths between 2 and 250\mm at high mass ($m_{\mathrm{X}}>500\GeV$). For a model where the standard model-like 
Higgs boson decays to two long-lived scalar particles and each long-lived scalar particle decays to a down (bottom) quark-antiquark pair, 
branching fractions for the exotic Higgs boson decay larger than 1\% (10\%) are excluded for mean proper decay lengths between 1 and 
340\mm (530\mm) when the scalar particle mass is larger than 40\GeV. For a supersymmetric (SUSY) model in 
the general gauge mediation scenario, where the long-lived gluino decays to a gluon and a lightest SUSY particle, gluino masses 
up to 2450\GeV are excluded for mean proper decay lengths between 6 and 550\mm. For another SUSY model in the mini-split 
scenario, where the long-lived gluino can decay to a quark-antiquark pair and the lightest neutralino, gluino masses up to 2500\GeV are 
excluded for mean proper decay lengths between 7 and 360\mm. An $R$-parity violating (RPV) SUSY model is also tested, where the 
long-lived gluino can decay to top, bottom, and strange antiquarks, and gluino masses up to 2500\GeV are excluded 
for mean proper decay lengths between 3 and 1000\mm. Another RPV SUSY model is studied, where the long-lived top squark can 
decay to a bottom quark and a charged lepton, and top squark masses up to 1600\GeV are excluded 
for mean proper decay lengths 
between 5 and 240\mm. For an RPV SUSY model, where the long-lived top squark can decay to a down quark and a charged 
lepton, top squark masses up to 1600\GeV are excluded for mean proper decay lengths between 3 and 360\mm. Finally, for a 
dynamical-RPV SUSY model, where the long-lived top squark can decay to two down antiquarks, top squark masses up to 
1600\GeV are excluded for mean proper decay lengths between 2 and 1320\mm. These are the most stringent limits to date 
on these models.   

\begin{acknowledgments}
  We congratulate our colleagues in the CERN accelerator departments for the excellent performance of the LHC and thank the technical and administrative staffs at CERN and at other CMS institutes for their contributions to the success of the CMS effort. In addition, we gratefully acknowledge the computing centers and personnel of the Worldwide LHC Computing Grid for delivering so effectively the computing infrastructure essential to our analyses. Finally, we acknowledge the enduring support for the construction and operation of the LHC and the CMS detector provided by the following funding agencies: BMBWF and FWF (Austria); FNRS and FWO (Belgium); CNPq, CAPES, FAPERJ, FAPERGS, and FAPESP (Brazil); MES (Bulgaria); CERN; CAS, MoST, and NSFC (China); COLCIENCIAS (Colombia); MSES and CSF (Croatia); RIF (Cyprus); SENESCYT (Ecuador); MoER, ERC PUT and ERDF (Estonia); Academy of Finland, MEC, and HIP (Finland); CEA and CNRS/IN2P3 (France); BMBF, DFG, and HGF (Germany); GSRT (Greece); NKFIA (Hungary); DAE and DST (India); IPM (Iran); SFI (Ireland); INFN (Italy); MSIP and NRF (Republic of Korea); MES (Latvia); LAS (Lithuania); MOE and UM (Malaysia); BUAP, CINVESTAV, CONACYT, LNS, SEP, and UASLP-FAI (Mexico); MOS (Montenegro); MBIE (New Zealand); PAEC (Pakistan); MSHE and NSC (Poland); FCT (Portugal); JINR (Dubna); MON, RosAtom, RAS, RFBR, and NRC KI (Russia); MESTD (Serbia); SEIDI, CPAN, PCTI, and FEDER (Spain); MOSTR (Sri Lanka); Swiss Funding Agencies (Switzerland); MST (Taipei); ThEPCenter, IPST, STAR, and NSTDA (Thailand); TUBITAK and TAEK (Turkey); NASU (Ukraine); STFC (United Kingdom); DOE and NSF (USA).
 
  \hyphenation{Rachada-pisek} Individuals have received support from the Marie-Curie program and the European Research Council and Horizon 2020 Grant, contract Nos.\ 675440, 724704, 752730, and 765710 (European Union); the Leventis Foundation; the A.P.\ Sloan Foundation; the Alexander von Humboldt Foundation; the Belgian Federal Science Policy Office; the Fonds pour la Formation \`a la Recherche dans l'Industrie et dans l'Agriculture (FRIA-Belgium); the Agentschap voor Innovatie door Wetenschap en Technologie (IWT-Belgium); the F.R.S.-FNRS and FWO (Belgium) under the ``Excellence of Science -- EOS" -- be.h project n.\ 30820817; the Beijing Municipal Science \& Technology Commission, No. Z191100007219010; the Ministry of Education, Youth and Sports (MEYS) of the Czech Republic; the Deutsche Forschungsgemeinschaft (DFG) under Germany's Excellence Strategy -- EXC 2121 ``Quantum Universe" -- 390833306; the Lend\"ulet (``Momentum") Program and the J\'anos Bolyai Research Scholarship of the Hungarian Academy of Sciences, the New National Excellence Program \'UNKP, the NKFIA research grants 123842, 123959, 124845, 124850, 125105, 128713, 128786, and 129058 (Hungary); the Council of Science and Industrial Research, India; the HOMING PLUS program of the Foundation for Polish Science, cofinanced from European Union, Regional Development Fund, the Mobility Plus program of the Ministry of Science and Higher Education, the National Science Center (Poland), contracts Harmonia 2014/14/M/ST2/00428, Opus 2014/13/B/ST2/02543, 2014/15/B/ST2/03998, and 2015/19/B/ST2/02861, Sonata-bis 2012/07/E/ST2/01406; the National Priorities Research Program by Qatar National Research Fund; the Ministry of Science and Higher Education, project no. 0723-2020-0041 (Russia); the Tomsk Polytechnic University Competitiveness Enhancement Program; the Programa Estatal de Fomento de la Investigaci{\'o}n Cient{\'i}fica y T{\'e}cnica de Excelencia Mar\'{\i}a de Maeztu, grant MDM-2015-0509 and the Programa Severo Ochoa del Principado de Asturias; the Thalis and Aristeia programs cofinanced by EU-ESF and the Greek NSRF; the Rachadapisek Sompot Fund for Postdoctoral Fellowship, Chulalongkorn University and the Chulalongkorn Academic into Its 2nd Century Project Advancement Project (Thailand); the Kavli Foundation; the Nvidia Corporation; the SuperMicro Corporation; the Welch Foundation, contract C-1845; and the Weston Havens Foundation (USA).\end{acknowledgments}

\bibliography{auto_generated}   
\clearpage
\appendix
\numberwithin{table}{section}
\numberwithin{figure}{section}
\section{Signal efficiencies for representative signal points in different models}\label{sec: appen}

\begin{table*}[hbt]
  \topcaption{ Signal efficiencies for the jet-jet model in the 2017 and 2018 analysis at different mean proper decay
lengths $c\tau_{0}$ and different masses $m_{\mathrm{X}}$. Selection requirements are cumulative from
    the first row to the last for each value of $m_{\mathrm{X}}$. Uncertainties are statistical only.}
\label{tab: eff_JetJet}
\centering

\cmsTable{
\begin{scotch}{lcccccc}
\multirow{2}*{Efficiency (\%)}&\multirow{2}*{$m_{\mathrm{X}}$ ({\GeVns})} & \multicolumn{5}{c}{$c\tau_{0}$}\\
 & & 1\mm & 10\mm & 30\mm & 100\mm & 1000\mm \\\hline
Trigger & \multirow{3}*{1000} & 29.47 $\pm$ 0.38 & 89.98 $\pm$ 0.67 & 91.16 $\pm$ 0.68 &  84.41 $\pm$ 0.65 & 71.72 $\pm$ 0.66 \\
Preselection &                & 22.56 $\pm$ 0.34 & 85.22 $\pm$ 0.65 & 84.92 $\pm$ 0.65 &  73.83 $\pm$ 0.61 & 27.47 $\pm$ 0.41 \\
Final selection &             & 16.27 $\pm$ 0.29 & 73.63 $\pm$ 0.61 & 73.51 $\pm$ 0.61 &  61.51 $\pm$ 0.55 & 20.13 $\pm$ 0.35 \\[\cmsTabSkip]
Trigger & \multirow{3}*{300} & 25.05 $\pm$ 0.35  & 70.50 $\pm$ 0.59  & 68.19 $\pm$ 0.58 &  58.97 $\pm$ 0.54 & 30.22 $\pm$ 0.39 \\
Preselection &               & 17.42 $\pm$ 0.30  & 59.89 $\pm$ 0.55 & 55.40 $\pm$ 0.53 &    42.38 $\pm$ 0.46 & 9.11 $\pm$ 0.21 \\
Final selection &            & 12.06  $\pm$ 0.25 & 48.41 $\pm$ 0.49 & 45.13 $\pm$ 0.48 &    32.42 $\pm$ 0.40 & 5.87 $\pm$ 0.17 \\[\cmsTabSkip]
Trigger & \multirow{3}*{100} & 2.65 $\pm$ 0.12 & 6.97 $\pm$ 0.19 & 6.47 $\pm$ 0.18     & 4.87 $\pm$ 0.16 & 0.95 $\pm$ 0.07 \\
Preselection &               & 1.81 $\pm$ 0.10 & 4.94 $\pm$ 0.16 & 4.41 $\pm$ 0.15     & 2.59 $\pm$ 0.11 & 0.28 $\pm$ 0.04 \\
Final selection &            & 1.03 $\pm$ 0.07 & 3.47 $\pm$ 0.13 & 3.00 $\pm$ 0.12     & 1.64 $\pm$ 0.09 & 0.17 $\pm$ 0.03 \\
\end{scotch}
}
\end{table*}

\begin{table*}[hbt]
  \topcaption{ Signal efficiencies for the model where the SM-like Higgs boson decays to two long-lived scalar particles 
$\mathrm{S}$ in the 2017 and 2018 analysis at different mean proper decay
lengths $c\tau_{0}$ and with $m_{\mathrm{S}}=55\GeV$. The long-lived scalar particle is assumed to decay to a down 
quark-antiquark pair ($\mathrm{S}\to\PQd\cPaqd$). Selection requirements are cumulative from
    the first row to the last. Uncertainties are statistical only.}
\label{tab: eff_Higgsdd}
\centering
\cmsTable{
\begin{scotch}{lcccccc}
\multirow{2}*{Efficiency $\times 10^{4}$}&\multirow{2}*{$m_{\mathrm{S}}$ ({\GeVns})} & \multicolumn{5}{c}{c$\tau_{0}$}\\
 & & 1\mm & 10\mm & 30\mm & 100\mm & 1000\mm \\\hline
Trigger & \multirow{3}*{55} & 6.63 $\pm$ 0.13 & 32.07 $\pm$ 0.29 & 33.44 $\pm$ 1.16 &  25.25 $\pm$ 0.26 & 5.71 $\pm$ 0.12 \\
Preselection &                & 3.11 $\pm$ 0.09 & 13.61 $\pm$ 0.19 & 13.72 $\pm$ 0.75 &  9.39 $\pm$ 0.16 & 1.36 $\pm$ 0.06 \\
Final selection &             & 0.95 $\pm$ 0.05 & 6.12 $\pm$ 0.13 & 6.34 $\pm$ 0.42 &  4.46 $\pm$ 0.11 & 0.64 $\pm$ 0.04 \\
\end{scotch}
}
\end{table*}

\begin{table*}[hbt]
  \topcaption{ Signal efficiencies for the model where the SM-like Higgs boson decays to two long-lived scalar particles
$\mathrm{S}$ in the 2017 and 2018 analysis at different mean proper decay
lengths $c\tau_{0}$ and with $m_{\mathrm{S}}=55\GeV$. The long-lived scalar particle is assumed to decay to a bottom 
quark-antiquark pair ($\mathrm{S}\to\PQb\PAQb$). Selection requirements are cumulative from
    the first row to the last. Uncertainties are statistical only.}
\label{tab: eff_Higgsbb}
\centering
\cmsTable{ 
\begin{scotch}{lcccccc}
\multirow{2}*{Efficiency $\times 10^{4}$}&\multirow{2}*{$m_{\mathrm{S}}$ ({\GeVns})} & \multicolumn{5}{c}{c$\tau_{0}$}\\
 & & 1\mm & 10\mm & 30\mm & 100\mm & 1000\mm \\\hline
Trigger & \multirow{3}*{55} & 4.30 $\pm$ 0.11 & 22.56 $\pm$ 0.24 & 24.45 $\pm$ 0.48 &  17.78 $\pm$ 0.21 & 4.05 $\pm$ 0.10 \\
Preselection &                & 0.66 $\pm$ 0.04 & 3.30 $\pm$ 0.09 & 3.97 $\pm$ 0.19 &  3.37 $\pm$ 0.09 & 0.57 $\pm$ 0.04 \\
Final selection &             & 0.08 $\pm$ 0.01 & 0.96 $\pm$ 0.05 & 1.17 $\pm$ 0.10 &  1.09 $\pm$ 0.05 & 0.19 $\pm$ 0.02 \\
\end{scotch}
}
\end{table*}

\begin{table*}[hbt]
  \topcaption{ Signal efficiencies for the $\sGlu\to\Glu\sGra$ model in the 2017 and 2018 analysis at different mean proper decay
lengths $c\tau_{0}$ and different masses $m_{\sGlu}$. Selection requirements are cumulative from
    the first row to the last for each value of $m_{\sGlu}$. Uncertainties are statistical only.}
\label{tab: eff_GMSB}
\centering
\cmsTable{
\begin{scotch}{lcccccc}
\multirow{2}*{Efficiency (\%)}&\multirow{2}*{$m_{\sGlu}$ ({\GeVns})} & \multicolumn{5}{c}{c$\tau_{0}$}\\
 & & 1\mm & 10\mm & 30\mm & 100\mm & 1000\mm \\\hline
Trigger & \multirow{3}*{2400} & 10.69 $\pm$ 0.24 & 69.70 $\pm$ 0.60 & 81.86 $\pm$ 0.65 &  82.86 $\pm$ 0.66 & 73.07 $\pm$ 0.38 \\
Preselection &                & 7.03 $\pm$ 0.19 & 63.62 $\pm$ 0.57 & 72.80 $\pm$ 0.61 &  69.67 $\pm$ 0.60 & 35.54 $\pm$ 0.27 \\
Final selection &             & 3.84 $\pm$ 0.14 & 44.20 $\pm$ 0.48 & 52.69 $\pm$ 0.52 &  50.84 $\pm$ 0.52 & 24.19 $\pm$ 0.22 \\[\cmsTabSkip]
Trigger & \multirow{3}*{1600} & 12.04 $\pm$ 0.25 & 68.36 $\pm$ 0.59 & 79.64 $\pm$ 0.64 &  79.96 $\pm$ 0.64 & 67.49 $\pm$ 0.38 \\
Preselection &               & 7.52 $\pm$ 0.20  & 60.55 $\pm$ 0.56 & 68.74 $\pm$ 0.59 &    63.47 $\pm$ 0.57 & 29.57 $\pm$ 0.25 \\
Final selection &            & 4.02  $\pm$ 0.15 & 41.33 $\pm$ 0.46 & 47.67 $\pm$ 0.49 &    43.16 $\pm$ 0.47 & 19.53 $\pm$ 0.20 \\[\cmsTabSkip]
Trigger & \multirow{3}*{1000} & 12.14$\pm$ 0.25 & 62.77 $\pm$ 0.56 & 72.44 $\pm$ 0.60    & 71.90 $\pm$ 0.60 & 52.77 $\pm$ 0.51 \\
Preselection &               & 7.29 $\pm$ 0.19 & 53.19 $\pm$ 0.52 & 57.56 $\pm$ 0.54     & 52.25 $\pm$ 0.51 & 21.10 $\pm$ 0.32 \\
Final selection &            & 3.75 $\pm$ 0.14 & 34.57 $\pm$ 0.42 & 37.55 $\pm$ 0.43     & 33.84 $\pm$ 0.41 & 12.88 $\pm$ 0.25 \\
\end{scotch}
}
\end{table*}

\begin{table*}[hbt]
  \topcaption{ Signal efficiencies for the $\sGlu\to\cPq\cPaq\widetilde{\chi}^{0}_{1}$ model ($m_{\widetilde{\chi}^{0}_{1}}=100\GeV$) in the 2017 and 2018 analysis at different mean proper decay
lengths $c\tau_{0}$ and different masses $m_{\sGlu}$. Selection requirements are cumulative from
    the first row to the last for each value of $m_{\sGlu}$. Uncertainties are statistical only.}
\label{tab: eff_Split}
\centering

\cmsTable{
\begin{scotch}{lcccccc}
\multirow{2}*{Efficiency (\%)}&\multirow{2}*{$m_{\sGlu}$ ({\GeVns})} & \multicolumn{5}{c}{c$\tau_{0}$}\\
 & & 1\mm & 10\mm & 30\mm & 100\mm & 1000\mm \\\hline
Trigger & \multirow{3}*{2600} & 15.08 $\pm$ 0.28 & 82.32 $\pm$ 0.66 & 90.46 $\pm$ 0.69 &  87.65 $\pm$ 0.68 & 78.78 $\pm$ 0.65 \\
Preselection &                & 9.91 $\pm$ 0.23 & 77.60 $\pm$ 0.64 & 85.82 $\pm$ 0.67 &  80.22 $\pm$ 0.65 & 43.24 $\pm$ 0.48 \\
Final selection &             & 5.75 $\pm$ 0.17 & 59.12 $\pm$ 0.56 & 68.44 $\pm$ 0.60 &  63.26 $\pm$ 0.58 & 30.84 $\pm$ 0.41 \\[\cmsTabSkip]
Trigger & \multirow{3}*{2000} & 17.70 $\pm$ 0.31 & 83.21 $\pm$ 0.67 & 90.46 $\pm$ 0.69 &  87.66 $\pm$ 0.68 & 79.04 $\pm$ 0.65 \\
Preselection &               & 11.20 $\pm$ 0.24  & 77.74 $\pm$ 0.65 & 85.04 $\pm$ 0.67 &    79.45 $\pm$ 0.65 & 39.69 $\pm$ 0.46 \\
Final selection &            & 6.57  $\pm$ 0.19 & 58.36 $\pm$ 0.56 & 66.29 $\pm$ 0.59 &    60.96 $\pm$ 0.57 & 27.12 $\pm$ 0.38 \\[\cmsTabSkip]
Trigger & \multirow{3}*{1600} & 19.39$\pm$ 0.32 & 82.56 $\pm$ 0.66 & 90.27 $\pm$ 0.69    & 87.70 $\pm$ 0.68 & 78.31 $\pm$ 0.64 \\
Preselection &               & 12.16 $\pm$ 0.26 & 76.59 $\pm$ 0.64 & 84.13 $\pm$ 0.59     & 77.71 $\pm$ 0.64 & 37.41 $\pm$ 0.44 \\
Final selection &            & 6.76 $\pm$ 0.19 & 57.19 $\pm$ 0.55 & 64.37 $\pm$ 0.58     & 57.98 $\pm$ 0.55 & 25.02 $\pm$ 0.36 \\
\end{scotch}
}
\end{table*}

\begin{table*}[hbt]
  \topcaption{ Signal efficiencies for the $\sGlu\to\PQt\PQb\cPqs$ model in the 2017 and 2018 analysis at different mean proper decay
lengths $c\tau_{0}$ and different masses $m_{\sGlu}$. Selection requirements are cumulative from
    the first row to the last for each value of $m_{\sGlu}$. Uncertainties are statistical only.}
\label{tab: eff_GGToNN}
\centering

\cmsTable{
\begin{scotch}{lcccccc}
\multirow{2}*{Efficiency (\%)}&\multirow{2}*{$m_{\sGlu}$ ({\GeVns})} & \multicolumn{5}{c}{c$\tau_{0}$}\\
 & & 1\mm & 10\mm & 30\mm & 100\mm & 1000\mm \\\hline
Trigger & \multirow{3}*{2600} & 25.14 $\pm$ 0.39 & 90.65 $\pm$ 0.70 & 95.78 $\pm$ 0.73 &  91.42 $\pm$ 0.73 & 83.61 $\pm$ 0.67 \\
Preselection &                & 16.26 $\pm$ 0.31 & 87.43 $\pm$ 0.69 & 94.08 $\pm$ 0.73 &  89.30 $\pm$ 0.72 & 56.49 $\pm$ 0.55 \\
Final selection &             & 9.60 $\pm$ 0.24 & 71.09 $\pm$ 0.62 & 81.12 $\pm$ 0.67 &  77.55 $\pm$ 0.67 & 42.62 $\pm$ 0.48 \\[\cmsTabSkip]
Trigger & \multirow{3}*{2000} & 29.89 $\pm$ 0.42 & 91.95 $\pm$ 0.71 & 95.53 $\pm$ 0.73 &    91.58 $\pm$ 0.72 & 83.38 $\pm$ 0.68 \\
Preselection &               & 18.93 $\pm$ 0.34  & 88.27 $\pm$ 0.70 & 93.55 $\pm$ 0.72 &    89.17 $\pm$ 0.71 & 52.54 $\pm$ 0.54 \\
Final selection &            & 11.29  $\pm$ 0.26 & 72.14 $\pm$ 0.63 & 79.58 $\pm$ 0.67 &    75.52 $\pm$ 0.65 & 38.12 $\pm$ 0.46 \\[\cmsTabSkip]
Trigger & \multirow{3}*{1600} & 31.61$\pm$ 0.43 & 92.16 $\pm$ 0.71 & 95.58 $\pm$ 0.76    & 92.03 $\pm$ 0.72 & 84.50 $\pm$ 0.67 \\
Preselection &               & 20.06 $\pm$ 0.34 & 88.40 $\pm$ 0.70 & 93.46 $\pm$ 0.75     & 89.04 $\pm$ 0.71 & 50.46 $\pm$ 0.52 \\
Final selection &            & 11.71 $\pm$ 0.26 & 70.85 $\pm$ 0.63 & 78.52 $\pm$ 0.69     & 73.77 $\pm$ 0.65 & 35.91 $\pm$ 0.44 \\
\end{scotch}
}
\end{table*}

\begin{table*}[hbt]
  \topcaption{ Signal efficiencies for the $\sTop\to\PQb\ell$ model in the 2017 and 2018 analysis at different mean proper decay
lengths $c\tau_{0}$ and different masses $m_{\sTop}$. Selection requirements are cumulative from the first row to the last for each 
value of $m_{\sTop}$. Uncertainties are statistical only.}
\label{tab: eff_StopToLB}
\centering

\cmsTable{
\begin{scotch}{lcccccc}
\multirow{2}*{Efficiency (\%)}&\multirow{2}*{$m_{\sTop}$ ({\GeVns})} & \multicolumn{5}{c}{c$\tau_{0}$}\\
 & & 1\mm & 10\mm & 30\mm & 100\mm & 1000\mm \\\hline
Trigger & \multirow{3}*{1600} & 18.00 $\pm$ 0.30 & 66.30 $\pm$ 0.58 & 74.36 $\pm$ 0.62 &  74.48 $\pm$ 0.62 & 66.67 $\pm$ 0.59 \\
Preselection &                & 8.45 $\pm$ 0.21 & 54.41 $\pm$ 0.53 & 59.13 $\pm$ 0.56 &  52.63 $\pm$ 0.52 & 21.00 $\pm$ 0.33 \\
Final selection &             & 5.03 $\pm$ 0.16 &  39.77 $\pm$ 0.45 & 42.85 $\pm$ 0.48 &  37.02 $\pm$ 0.44 & 12.58 $\pm$ 0.25 \\[\cmsTabSkip]
Trigger & \multirow{3}*{1000} & 20.13 $\pm$ 0.32 & 65.37 $\pm$ 0.58 &    73.13 $\pm$ 0.61 & 73.74 $\pm$ 0.61 & 61.20 $\pm$ 0.56\\
Preselection &               & 8.32 $\pm$ 0.21   & 48.92 $\pm$ 0.49 &    53.45 $\pm$ 0.53 & 47.68 $\pm$ 0.49 & 16.99 $\pm$ 0.30\\
Final selection &            & 4.65  $\pm$ 0.16  & 33.46 $\pm$ 0.41 &    37.56 $\pm$ 0.44 & 31.81 $\pm$ 0.40 & 10.12 $\pm$ 0.23\\[\cmsTabSkip]
Trigger & \multirow{3}*{600} & 19.96$\pm$ 0.32 & 58.90 $\pm$ 0.54 & 64.58 $\pm$ 0.58    & 64.93 $\pm$ 0.58 & 48.32 $\pm$ 0.50 \\
Preselection &               & 6.85 $\pm$ 0.19 & 39.15 $\pm$ 0.44 & 42.25 $\pm$ 0.47    & 36.80 $\pm$ 0.44 & 11.59 $\pm$ 0.25 \\
Final selection &            & 3.33 $\pm$ 0.13 & 24.44 $\pm$ 0.35 & 27.69 $\pm$ 0.38     & 23.12 $\pm$ 0.35 & 6.47 $\pm$ 0.18 \\
\end{scotch}
}
\end{table*}

\begin{table*}[hbt]
  \topcaption{ Signal efficiencies for the $\sTop\to\PQd\ell$ model in the 2017 and 2018 analysis at different mean proper decay
lengths $c\tau_{0}$ and different masses $m_{\sTop}$. Selection requirements are cumulative from
    the first row to the last for each value of $m_{\sTop}$. Uncertainties are statistical only.}
\label{tab: eff_StopToLD}
\centering
\cmsTable{
\begin{scotch}{lcccccc}
\multirow{2}*{Efficiency (\%)}&\multirow{2}*{$m_{\sTop}$ ({\GeVns})} & \multicolumn{5}{c}{c$\tau_{0}$}\\
 & & 1\mm & 10\mm & 30\mm & 100\mm & 1000\mm \\\hline
Trigger & \multirow{3}*{1600} & 18.58 $\pm$ 0.31 & 66.73 $\pm$ 0.59 & 74.69 $\pm$ 0.63 &  74.59 $\pm$ 0.64 & 67.27 $\pm$ 0.60 \\
Preselection &                & 10.72 $\pm$ 0.24 & 57.96 $\pm$ 0.55 & 63.08 $\pm$ 0.58 &  56.77 $\pm$ 0.56 & 22.79 $\pm$ 0.35 \\
Final selection &             & 6.62 $\pm$ 0.19 &  43.88 $\pm$ 0.48 & 47.92 $\pm$ 0.51 &  41.09 $\pm$ 0.47 & 14.33 $\pm$ 0.28 \\[\cmsTabSkip]
Trigger & \multirow{3}*{1000} & 19.95 $\pm$ 0.33 & 66.23 $\pm$ 0.59 & 73.87 $\pm$ 0.63 &    73.99 $\pm$ 0.64 & 62.66 $\pm$ 0.57 \\
Preselection &               & 9.71 $\pm$ 0.23  & 55.31 $\pm$ 0.54 & 59.93 $\pm$ 0.57 &    53.79 $\pm$ 0.54 & 19.80 $\pm$ 0.32 \\
Final selection &            & 5.60  $\pm$ 0.17 & 41.22 $\pm$ 0.47 & 44.38 $\pm$ 0.49 &    38.07 $\pm$ 0.46 & 12.34 $\pm$ 0.25 \\[\cmsTabSkip]
Trigger & \multirow{3}*{600} & 20.27$\pm$ 0.33 & 60.49 $\pm$ 0.58 & 66.86 $\pm$ 0.61    & 66.85 $\pm$ 0.61 & 49.73 $\pm$ 0.52 \\
Preselection &               & 8.83 $\pm$ 0.22 & 47.48 $\pm$ 0.51 & 50.55 $\pm$ 0.53     & 44.37 $\pm$ 0.50 & 13.46 $\pm$ 0.27 \\
Final selection &            & 5.14 $\pm$ 0.17 & 33.96 $\pm$ 0.43 & 36.03 $\pm$ 0.45     & 30.16 $\pm$ 0.41 & 8.19 $\pm$ 0.21 \\
\end{scotch}
}
\end{table*}

\begin{table*}[hbt]
  \topcaption{ Signal efficiencies for the $\sTop\to\cPaqd\cPaqd$ model in the 2017 and 2018 analysis at different mean proper decay
lengths $c\tau_{0}$ and different masses $m_{\sTop}$. Selection requirements are cumulative from
    the first row to the last for each value of $m_{\sTop}$. Uncertainties are statistical only.}
\label{tab: eff_StopToDD}
\centering

\cmsTable{
\begin{scotch}{lcccccc}
\multirow{2}*{Efficiency (\%)}&\multirow{2}*{$m_{\sTop}$ ({\GeVns})} & \multicolumn{5}{c}{c$\tau_{0}$}\\
 & & 1\mm & 10\mm & 30\mm & 100\mm & 1000\mm \\\hline
Trigger & \multirow{3}*{1600} & 22.10 $\pm$ 0.35 & 87.15 $\pm$ 0.71 & 92.19 $\pm$ 0.72 &  87.73 $\pm$ 0.71 & 79.12 $\pm$ 0.67 \\
Preselection &                & 17.06 $\pm$ 0.31 & 84.06 $\pm$ 0.70 & 88.49 $\pm$ 0.71 &  81.10 $\pm$ 0.68 & 39.89 $\pm$ 0.48 \\
Final selection &             & 11.78 $\pm$ 0.26 &  71.44 $\pm$ 0.65 & 76.02 $\pm$ 0.66 &  67.02 $\pm$ 0.62 & 28.52 $\pm$ 0.40 \\[\cmsTabSkip]
Trigger & \multirow{3}*{1000} & 23.86 $\pm$ 0.36 & 86.99 $\pm$ 0.72 & 91.58 $\pm$ 0.77 &    88.19 $\pm$ 0.70 & 78.31 $\pm$ 0.66 \\
Preselection &               & 16.96 $\pm$ 0.30  & 82.33 $\pm$ 0.70 & 86.73 $\pm$ 0.75 &    79.72 $\pm$ 0.67 & 35.91 $\pm$ 0.45 \\
Final selection &            & 11.91  $\pm$ 0.26 & 69.01 $\pm$ 0.64 & 73.52 $\pm$ 0.69 &    64.23 $\pm$ 0.60 & 25.52 $\pm$ 0.38 \\[\cmsTabSkip]
Trigger & \multirow{3}*{600} & 24.51$\pm$ 0.37 & 84.75 $\pm$ 0.68 & 89.27 $\pm$ 0.70    & 85.55 $\pm$ 0.68 & 69.66 $\pm$ 0.61 \\
Preselection &               & 15.80 $\pm$ 0.29 & 78.05 $\pm$ 0.66 & 81.84 $\pm$ 0.67     & 74.51 $\pm$ 0.64 & 29.37 $\pm$ 0.39 \\
Final selection &            & 11.15 $\pm$ 0.25 & 64.32 $\pm$ 0.60 & 67.68 $\pm$ 0.61     & 58.85 $\pm$ 0.57 & 20.66 $\pm$ 0.33 \\
\end{scotch}
}
\end{table*}

\begin{figure*}[tbp]
\centering
\includegraphics[width=0.44\textwidth]{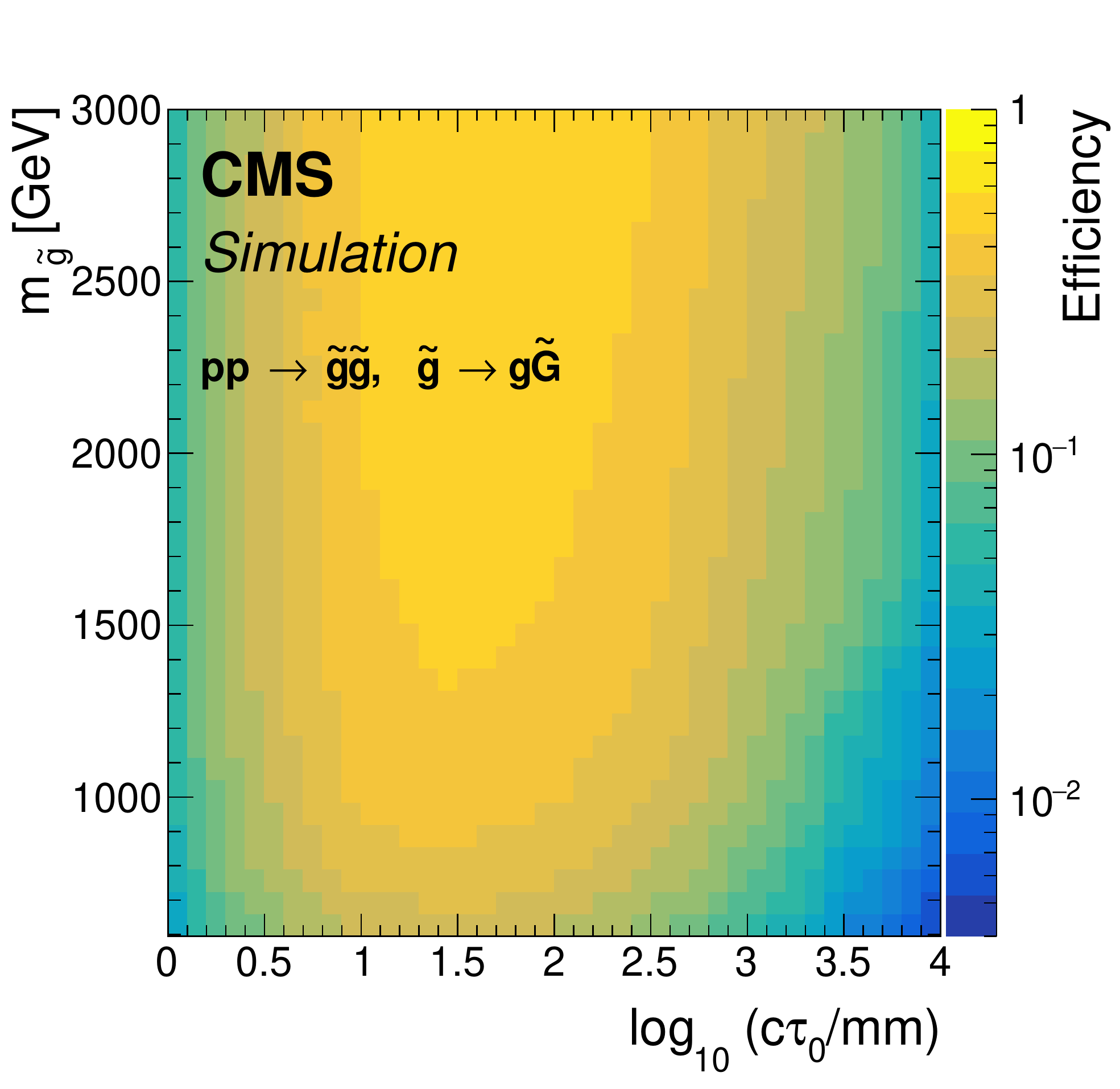}
\includegraphics[width=0.44\textwidth]{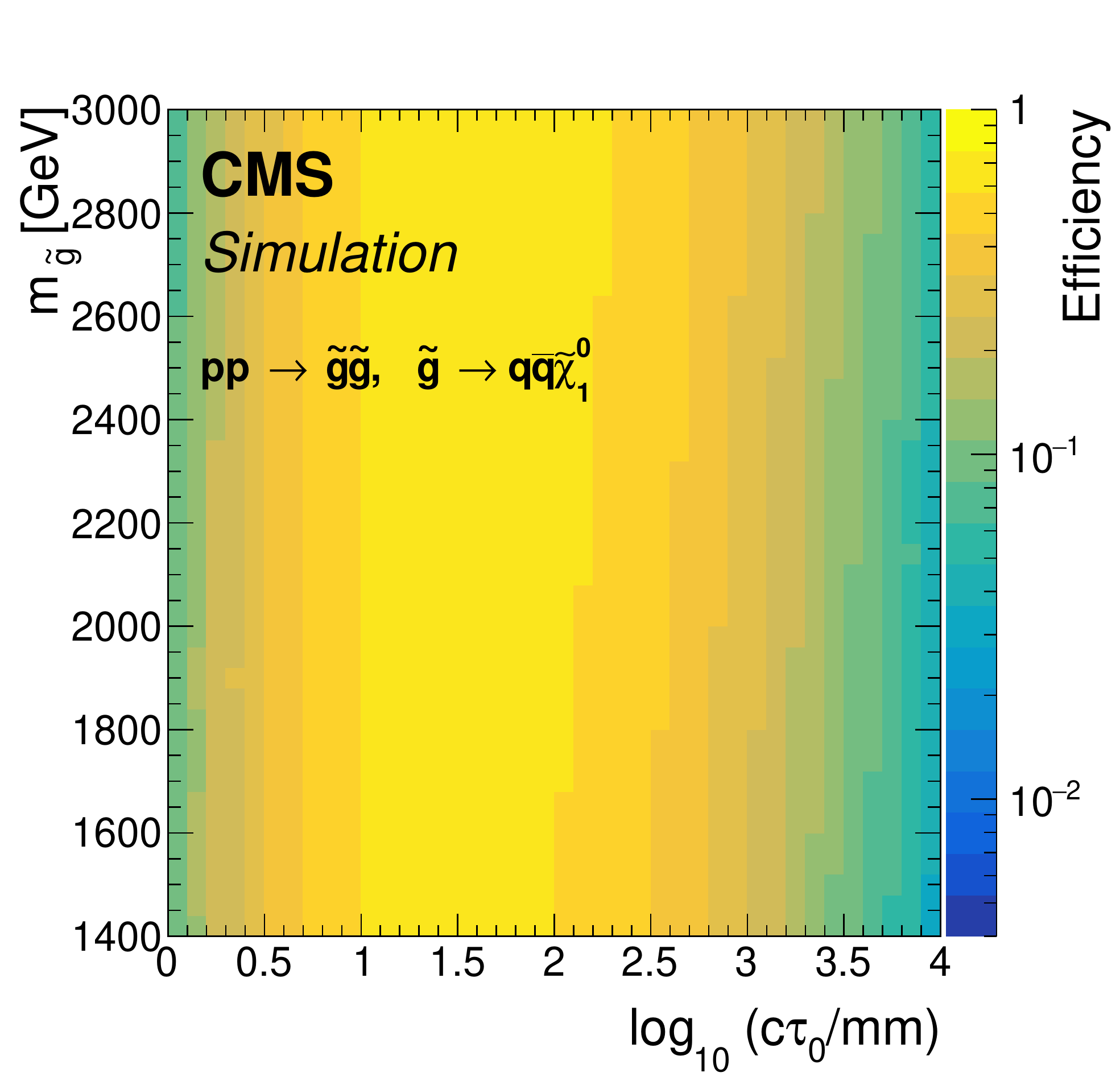}\\
\includegraphics[width=0.44\textwidth]{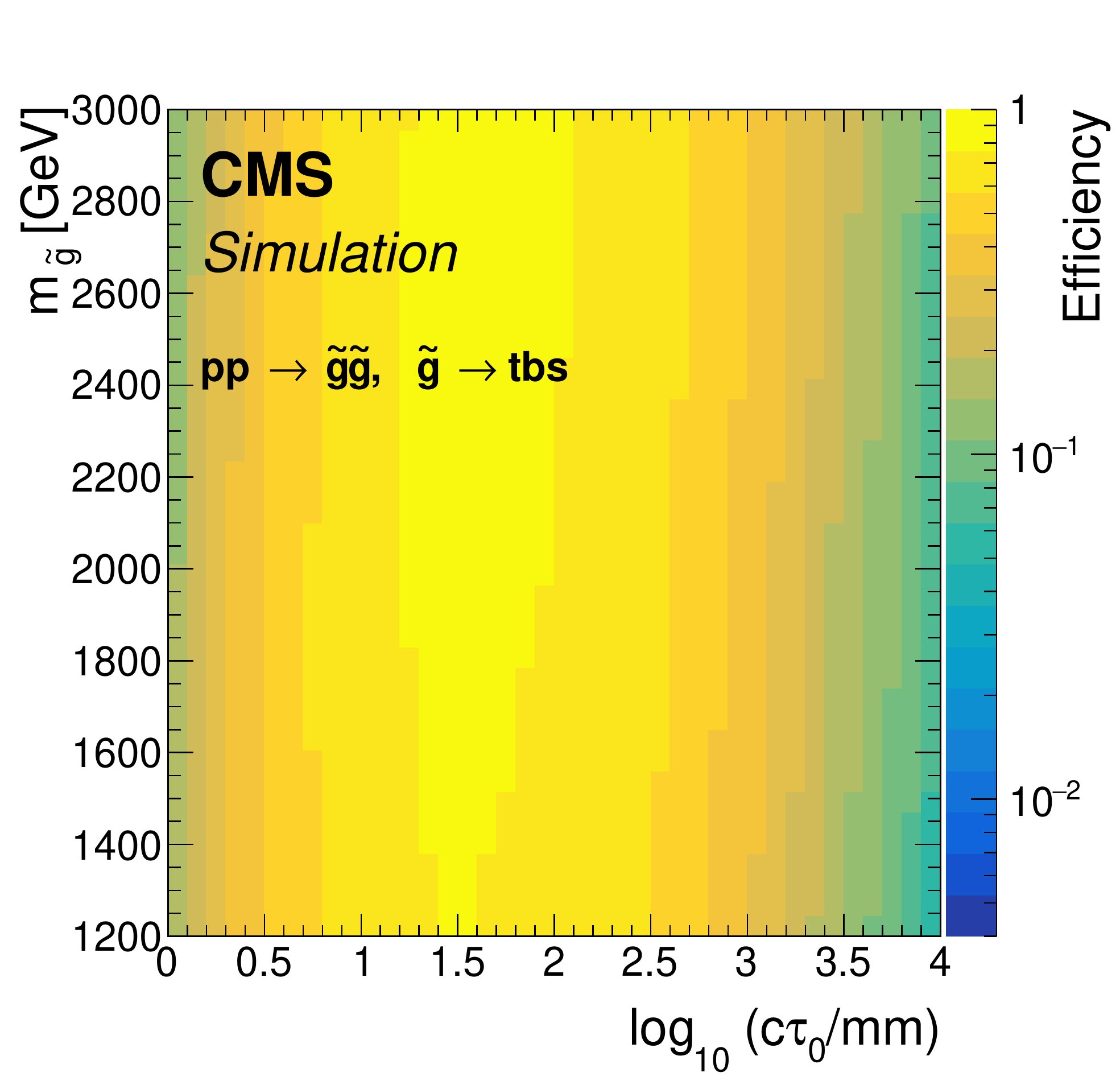}
\includegraphics[width=0.44\textwidth]{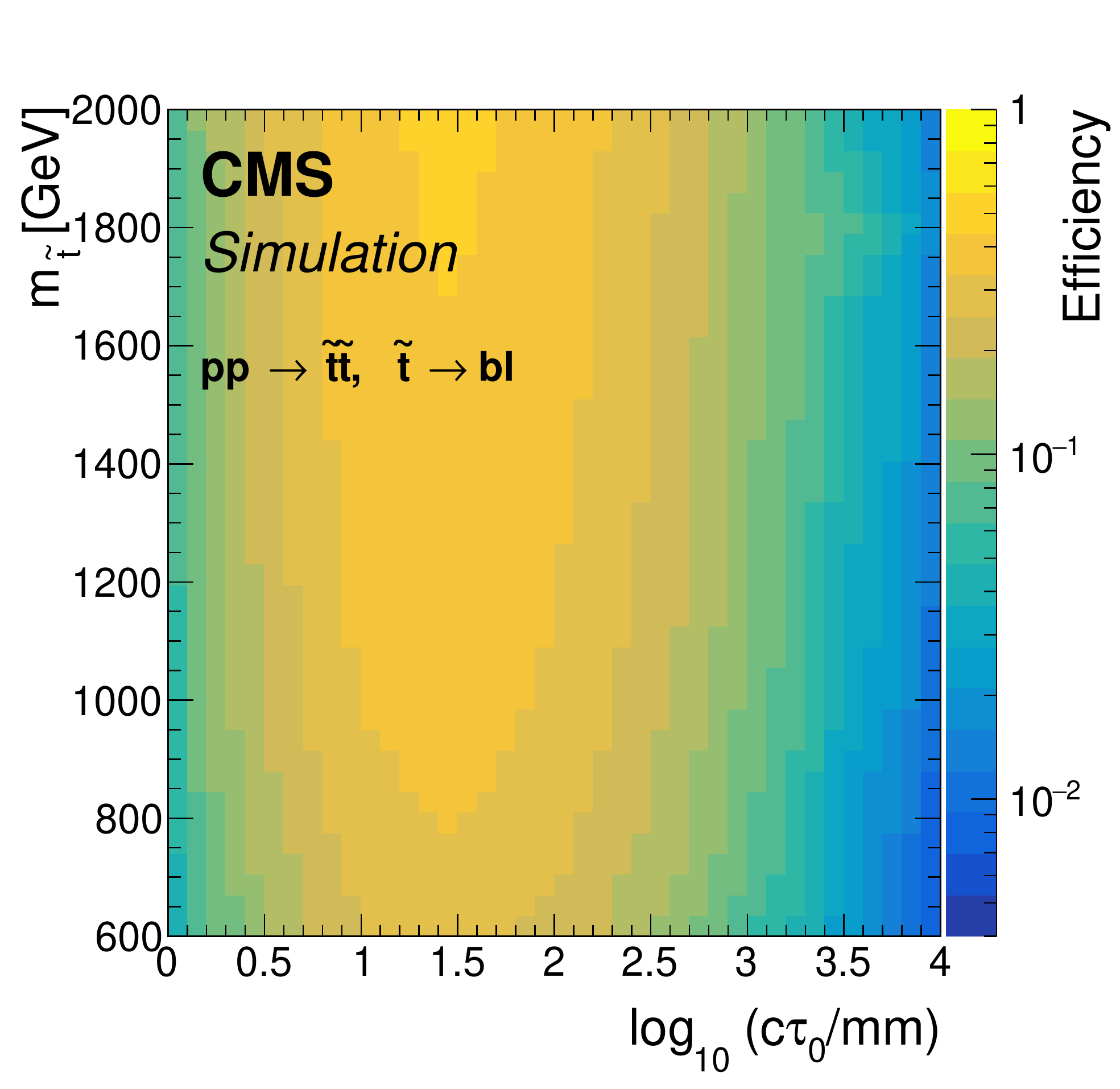}\\
\includegraphics[width=0.44\textwidth]{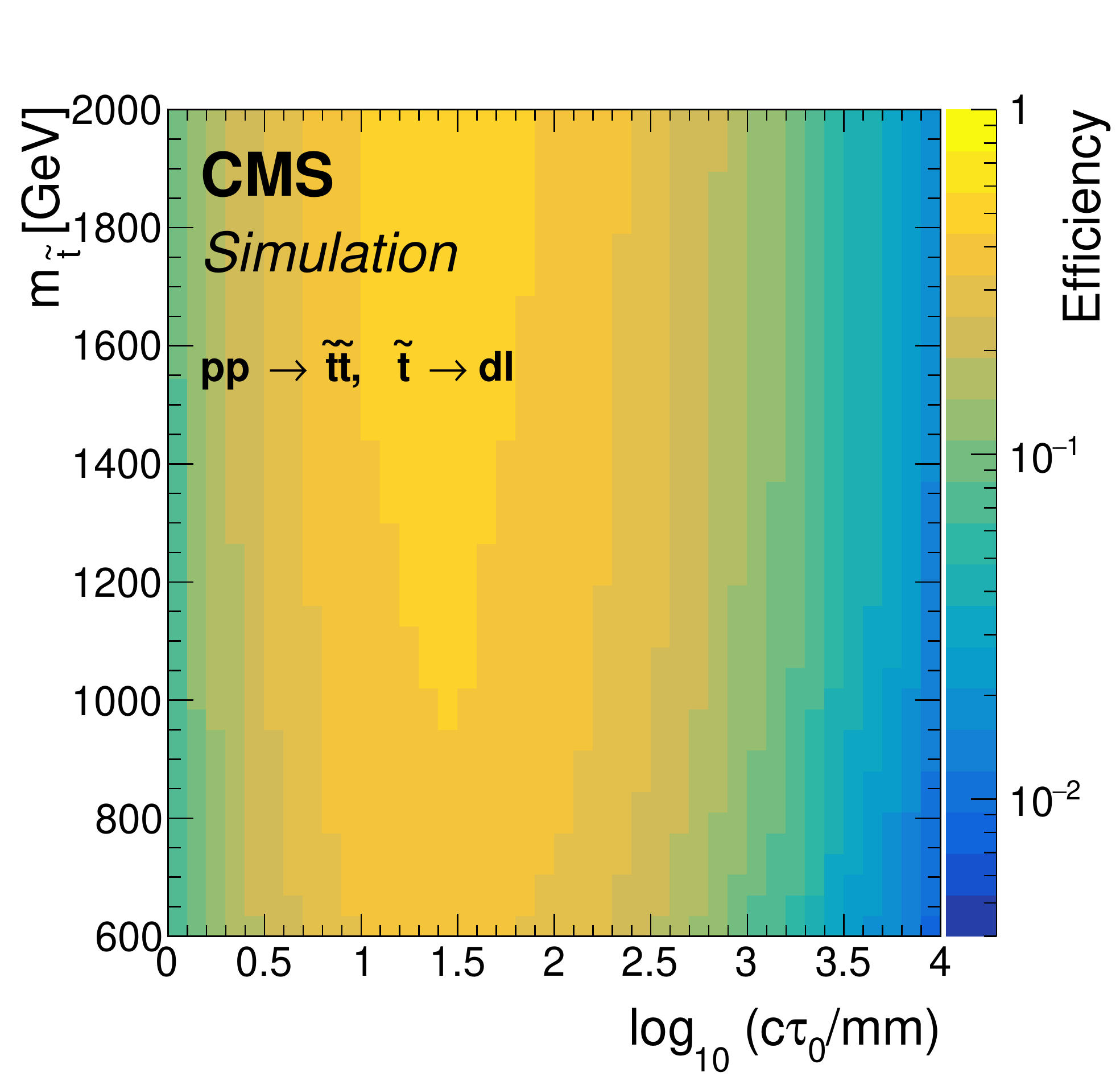}
\includegraphics[width=0.44\textwidth]{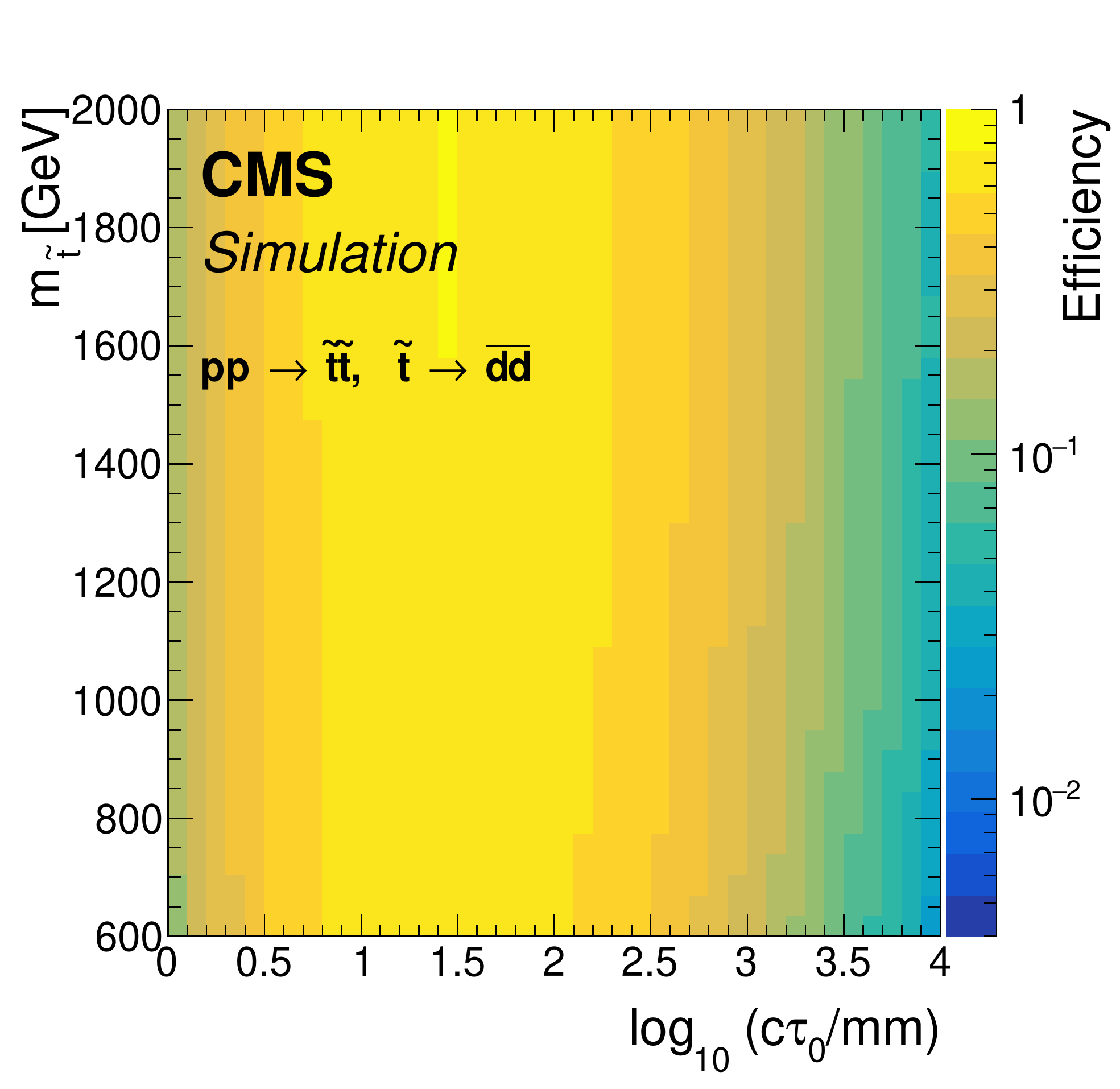}
\caption{The signal efficiencies as functions of the long-lived particle mass and mean proper decay length in the 2017 and 2018 analysis, for the $\sGlu\to\Glu\sGra$ model (upper left), the $\sGlu\to\cPq\cPaq\widetilde{\chi}^{0}_{1}$ 
model (upper right), the $\sGlu\to\PQt\PQb\cPqs$ model (middle left), the $\sTop\to\PQb\ell$ model (middle right), the $\sTop\to\PQd\ell$ model (lower left), and the $\sTop\to\cPaqd\cPaqd$ 
model (lower right).}
\label{fig: eff_maps}
\end{figure*}
\cleardoublepage \section{The CMS Collaboration \label{app:collab}}\begin{sloppypar}\hyphenpenalty=5000\widowpenalty=500\clubpenalty=5000\vskip\cmsinstskip
\textbf{Yerevan Physics Institute, Yerevan, Armenia}\\*[0pt]
A.M.~Sirunyan$^{\textrm{\dag}}$, A.~Tumasyan
\vskip\cmsinstskip
\textbf{Institut f\"{u}r Hochenergiephysik, Wien, Austria}\\*[0pt]
W.~Adam, F.~Ambrogi, T.~Bergauer, M.~Dragicevic, J.~Er\"{o}, A.~Escalante~Del~Valle, R.~Fr\"{u}hwirth\cmsAuthorMark{1}, M.~Jeitler\cmsAuthorMark{1}, N.~Krammer, L.~Lechner, D.~Liko, T.~Madlener, I.~Mikulec, F.M.~Pitters, N.~Rad, J.~Schieck\cmsAuthorMark{1}, R.~Sch\"{o}fbeck, M.~Spanring, S.~Templ, W.~Waltenberger, C.-E.~Wulz\cmsAuthorMark{1}, M.~Zarucki
% ...
\vskip\cmsinstskip
\dag: Deceased\\
1:  Also at Vienna University of Technology, Vienna, Austria\\
% ...
\end{sloppypar}
\end{document}